\documentclass[a4paper,11pt]{article}
\usepackage{jheppub} 
\usepackage{lineno}
\usepackage{macros}
\usepackage{tikz}
\usetikzlibrary{positioning,arrows.meta}
\usepackage{mathrsfs}
\usepackage{amsthm}
\usepackage{mathtools}
\usepackage[T1]{fontenc}
\usepackage[utf8]{inputenc}
\usepackage{amsmath,tikz-cd}
\usepackage{graphicx}
\usepackage{tikz-cd}
\usepackage{subcaption} 
\DeclareUnicodeCharacter{2010}{-}

\usepackage{color}
\definecolor{dark-gray}{gray}{0.20}
\definecolor{gray}{gray}{0.30}
\definecolor{light-gray}{gray}{0.80}
\definecolor{dark-red}{rgb}{0.7,0,0}
\definecolor{dark-green}{rgb}{0.1,0.4,0}
\definecolor{dark-blue}{rgb}{0.3,0.3,0.7}
\definecolor{dark-blue2}{RGB}{0, 114,178}
\definecolor{light-blue}{rgb}{0.8,0.8,1}
\definecolor{swamp}{RGB}{240, 199, 197}
\definecolor{antiquefuchsia}{rgb}{0.57, 0.36, 0.51}

\definecolor{torusedit}{RGB}{0,128,128}

\usepackage{slashed}
\usepackage{hyperref}
\usepackage{cleveref}

\usepackage{makecell}
\hypersetup{
	colorlinks=true,
	linkcolor=dark-blue2,
	citecolor=dark-blue2,
	urlcolor=dark-blue2,
	linktoc=section
}

\usepackage{tocloft}

\title{Black Hole Probes under T-duality}

\author{Matilda Delgado$^{a,b}$, }
\author{Dieter L\"ust$^{b,c}$, }
\author{Giulia Tazzoli$^b$, }
\author{Matteo Zatti$^b$ }
\affiliation[a]{Jefferson Physical Laboratory, Harvard University\\ 17 Oxford St, Cambridge, MA 02138, United States of America}
\affiliation[b]{Max-Planck-Institut f\"ur Physik (Werner-Heisenberg-Institut)\\
Boltzmannstr. 8, 85748 Garching, Germany}
\affiliation[c]{Arnold Sommerfeld Center for Theoretical Physics,\\
Ludwig-Maximilians-Universit\"at M\"unchen, 80333 M\"unchen, Germany}

\emailAdd{matildadelgado@fas.harvard.edu}
\emailAdd{luest@mpp.mpg.de}
\emailAdd{tazzoli@mpp.mpg.de}
\emailAdd{zatti@mpp.mpg.de}

\abstract{Attractor black holes are finite-size probes of infinite-distance limits in the vector-multiplet moduli space of type IIA Calabi--Yau compactifications. We ask whether the hypermultiplet moduli space admits analogous probes. T-dualizing BPS black holes along a Euclidean time circle and along a spatial circle, which realizes the temporal and spatial c-map, we construct the dual type IIB D-instantons and axionic strings, on $T^6/(\mathbb{Z}_2\times\mathbb{Z}_2)$ and on a generic Calabi--Yau threefold. The duality maps the ADM mass of the black hole to the instanton action and to the string tension, its entropy to the instanton measure through the BPS index, and its warp factor to the four-dimensional dilaton, which therefore always runs along the dual flows. The instanton profiles are null geodesics in the complexified moduli space, and their action is bounded from below by the square root of the black-hole entropy. Hence, at fixed four-dimensional dilaton, only instantons dual to small black holes can become unsuppressed. The strings are bound states of EFT strings. Those dual to finite-size black holes, all probe the same emergent fundamental-string limit, which survives quantum corrections to the hypermultiplet moduli space. Those dual to small black holes probe other emergent-string and decompactification limits and we discuss which of these limits survive quantum corrections.}

\begin{document}\emergencystretch 3em
\hypersetup{pageanchor=false}
\makeatletter
\let\old@fpheader\@fpheader
\preprint{MPP-2026-215}

\maketitle

\flushbottom

\section{Introduction and summary}

In string compactifications, the masses and couplings of the effective theory are determined by the vacuum expectation values of scalar fields. As these expectation values vary, the spectrum and interactions of the theory change, sometimes dramatically. This is particularly striking at large distances in moduli space, where the Swampland Distance Conjecture predicts the appearance of an infinite tower of parametrically light states and a corresponding breakdown of the original effective field theory description \cite{Ooguri:2006in}. It is therefore natural to ask whether finite-size physical experiments can be used to explore regions of moduli space away from the asymptotic vacuum in which they are prepared. 

Attractor mechanisms provide a natural answer to this question. Charged objects source scalar fields and generate radial flows in moduli space, with the charges acting as an effective potential for the moduli. In supersymmetric settings, these flows are highly constrained \cite{Ferrara:1995ih,Ferrara:1996dd,Ferrara:1996um,Ferrara:1997tw}: the scalar fields evolve toward charge-dependent values near the core of the object, while its tension or mass is determined by the corresponding central charge. Attractor phenomena have consequently played an important role in the Swampland program, both as a tool for relating charge spectra to the geometry of field space and as a concrete realization of trajectories toward asymptotic regions of moduli space \cite{Bonnefoy:2019nzv,Cribiori:2022nke,Cribiori:2023ffn,Calderon-Infante:2023uhz,Angius:2023xtu,Basile:2024dqq,Basile:2023blg,Calderon-Infante:2025pls,Delgado:2025crl}. 

\begin{figure}[]
    \centering
    \includegraphics[width=1\linewidth]{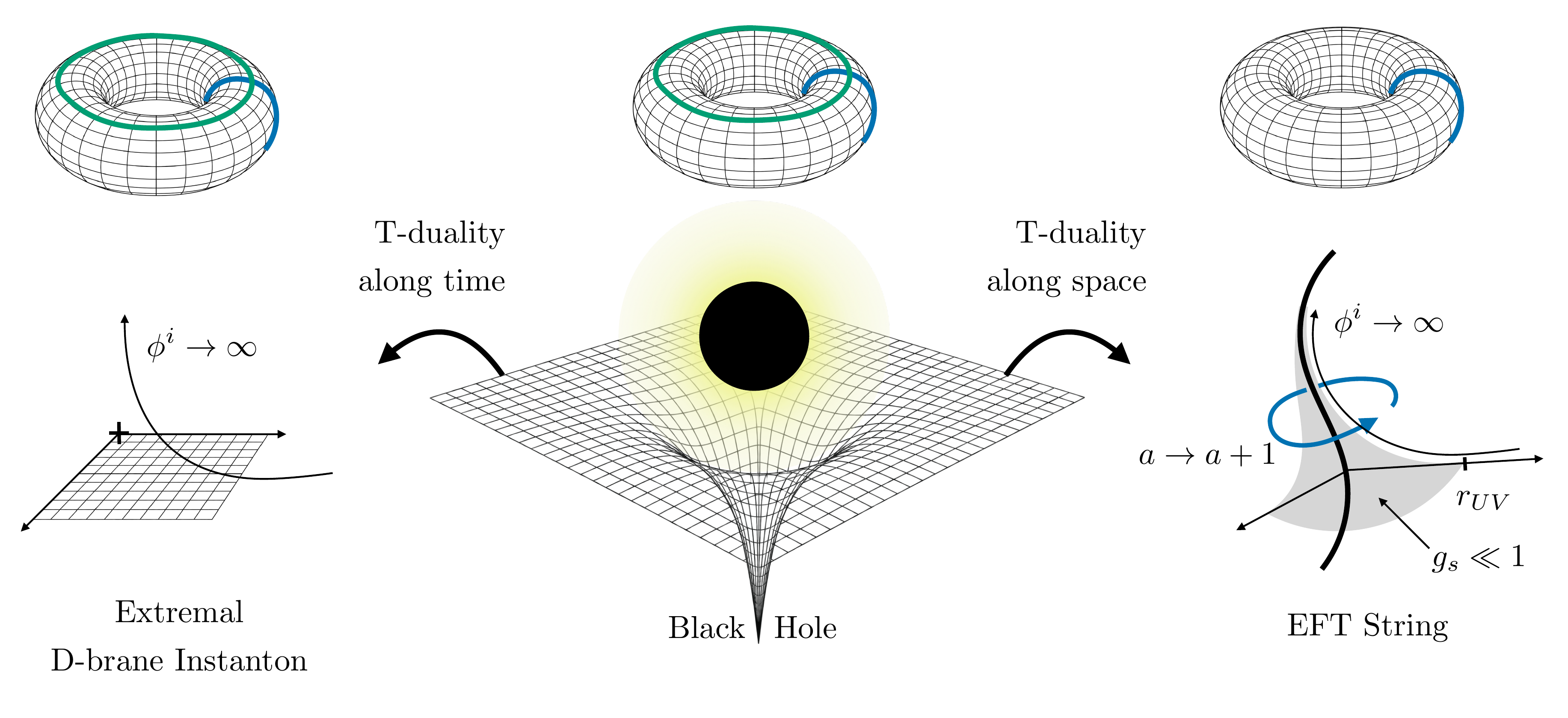}
    \caption{A depiction of how strings and D-brane instantons can be obtained from T-dualizing finite-size $4d \;\mathcal{N}=2$ black holes along space and time.}
    \label{fig:placeholder}
\end{figure}

The best-understood example is furnished by extremal black holes in four-dimensional $\mathcal{N}=2$ compactifications of string theory \cite{Ferrara:1995ih,Ferrara:1996dd,Ferrara:1996um,Behrndt:1996jn,Ferrara:1997tw,Moore:1998pn,Moore:1998zu,Strominger:1996kf,Denef:1998sv,Denef:2001xn}. Their electric and magnetic charges induce an attractor flow on the vector-multiplet moduli space, which fixes the scalars at the horizon to charge-dependent values. For suitable families of charges, these horizon values lie arbitrarily close to the boundary of moduli space. The black hole then provides a finite-size spacetime realization of an infinite field-space excursion: ultraviolet information about the underlying Calabi--Yau compactification is encoded in infrared observables such as its mass, entropy, and near-horizon geometry \cite{Delgado:2022dkz}. More generally, classical charged solutions can retain information about parametrically low ultraviolet scales, including Kaluza--Klein scales and the species scale \cite{Dvali:2007hz,Dvali:2007wp,Dvali:2009ks,Castellano:2022bvr,vandeHeisteeg:2022btw,Cribiori:2022nke}, even when these scales are not manifest in the low-energy two-derivative action \cite{Calderon-Infante:2025pls}. 

This raises the broader question of how general this mechanism is. Are all infinite-distance limits accessible through controlled spacetime probes, or is the black-hole attractor mechanism special to the vector-multiplet sector? The problem is particularly difficult for hypermultiplet moduli. Unlike the vector-multiplet metric in type IIB, which is independent of the string coupling and is exact, the hypermultiplet metric depends on the dilaton and receives both perturbative and non-perturbative corrections in $g_s$ and in $\alpha'$ (see \cite{Ooguri:1996me,deWit:1999fp, deWit:2001brd, deWit:2006gn, Robles-Llana:2006vct, RoblesLlana:2006is, Saueressig:2007dr, Alexandrov:2008gh, Alexandrov:2009zh, Alexandrov:2009qq, Pioline:2009ia, Alexandrov:2011va, Alexandrov:2013yva, Alexandrov:2014sya} for early works, and \cite{Marchesano:2019ifh,Baume:2019sry,Alvarez-Garcia:2021pxo} for recent works in the context of the swampland program). Its asymptotic regions are therefore substantially harder to analyze, and it is not evident which physical objects, if any, can generate controlled flows toward them. It is also unclear whether black holes exhaust the relevant possibilities, or whether strings and other extended objects can play an analogous role. Cosmic string solutions in 4d were in fact observed to probe infinite distance limits in a controlled setting (see \cite{Greene:1989ya,Green:1993zr} for early work and \cite{Lanza:2021udy,Marchesano:2022avb,Marchesano:2022axe,Angius:2022aeq,Grieco:2025bjy,Hassfeld:2025uoy, Grieco:2026yip,Etheredge:2026jun,Collazuol:2026xuz,Kaufmann:2026rcn} for recent work in the context of the swampland program). 

In this paper, we use T-duality and the c-map \cite{Ferrara:1989ik} to relate the attractor mechanism of four-dimensional black holes to string and instanton solutions in $\mathcal{N}=2$ Calabi--Yau compactifications. Starting from black-hole configurations whose attractor flows probe infinite-distance limits in vector-multiplet moduli space, we follow their duality images into the hypermultiplet sector and determine how the corresponding asymptotic behaviour is encoded in the T-dual solutions. Our aim is to establish in what sense these strings and instantons remain sensitive to the ultraviolet physics associated with infinite-distance limits, and whether they can be regarded as controlled physical probes of the hypermultiplet moduli space. 

We illustrate this with a specific class of solutions for a $\mathbb{T}^6/(\mathbb{Z}_2 \times \mathbb{Z}_2)$ compactification. We then construct the explicit T-dual solutions of attractor black holes in general Calabi--Yau compactifications, establishing a solution generating technique. T-duality along the Euclidean time direction yields D-brane instantons,\footnote{Throughout this paper, we perform a T-duality along the Euclidean time circle, obtained after a Wick-rotation. This relates two Euclidean supergravity theories. We are not T-dualizing along the Lorentzian time circle. } while T-duality along a spatial circle yields realizations of the EFT strings of \cite{Lanza:2021udy}. We establish how the physical quantities characterizing each solution are mapped into one another under the duality; this dictionary is summarized in table \ref{tab:dictionary}. We also show that the duality map necessarily intertwines the moduli-space flow with the spacetime geometry of the original black-hole solution. 

Upon compactifying a type IIA black-hole attractor on a circle to three dimensions, the warp factor of the four-dimensional metric induces a nontrivial radial profile for the Kaluza--Klein radion. After T-duality along the compactification circle, this radion becomes the four-dimensional type IIB dilaton. Consequently, the dual hypermultiplet flow does not merely vary the K\"ahler moduli: it necessarily also drives the type IIB dilaton. This is precisely what distinguishes probing moduli space with attractor solutions from adiabatically varying the moduli vevs, as in e.g. \cite{Marchesano:2019ifh}. In the attractor setting, the gravitational backreaction of the black hole is itself incorporated into the scalar trajectory after duality. Consequently, even when the original type IIA flow approaches an infinite-distance boundary entirely within the vector-multiplet moduli space, its type IIB dual cannot realize a purely K\"ahler infinite-distance limit in the hypermultiplet sector. The dilaton necessarily runs, because the T-dual image of the black-hole warp factor contributes directly to its profile. This restriction substantially reduces the range of asymptotic limits that can be realized by the dual solutions, as we discuss now. When discussing which limits are probed, we mostly focus on the strings, whose scalar profiles define trajectories in the moduli space of the Lorentzian theory. The profiles of the instantons, instead, define trajectories in the complexified moduli space of the Euclidean theory.

On the type IIA side, different charge scalings of finite-size black holes lead to qualitatively distinct infinite-distance regimes, including decompactification to M-theory on the Calabi--Yau threefold and emergent-string limits. We argue that, for finite-size black holes, this distinction is lost after the duality transformation. 
Indeed, since the type IIB dilaton necessarily participates in the asymptotic trajectory, the resulting limits are universally of emergent F1-string type

\begin{equation*}
\begin{tikzpicture}[
  box/.style={draw, rounded corners=3pt, align=center, minimum width=5cm,
              minimum height=1.3cm, inner sep=5pt, font=\small},
  >={Stealth[length=2.5mm]}
]
  \node[box] (A) {Finite-size BH probing\\[2pt] any VM infinite-distance limit};
  \node[box, right=1.8cm of A] (B) {EFT string bound state \\[2pt] probing F1-string limit};
  \draw[->, thick] (A) -- node[above, font=\small] {T-duality} (B);
\end{tikzpicture}
\end{equation*}
This conclusion, however, relies on the black hole having a finite-size horizon, and it is useful to distinguish two qualitatively different classes of black-hole probes. The first class consists of \emph{finite-size} black holes, whose horizon area is non-vanishing already in the two-derivative theory. For them, the attractor mechanism fixes all vector-multiplet moduli, including the axions, to finite, charge-dependent values at the horizon, and infinite-distance limits are approached by scaling the charges. Among them are the \emph{minimal} black holes of \cite{Cribiori:2022nke,Cribiori:2023ffn,Basile:2024dqq,Basile:2023blg,Calderon-Infante:2025pls}, for which only some of the charges are taken large while the others remain of order one. Minimal black holes are large in Planck units but of species-scale size, so their near-horizon region is not under parametric control in the two-derivative theory. In generic Calabi--Yau compactifications, their entropy also receives order-one corrections from higher-curvature terms. The statements above apply to this first class. Since the T-dual flows inherit the attractor values of the K\"ahler moduli at the core, the only field that runs to infinite distance on the type IIB side is the dilaton, and all these black holes are mapped to probes of the same weak-coupling limit, with an emergent fundamental string, which, as shown in section \ref{sec:QCtoF1}, survives quantum corrections.

The second class consists of \emph{small} black holes, which do not carry enough charges to develop a classical horizon. Their attractor flows do not fix all the moduli, some of which run off to infinite distance at a singular core whose size is set by the UV cutoff. Their T-dual images are qualitatively different, and can classically probe infinite-distance limits of the hypermultiplet moduli space other than the weak-coupling limit, namely emergent-string limits in which the T-dual string itself becomes the critical string, and decompactification limits. Whether these classical limits survive quantum corrections has to be decided case by case: in some cases the corrections can be computed and the limits shown to survive, while in others the answer requires a more complete knowledge of the stringy corrections to the hypermultiplet moduli space

\begin{equation*}
\begin{tikzpicture}[
  box/.style={draw, rounded corners=3pt, align=center, minimum width=5cm,
              minimum height=1.3cm, inner sep=5pt, font=\small},
  >={Stealth[length=2.5mm]}
]
  \node[box] (A) {Small BH probing\\[2pt] VM infinite-distance limit};
  \node[box, right=1.8cm of A] (B) {String probing \emph{classical} \\[2pt] HM infinite-distance limit \\[2pt] $+$ quantum corrections};
  \draw[->, thick] (A) -- node[above, font=\small] {T-duality} (B);
\end{tikzpicture}
\end{equation*}

We now summarize our main results. In the instanton side, the Euclidean Einstein-frame metric is flat, the four-dimensional dilaton runs to strong coupling at the core, and the K\"ahler moduli are attracted to the horizon values of the dual black hole. The RR axions, T-dual to the electrostatic potentials of the black hole, acquire imaginary profiles, so that the solution is a null geodesic in the complexified moduli space. The instanton action is set by the ADM mass of the dual black hole and is bounded from below by the square root of its entropy, while the entropy itself reappears in the instanton measure correcting the hypermultiplet action. Hence, only instantons dual to small black holes can become unsuppressed along infinite-distance limits at fixed four-dimensional dilaton. Near the core of the instantons, NS5-instantons and mutually non-local D-instantons become unsuppressed, and so we lose complete control over the description. We explore how instanton corrections can modify trajectories in complexified moduli space in Appendix \ref{sec:appendixB}.

In the string side, the solutions are axionic strings, bound states of the EFT strings of \cite{Lanza:2021udy} in the hypermultiplet sector, around which the RR axions wind with monodromies given by the charges, while their saxionic partners run logarithmically to infinite distance at the core. For finite-size black holes, the K\"ahler moduli freeze at the core and the four-dimensional dilaton drives an emergent fundamental-string limit, which survives quantum corrections in a generic Calabi--Yau. For small black holes, other limits appear: on $T^6/(\mathbb{Z}_2\times\mathbb{Z}_2)$ emergent D-brane string limits and decompactifications, which also survive quantum corrections, while on a generic Calabi--Yau most of these flows leave the large-volume regime, and their fate requires a better knowledge of the stringy corrections to the hypermultiplet moduli space.  To our knowledge, EFT strings in the hypermultiplet sector of $\mathcal N=2$ compactifications have not been studied systematically before.\footnote{On the truncation, some of our flows coincide with D1-, D3-, D5- and D7-string flows of the $\mathcal N=1$ orientifolds of \cite{Lanza:2021udy}, see appendix~\ref{app:singleDbranes}; in type IIA on the quintic, the F1 and a D4-string were identified as EFT strings in \cite[sec.~4.4]{Etheredge:2026jun}.} The hypermultiplet sector contains D-brane, world-sheet and NS5-instantons, which are not electric-magnetic dual to the strings and can become unsuppressed along their flows, so that we pay particular attention to quantum corrections. Moreover, the strings dual to finite-size black holes are bound states of EFT strings with the properties recently found in \cite{Etheredge:2026jun}: the F1 tower has scaling weight $w=\frac12$ along each elementary flow, and these add up to $w=2$ along the bound-state flow, where the F1 is an emergent critical string parametrically lighter than the EFT string itself.

This paper is organized as follows. In section \ref{sec:torus}, we illustrate our results in type II compactifications on $T^6/(\mathbb{Z}_2\times \mathbb{Z}_2)$, which we use as a toy model.  We begin in section \ref{sec:BHprobes} by discussing finite-size, minimal and small black holes. We then review the c-map as a T-duality on $T^6/(\mathbb{Z}_2\times \mathbb{Z}_2)$ in sections \ref{sec:cmap} and \ref{sec:runningcmap}. This allows us to study the T-dual counterparts of the black holes discussed above: instantons in section \ref{sec:instantonstorus}, and strings in section \ref{sec:EFTstringtorus}. In this context, we compare our instantons with the imaginary distance bound of \cite{Maldacena:2026jqd} and with recent bounds on instanton actions \cite{ArkaniHamed:2006dz, Reece:2025thc, Reece:2026hmp, Artime:2026kfq}. In section \ref{sec:CY}, we follow a similar structure, but constructing solutions associated with black holes and their T-duals in general Calabi--Yau compactifications. Both section \ref{sec:torus} and \ref{sec:CY} focus on classical trajectories and infinite distance limits. However, quantum corrections may modify this picture: on the type IIB side, the hypermultiplet metric receives perturbative and non-perturbative quantum corrections. The latter can modify both the solutions and the infinite-distance limits they probe. We analyze them in section \ref{sec:quantumCorrections}, mostly for the toroidal orbifold, and conclude in section \ref{sec:conclusions}. In appendix~\ref{app:singleDbranes}, we construct in detail the classical flows of the strings T-dual to small black holes made of a single type of D-brane on a generic Calabi--Yau threefold. In appendix~\ref{sec:appendixB}, we take an exploratory approach to possible deflections of flows in the complexified moduli space by integrating in instantons.

\begin{table}[t]
    \centering
    \renewcommand{\arraystretch}{1.4}
    \begin{tabular}{l|c|c|c}
      & Black holes & Instantons & Strings  \\ \hline
      Charges & $\{q_I, p^I\}$ & $\{q_I, p^I\}$ & $\{q_I, p^I\}$ \\
      Mass/action/tension & $\dfrac{M_{\rm ADM}}{M_p}$ & $S_{\rm inst} = \dfrac{\pi}{g^\infty_{4,s}}\dfrac{M_{\rm ADM}}{M_p}$ & $\dfrac{\mathcal{T}(r_0)}{M_p^2}=\dfrac{g_{4,s}(r_0)}{2}\dfrac{M_{\rm ADM}}{M_p}$ \\
    \end{tabular}
    \caption{Dictionary between the physical quantities characterizing the three T-dual descriptions of the same charged configuration, see eqs.~\eqref{eq:actionADMmass} and \eqref{eq:TensionADMmass}, where $g^\infty_{4,s}$ and $g_{4,s}(r_0)$ are the type IIB four-dimensional string couplings at infinity and at the reference radius $r_0$.}
    \label{tab:dictionary}
\end{table}

\section{Toy model: type II string theory on $T^6/(\mathbb{Z}_2\times \mathbb{Z}_2)$}\label{sec:torus}

In this section, we illustrate the T-duality between black holes, instantons and strings in the context of $\mathcal{N}=2$ supergravity. To do so, we select the simplest possible setting: type II string theory compactified on the STU toroidal orbifold. We consider the four-charge D0--D4--D4--D4 black hole, which, depending on its charges, is a finite-size, a minimal, or a small black hole (section \ref{sec:BHprobes}). We briefly review how black hole attractors can be used as probes of infinite-distance limits in the moduli space (section \ref{sec:blackholestorus}). We then T-dualize the brane content along Euclidean time and along a spatial direction obtaining the D$(-1)$--ED3--ED3--ED3 instantons (section \ref{sec:instantonstorus}) and D1--D5--D5--D5 strings (section \ref{sec:EFTstringtorus}), respectively. The solutions of type IIB equations of motion for such configurations show that the spacetime profile of type IIA vector-multiplets is replaced by a spacetime profile for the IIB hypermultiplets. We can interpret this feature of the T-dual supergravity configurations as a consequence of the duality between the type IIA vector-multiplet sector and the type IIB hypermultiplet sector  when compactified on a circle times a threefold (section \ref{sec:cmap}). Such duality is realized as a combination of the c-map and T-duality and we will use it to establish a solution generating technique which explicitly maps type IIA vector multiplet attractor flows into type IIB hypermultiplet attractor flows (in section \ref{sec:runningcmap} we describe the main idea, leaving the full proof to section \ref{sec:CY}). With such result in mind, we will present the content of this section in four-dimensional $\mathcal{N}=2$ language and we use the notion of T-duality and c-map of a black hole interchangeably. We keep track of the structure inherited by instantons and strings from black holes attractors analyzing three things: how the scalar fields, including the axions, behave at the core of each solution, which physical quantities are mapped into one another by the duality, which are the infinite-distance limits probed by the new solutions. We focus on the features which are not special to the torus but valid in a generic Calabi--Yau compactification. In particular, we study the relation between black-hole entropy, the instanton action, and the string tension, and we analyze whether the observables of the type IIB configurations retain some UV sensitivity due to the duality with black holes. All the analysis in this section is classical, we will check in section \ref{sec:quantumCorrections} the effect of quantum corrections on the infinite-distance limits.

\subsection{Black hole probes}\label{sec:BHprobes} 

We start selecting a system that allows us to easily build four-dimensional black holes out of bound-states of D-branes. To do so we might consider Type IIA string theory on $T^6$, but the resulting four-dimensional theory is $\mathcal{N}=8$ supergravity. To make contact with $\mathcal{N}=2$ compactifications, and with the Calabi--Yau case of section \ref{sec:CY}, we consider the STU model. The internal manifold is an orbifold projection of the torus and supersymmetry is reduced to $\mathcal{N}=2$. In this setup we can consider bound-states of D0 and D4 branes. Finite-size four-dimensional black holes can be constructed wrapping D4 branes on each one of the three four-cycles. We indicate such brane configuration as D0--D4--D4--D4. 

\paragraph{The STU model.}  We take type IIA string theory on $T^6=T^2_1\times T^2_2\times T^2_3$ and keep only the fields that are invariant under the $\mathbb{Z}_2\times\mathbb{Z}_2$ that reflects pairs of two-tori \cite{Vafa:1994rv}.\footnote{The orbifold action is $g:(z^1,z^2, z^3)\to ( z^1, - z^2, - z^3)$ and $h:(z^1,z^2, z^3)\to ( -z^1, - z^2, z^3)$, with $z^i$ the complex coordinates on the two-torus $T^2_i$.\label{fn:Footnote1}} This is a consistent truncation to a four-dimensional $\mathcal{N}=2$ supergravity, the STU model coupled to four hypermultiplets \cite{Duff:1995sm}, which can be described in the same language as a Calabi--Yau compactification with $h^{1,1}=h^{2,1}=3$.\footnote{There are also twisted cycles in the resolved Calabi--Yau. They will play no role in this work. } The relevant two-cycles are the three two-tori $T^2_i$, $i=1,2,3$, whose dual four-cycles are $T^2_j\times T^2_k$, with $(i,j,k)$ a permutation of $(1,2,3)$. The only non-vanishing triple intersection number is $\mathcal{K}_{123}=1$. The three hypermultiplets contain the complex-structure moduli of the three two-tori and the periods of the RR 3-form over the eight invariant three-cycles, together with the universal hypermultiplet, which contains the dilaton. The vector multiplet sector is fully characterized by the prepotential
\begin{equation}\label{eq:STUprepotential}
    \mathcal{F}=-\frac{X^1X^2X^3}{X^0}\,,
\end{equation}
where $X^I$, $I=0,\dots,3$, are the projective coordinates related to the complexified K\"ahler moduli via $z^i=X^i/X^0$. The saxions $t^i ={\rm Im}z^i =   l_s^{-2}\int _{T^2_i} J$ measure the volumes of the two-tori in string units; $l_s\equiv M_s^{-1}$ is the string length and $J$ the K\"ahler form. The volume of $T^6$ in string units is then $\mathcal{V}=t^1t^2t^3$. There are four gauge fields $A^I$ arising from the RR 1-form and from the RR 3-form on the $T^2_i$. Truncating the three hypermultiplets and the $B$-field axions  we obtain the four-dimensional action
\begin{equation} 
S_4
=\frac{M_p^2}{2}\int d^4x\,\sqrt{-g}\Bigg[R
-\frac{1}{2\Phi^2}(\partial \Phi)^2
-\sum_{i=1}^3\frac{1}{2(t^i)^2}(\partial t^i)^2 -\frac{1}{2\cdot 2!}\,\sum_{I=0}^3 f_I |F^I|^2
\Bigg]\,,
\label{eq:BH-4d-action}
\end{equation}
where $|F_p|^2\equiv F_{\mu_1\dots\mu_p}F^{\mu_1\dots\mu_p}$, $\Phi=e^{-2 \varphi_4}$ with $\varphi_4$ the four-dimensional dilaton, which also gives the four-dimensional string coupling $e^{\varphi_4}=g_{4,s}$, and it is related to the ten dimensional one by $e^{\phi}=g_s=\mathcal{V}^{1/2}g_{4,s}$. The gauge kinetic functions are
\begin{equation}
f_0=t^1t^2t^3,
\quad
f_1=\frac{t^2t^3}{t^1},
\quad
f_2=\frac{t^1t^3}{t^2},
\quad
f_3=\frac{t^1t^2}{t^3}.
\label{eq:gauge-kinetic-STU}
\end{equation}
$M_p$ is the reduced four-dimensional Planck mass, which is related to the unreduced Planck mass by $m_p^2 = G_N^{-1} = 8\pi M_p^2$, where $G_N$ is the four-dimensional Newton constant. In the following, we will use the Planck length in terms of the reduced Planck mass, i.e. $l_p\equiv M_p^{-1}$, so that $8\pi G_N=l_p^2$. 

\paragraph{The D0--D4--D4--D4 black hole.} The D0--D4--D4--D4 configuration in ten dimensions is the one studied in \cite{Behrndt:1997ch}.
After compactifying on $T^6$, we obtain that the equations of motion of \eqref{eq:BH-4d-action} are solved by the static Israel–Wilson–Perj\'es (IWP) metric
\begin{equation}
ds^2 = -e^{2U} dt^2 + e^{-2U}(dr^2+r^2d\Omega_2^2),
\qquad
e^{-4U} = H_0H^1H^2H^3,
\label{eq:4dMetricBH}
\end{equation}
together with a constant four-dimensional dilaton $\Phi=1$, the moduli
\begin{equation}\label{eq:moduli}
t^1 = \sqrt{\frac{H_0H^1}{H^2H^3}},
\qquad
t^2 = \sqrt{\frac{H_0H^2}{H^1H^3}},
\qquad
t^3 = \sqrt{\frac{H_0H^3}{H^1H^2}},
\end{equation}
and the field strengths
\begin{equation} \label{eq:GaugeFields}
    F^0=d(H_0^{-1})\wedge dt \,, \qquad F^i=f_i^{-1}\star\big(d(1/H^i)\wedge dt\big) \,, \qquad i=1,2,3 \,.
\end{equation}
The $H$ functions are harmonic in $\mathbb{R}^3$
\begin{equation}
    H_0 = h_0 + \frac{q_0 l_p}{r} \,, \qquad H^i=h^i+\frac{p^i l_p}{r} \,,
\end{equation}
The harmonic functions have poles at $r=0$, which is the location of the horizon, and their residues encodes the D-brane charges. This is the standard extremal four-charge black hole of the STU model, in a duality frame with one electric and three magnetic charges. The constants $h$ are not independent but fixed by asymptotic flatness\footnote{Requiring $e^{-4U}\to1$ as $r\to \infty$, imposes $h_0h^1h^2h^3=1$.} and the asymptotic values of the moduli $(t^i)_\infty$ through \eqref{eq:moduli}.

The ADM mass can be read off from the large-$r$ expansion of the metric
\begin{equation}\label{eq:4dMetricBHADM}
    M_{\rm ADM}=2\pi\Big(\frac{q_0 }{h_0}+\sum_i\frac{p^i}{h^i}\Big)M_p\,.
\end{equation}
The Bekenstein-Hawking entropy of this black hole is instead controlled by the product of the charges. Using that the horizon area is $A_h=4\pi\lim_{r\to0}r^2e^{-2U}$, one finds
\begin{equation}\label{eq:firstentropy}
    S_{BH}=\frac{A_h}{4G_N}=8\pi^2\sqrt{q_0p^1p^2p^3} = 2\pi \sqrt{N_0 N^1 N^2 N^3}\,,
\end{equation}
where we  identified the integer-normalized charges with $N_I = q_I(2\sqrt\pi)$ and $q_I\equiv(q_0,p^i)$.

\subsubsection{Black holes probing the boundary of moduli space }\label{sec:blackholestorus}
Suppose we want to use such a black hole to push the moduli $t^i$ to the boundary of moduli space at its horizon. The values of the moduli at the horizon are obtained by evaluating \eqref{eq:moduli} at $r\to0$
\begin{equation}\label{eq:attractedvalue}
    (t^1)_h= \sqrt{\frac{q_0 p^1}{p^2p^3}}\,,\qquad (t^2)_h= \sqrt{\frac{q_0 p^2}{p^1p^3}}\,,\qquad (t^3)_h= \sqrt{\frac{q_0 p^3}{p^1p^2}}\,.
\end{equation}
Notice that they are independent of the asymptotic values of the scalar fields $(t^i)_\infty$. This is a consequence of the fact that the scalar fields obey first order differential equations called attractor equations. In ungauged four-dimensional $\mathcal{N}=2$ supergravity, the scalar profile of a static, spherically symmetric BPS black hole in the vector-multiplet moduli space satisfies
\begin{equation}
\dot U = -e^U |Z|,
\qquad
\dot z^i
=
-2e^U g^{i\bar j}\,\partial_{\bar j}|Z|,
\end{equation}
with $D_i Z = 0$ at a regular supersymmetric attractor;  $e^{2U}$ is the warping factor of the IWP black hole metric, $Z$ is the central charge of the background and $g_{i\bar j}$ the metric of the K\"ahler moduli. $(t^i)_h$ is then called the attractor value of the moduli and the spacetime radial profile of $t^i$ is called the attractor flow. Notice that the trajectory of the complexified K\"ahler moduli $z^i$ is not a geodesic of the moduli space. They experience a charge-dependent force corresponding to a radial potential induced by the black hole
\begin{equation}
D_\tau \dot z^i
=
e^{2U} g^{i\bar j}\,\partial_{\bar j} V_{\mathrm{BH}},
\qquad
 V_{\mathrm{BH}}
=
|Z|^2
+ g^{i\bar j} D_i Z\,\overline{D_j Z} \,.
\end{equation}
There are two ways to make the attractor values $(t^i)_h$ large, and thereby drag (some of) the scalars towards the boundary of moduli space. The first is to make some of the charges large; the smallest black holes of this type are the so-called \emph{minimal} black holes. The second is to set some of the charges to zero, so that the black hole becomes singular, i.e.\ a \emph{small} black hole. We now discuss these two options in turn. 

\paragraph{Minimal black holes.} By choosing some of the charges large, for instance $q_0$, we can push the horizon values of the moduli towards the boundary of moduli space. There is thus an infinite family of black-hole solutions whose horizon moduli approach the boundary of moduli space ever more closely as the charges are increased. For fixed boundary conditions at infinity, the larger the charges, the steeper the attractor flow and the farther the moduli are dragged towards the boundary. Of course, this also makes the horizon very large in Planck units, cf.\ \eqref{eq:firstentropy}. The smallest black holes in such a family are obtained by keeping the remaining charges of order one. These are the so-called \emph{minimal black holes} \cite{Cribiori:2022nke,Cribiori:2023ffn}, which were argued to be at least species-scale size in \cite{Calderon-Infante:2025pls}.\footnote{See \cite{Herraez:2024kux, Herraez:2025clp, Aparici:2026svo} for their use in the context of  for the black hole--tower correspondence.} Indeed, as one approaches the boundary of moduli space, UV length scales become exponentially large in Planck units, as per the Distance Conjecture \cite{Ooguri:2006in}. Hence, despite being large in Planck units, minimal black holes can be of order one in units of the UV length scales of the EFT. In \cite{Calderon-Infante:2025pls}, minimal black holes were shown to be of species-scale size in decompactification and emergent-string limits, and of KK-scale size when D6 charges are turned on. The fact that a minimal black hole with D6 charge can not be species-scale size can be understood in terms of a topological obstruction in the M-theory uplift. At the attractor locus the M-theory circle and the horizon $S^2$ factor combines in a Lens space $S^3/\mathbb{Z}_{p^0}$, with $p^0$ the amount of D6 charge.\footnote{See appendix B of \cite{Castellano:2025ljk} for more details.} Thus, the size of the smallest black hole in a family with D6 charge can not be smaller than the KK scale set by the M-theory circle. This suggests a refinement of the notion of minimal black holes. A species-scale size minimal black hole can be produced only as the smallest black hole of a family without topological obstructions. 

As an example, consider the large-volume limit at fixed four-dimensional dilaton, i.e.\ the analogue of the type IV limit of Calabi--Yau compactifications \cite{Corvilain:2018lgw}, for the family of D0--D4--D4--D4 black holes described above. The minimal classical black hole is obtained by taking the D0 charge large while keeping the D4 charges of order one: $q_0 \gg p^i$, with $p^1\sim p^2\sim p^3 \sim p \sim \mathcal{O}(1)$. The attractor values then scale as $t^1\sim t^2 \sim t^3\sim t \sim \sqrt{q_0}$, and the entropy as $S_{BH} \sim \sqrt{q_0} \sim t$. In this limit, which decompactifies to five dimensions, the species scale is the five-dimensional Planck scale $\Lambda_{\rm sp}/M_p\sim t^{-1/2}$,\footnote{At fixed four-dimensional dilaton, the D0-brane mass scales as $m_{\rm D0}/M_p\sim \mathcal{V}^{-1/2}\sim t^{-3/2}$, and for a decompactification of one dimension $\Lambda_{\rm sp}/M_p\sim (m_{D0}/M_p)^{1/3}$.} so that
\begin{equation}
    S_{BH} \sim \sqrt{q_0} \sim t \sim \Big(\frac{M_p}{\Lambda_{\rm sp}} \Big)^2\,,
\end{equation}
i.e.\ the minimal black hole is indeed of species-scale size in the five-dimensional decompactification limit. In ten-dimensional terms, the type IIA string coupling grows as $g_s\sim\mathcal{V}^{1/2}\sim t^{3/2}$, where $g_s$ is the ten-dimensional string coupling, and the extra dimension is the M-theory circle, whose KK modes are the D0-branes. Taking instead $q_0\sim p^1\gg p^2\sim p^3\sim\mathcal{O}(1)$, only $t^1$ grows, and the minimal black hole probes an emergent-string limit \cite{Calderon-Infante:2025pls}, cf.\ the small D0--D4 black hole discussed below.

Since these black holes are of a size comparable to the species length, their classical description should not be trusted all the way to the horizon: in this regime, the solution probes physics at, or beyond, the UV cutoff of the four-dimensional EFT. This is the string-theoretic realization of the well-known fact that, unless \emph{all} black-hole charges are made large, higher-derivative corrections need not be small at the horizon. Indeed, the standard large-charge expansion is defined by a uniform scaling of all charges with a common parameter, keeping the charge ratios fixed \cite{Mohaupt:2000mj,Banerjee:2010qc}. A limit in which only a subset of the charges becomes parametrically large, while the others remain \(\mathcal{O}(1)\), lies outside this expansion, and higher-derivative corrections are then not guaranteed to be parametrically suppressed. How severe this is depends on the compactification and the specific limit probed. In a generic Calabi--Yau threefold, the supersymmetric four-derivative correction shifts the entropy by a term proportional to $c_{2,i}p^i$ \cite{Maldacena:1997de,LopesCardoso:1998tkj}, which is of the same order as the two-derivative entropy for minimal black holes. They are then \emph{bad saddles} of the two-derivative action, whose entropy is modified at order one. However, although the classical solution cannot be trusted near the horizon, the microscopic degeneracy of minimal black holes can still be under control, and its fate in the infinite-distance limit can often be inferred from the nature of the limit itself. Microscopically, the black hole is realized as a bound state of D-branes wrapping internal cycles, and approaching the infinite-distance boundary takes us to a regime in which this brane configuration is more naturally described in terms of different variables. In a decompactification limit, for example, the four-dimensional black hole may lift to a higher-dimensional black object. In that case, even though the perturbative expansion of the four-dimensional entropy breaks down, the higher-derivative effects can be diluted once the relevant non-perturbative corrections are resummed, so that the classical scaling is preserved. This has been shown explicitly for a decompactification along the M-theory circle \cite{Castellano:2025ljk, Castellano:2025yur, Castellano:2025rvn, Castellano:2026ojb, Zatti:2026fso}. Alternatively, the limit may take us to a different duality frame, in which the same microscopic brane configuration is no longer naturally interpreted as a black hole/brane at all. That every infinite-distance limit is either a decompactification limit or an emergent weakly coupled string limit is precisely the content of the Emergent String Conjecture \cite{Lee:2019wij}.

\paragraph{Small black holes.} Equation \eqref{eq:attractedvalue} suggests a second way to push the moduli to the boundary of moduli space: sending some of the charges to zero. Since the charges are quantized, this cannot be done continuously. One can, however, solve the attractor equations with a smaller set of non-vanishing charges, and obtain a black hole whose horizon area vanishes classically (see \cite{Dabholkar:2012zz} and references therein). A simple example is the D0-brane of ten-dimensional type IIA, but all small black holes share the same features: their horizon sits at a curvature singularity and some scalars run off to infinite distance as one approaches the singular core (see \cite{Hamada:2021yxy,Angius:2022aeq,Angius:2023xtu} for other uses of small black holes to explore swampland constraints). Although the scalars travel an infinite distance, the singularity lies at finite proper distance, so that small black holes realize the local dynamical cobordisms, or end-of-the-world branes, of \cite{Angius:2022aeq}.

Within the D0--D4--D4--D4 system, small black holes are obtained by switching off some of the charges, i.e.\ by setting the corresponding harmonic functions to constants, so that $q_0p^1p^2p^3=0$. Up to permutations of the two-tori, this leaves six cases: the D0, D4, D0--D4, D4--D4, D0--D4--D4 and D4--D4--D4 black holes. Since the four-dimensional type IIA dilaton is constant, the ten-dimensional coupling follows the volume, $g_s\propto\mathcal{V}^{1/2}$, and the behaviour at the core follows from \eqref{eq:moduli}, with each charged harmonic function growing like $1/r$. Taking this into account, the six cases probe only two kinds of infinite-distance limits summarized in table \ref{tab:BHsmall}:
\begin{table}[t]
    \centering
    \small
    \renewcommand{\arraystretch}{1.3}
    \setlength{\tabcolsep}{4pt}
    \begin{tabular}{l|c|c|c|c|l}
      Black hole & $g_{4,s}$ & $g_s$ & $(t^1,t^2,t^3)$ & lightest, $m^2/M_p^2$ & limit \\ \hline
      D0 & const. & $r^{-3/4}$ & $(r^{-1/2},r^{-1/2},r^{-1/2})$ & D0: $r^{3/2}$ & decomp.\ to 5d \\
      D4 & const. & $r^{1/4}$ & $(r^{-1/2},r^{1/2},r^{1/2})$ & D4 on $T^2_2 \times T^2_3$: $r^{3/2}$  & decomp.\ to 5d \\
      D0--D4 & const. & $r^{-1/2}$ & $(r^{-1},1,1)$ & D0, D2, D4, KK, NS5: $r$ & emergent string \\
      D4--D4 & const. & $r^{1/2}$ & $(1,1,r)$ & $5^2_2$: $r$ & emergent string \\
      D0--D4--D4 & const. & $r^{-1/4}$ & $(r^{-1/2},r^{-1/2},r^{1/2})$ & D2: $r^{3/2}$ & decomp.\ to 5d \\
      D4--D4--D4 & const. & $r^{3/4}$ & $(r^{1/2},r^{1/2},r^{1/2})$ & D6: $r^{3/2}$ & decomp.\ to 5d \\ \hline
      D0--D4--D4--D4 & const. & $1$ & $(1,1,1)$ & --- & finite horizon
    \end{tabular}
        \caption{Behaviour at the core, $r\to0$, of the small black holes of the D0--D4--D4--D4 system on $T^6/(\mathbb{Z}_2\times \mathbb{Z}_2)$. The non-vanishing charges are $q_0$, $p^1$, $(q_0,p^1)$, $(p^1,p^2)$, $(q_0,p^1,p^2)$ and $(p^1,p^2,p^3)$, respectively; other choices follow by permuting the two-tori. The four-dimensional dilaton is constant throughout. The last line is the finite-size case, for which all moduli are attracted to the finite (large) values \eqref{eq:attractedvalue}. More generally, every $1$ in this table should be interpreted as a large but finite number, so as to stay in the large volume approximation where the cubic prepotential applies. }
    \label{tab:BHsmall}
\end{table}
\begin{itemize}
    \item \textbf{Decompactification to five dimensions:} D0, D4, D0--D4--D4 and D4--D4--D4. For the D0 black hole, the three volumes grow uniformly, $t^i\sim r^{-1/2}$, and $g_s\sim r^{-3/4}$. This is the same direction in moduli space as for the minimal D0--D4--D4--D4 black hole with large $q_0$ discussed above, except that it is now reached at the core. The lightest tower consists of D0-branes, $m_{\rm D0}/M_p\sim r^{3/4}$, signalling a decompactification of the M-theory circle. For the other three cases, some of the two-tori shrink instead: for a single D4 charge $p^1$, $t^1\sim r^{-1/2}$ while $t^2,t^3\sim r^{1/2}$; for D0--D4--D4 with charges $(q_0,p^1,p^2)$, $t^1,t^2\sim r^{-1/2}$ and $t^3\sim r^{1/2}$; and for D4--D4--D4, $t^i\sim r^{1/2}$. T-dualizing along the shrinking two-tori maps each of these profiles to the D0 one, $t^i\sim r^{-1/2}$ and $g_s\sim r^{-3/4}$. In that frame, all charges are carried by D0- and D2-branes, which lift to momentum and M2-branes in M-theory. It is worth noting that in general, the limits where $t^i\to 0$ are not in the large volume regime, and for a generic Calabi--Yau we expect quantum corrections to heavily affect these classical limits. On the orbifold, however, T-duality along both directions of a single two-torus is an exact symmetry mapping the geometric orbifold to itself \cite{Pradisi:2014fqa}, so the shrinking-torus profiles are equivalent to large-volume ones.

    \item \textbf{Emergent-string limit:} D0--D4 and D4--D4. For the D0--D4 black hole with charges $(q_0,p^1)$, $t^1\sim r^{-1}$ grows while $t^2$ and $t^3$ approach constant values, and $g_s\sim r^{-1/2}$. The D0-branes, the D2-branes on $T^2_2$ and $T^2_3$, the D4-branes on $T^2_2\times T^2_3$ and the KK modes on $T^2_1$ all become light at the same rate, $m^2/M_p^2\sim r$, as the tension of the string obtained by wrapping an NS5-brane on $T^2_2\times T^2_3$, $\mathcal{T}_{\rm NS5}/M_p^2\sim r$. Since the $\mathbb{Z}_2$ generator $g$ (cf. footnote \ref{fn:Footnote1}) fixes $T^2_1$ and acts as $-1$ on $T^2_2\times T^2_3$, the orbifold is a K3 fibration over $T^2_1/\mathbb{Z}_2\simeq\mathbb{P}^1$ with fibre $T^4/\mathbb{Z}_2$, and the divisor $T^2_2\times T^2_3$ is the K3 fibre. The limit is therefore an equi-dimensional emergent heterotic string limit \cite{Grimm:2018ohb,Corvilain:2018lgw,Lee:2019wij}: the NS5-brane on the K3 fibre is the critical heterotic string. The D4--D4 black hole with charges $(p^1,p^2)$ has $t^3\sim r\to0$ at fixed $t^1,t^2$ and $g_s\sim r^{1/2}$. T-duality along $T^2_3$ maps it to a D2--D2 black hole whose scalar profile coincides with that of the D0--D4 black hole upon exchanging $T^2_1\leftrightarrow T^2_3$, although the charges differ. In that frame, the emergent heterotic string is the NS5-brane on the K3 fibre $T^2_1\times T^2_2$; in the original frame it is an exotic $5^2_2$-brane wrapped on $T^2_1\times T^2_2$ \cite{deBoer:2012ma}.
\end{itemize} 

Note that in a generic Calabi--Yau threefold, this picture is less clean. Shrinking two-cycles cannot be T-dualized away, and they take the solution out of the large-volume regime in which the cubic prepotential is reliable. Moreover, which limit is probed depends on the fibration structure of the threefold, e.g.\ on the existence of a K3 or $T^4$ fibration supporting an emergent string. We discuss these subtleties in section \ref{sec:CY}.

In string theory, the singularity is expected to be resolved by higher-derivative corrections, which dress it with a stretched horizon at the string or species scale \cite{Sen:1995in,Dabholkar:2004yr,Dabholkar:2004dq,Dabholkar:2005by,Calderon-Infante:2023uhz}. The simplest example is the D0--D4 system on K3$\times T^2$, for which a four-derivative correction gives $S=4\pi\sqrt{q_0p^1}$, reproducing the microscopic degeneracy. In \cite{Calderon-Infante:2023uhz}, it was shown that the correction to the vanishing entropy are of the order of the species-scale  $S\sim(M_p/\Lambda_{\rm sp})^2$, and that can be understood as the effect of integrating out the tower of states that becomes light at the core. For the two-charge small black holes, which are U-dual to a fundamental string carrying momentum and winding, one nonetheless expects a stretched horizon at the string scale, as in the analysis of \cite{Sen:1995in} for the heterotic string. Small black holes are therefore intrinsically UV objects, whose properties are set by the tower of light states: they are bad saddles of the two-derivative action \emph{by definition}.

\subsubsection{c-map and T-duality for type II on $T^6/(\mathbb{Z}_2 \times \mathbb{Z}_2) \times S^1$}\label{sec:cmap}

We are now going to introduce the c-map for four-dimensional $\mathcal{N} = 2$ theories and explain its relation with T-duality using the STU model as example. The four-dimensional moduli space of type IIA on $T^6/(\mathbb{Z}_2 \times \mathbb{Z}_2)$ is the direct product
\begin{equation}
    \mathcal{M}_{4d}= \mathcal{M}^{\rm IIA}_{VM} \times \mathcal{M}^{\rm IIA}_{HM} \,,\quad \mathcal{M}^{\rm IIA}_{VM}=\left(\frac{SL(2,\mathbb{R})}{U(1)}\right)^3\,,\quad \mathcal{M}^{\rm IIA}_{HM}=\frac{SO(4,4)}{SO(4)\times SO(4)}\,,
\end{equation}
where the hypermultiplet moduli space is given at tree level. Compactifying further on a circle $S^1_A$ yields a three-dimensional theory with additional moduli. The reduction of the metric gives the radius $R_A$ of the circle and the Kaluza--Klein (KK) vector. The four gauge fields $A^I$ produce four three-dimensional gauge fields while the Wilson lines along the circle give four axions $\xi^I$. The five three-dimensional gauge fields can be dualized into five scalars $(\tilde{\xi}_I,\sigma)$. The moduli $(z^i,\xi^I,\tilde{\xi}_I,R_A,\sigma)$ parametrize a three-dimensional twisted-hypermultiplet moduli space $\mathcal{M}^{\rm IIA}_{twisted}$ of real dimension $4(h^{1,1}+1)=16$. The hypermultiplet moduli space $\mathcal{M}_{HM}^{\rm IIA}$ does not depend on the radius of the circle and is therefore the same in three and four spacetime dimensions. The three-dimensional moduli space is then\footnote{ For the STU model, we can interpret $\mathcal{M}_{3d}$ as the truncation of the moduli space $E_{8(8)}/SO(16)$ of maximal three-dimensional supergravity \cite{Marcus:1983hb}}
\begin{equation}
  \mathcal{M}_{3d}=\mathcal{M}^{\rm IIA}_{twisted} \times \mathcal{M}^{\rm IIA}_{HM} \,,\qquad \mathcal{M}^{\rm IIA}_{twisted}=\frac{SO(4,4)}{SO(4)\times SO(4)}\,.
\end{equation}
The classical spatial c-map is the map which associates to the projective special K\"ahler manifold $\mathcal{M}^{\rm IIA}_{VM}$ the quaternionic-K\"ahler manifold $\mathcal{M}^{\rm IIA}_{twisted}$ obtained by dimensional reduction and dualization.\footnote{The temporal c-map's target is instead the para-quaternionic-K\"ahler space $SO(4,4)/(SO(2,2)\times SO(2,2))$, with indefinite signature.} Fixing the additional scalars $(\xi^I,\tilde{\xi}_I,R_A,\sigma)$, the c-map  defines a local totally geodesic embedding of the $z^i$ K\"ahler moduli of $\mathcal{M}^{\rm IIA}_{VM}$ into $\mathcal{M}^{\rm IIA}_{twisted}$.

The hypermultiplet sector of type IIB on $T^6/(\mathbb{Z}_2 \times \mathbb{Z}_2)$ has the following scalar field content\footnote{Throughout, we adopt the convention of \cite{Marchesano:2019ifh} for the type IIB hypermultiplet sector.}
\begin{align}
    \text{universal hypermultiplet:} &\;\;\;\;\;\; \tau = C_0 + i e^{ -\phi}\,, \;\; b^0\,,\;\;c^0\,,\\
    \text{three hypermultiplets:} &\;\;\;\;\;\; z^i = b^i + i t^i\,, \;\; c^i\,,\;\;d^i\,,\qquad i=1,2,3\,.
    \label{eq:defHyermultiplets}
\end{align} The universal hypermultiplet contains the axio-dilaton $\tau$ and the scalars $b^0$ and $c^0$ dual to the four-dimensional NSNS and RR 2-forms. The remaining three hypermultiplets contain the complexified K\"ahler moduli $z^i$ of the three two-tori. The scalars $c^i$ and $d^i$ are the periods of the RR 2-form and the RR 4-form over two- and four-cycles. T-duality tells us that type IIB on $T^6/(\mathbb{Z}_2 \times \mathbb{Z}_2) \times S ^1_B$ along the circle is equivalent to IIA on $T^6/(\mathbb{Z}_2 \times \mathbb{Z}_2) \times S^1_A$ with $S^1_A$ the dual circle. The type IIB scalars obtained from the dimensional reduction of the hypermultiplet sector can be identified with the scalars of $\mathcal{M}^{\rm IIA}_{twisted}$. Writing the radii in string units as $L_{A,B}=R_{A,B} M_s$, the Buscher rules, in the conventions of \cite{Baume:2019sry}, give
\begin{equation} \qquad
    g_{4,B}^{-1} = g_{4,A}^{-1} L_A = R_A M_{p, A}\,, \qquad
    g_{4,A}^{-1}= g_{4,B}^{-1}L_B= R_BM_{p, B}\,. \label{eq:gsRA}
\end{equation}
Thus the radion $R_A$ arising from the reduction of the type IIA metric is mapped to the four-dimensional type IIB dilaton. The latter is related to the ten-dimensional dilaton through $ e^{-2\varphi_{4,B}}=g_{4,B}^{-2} = e^{-2\phi_B}\,\mathcal{V}$. The IIB RR axions get identified with the Wilson line moduli $\xi^I$ and scalar duals $\tilde \xi_I$ of the IIA graviphoton and vector multiplets \cite{Alexandrov:2009qq}.  In particular, the Wilson line $\xi^0$ of the type IIA graviphoton around the circle becomes the type IIB RR axion $C_0$, so the type IIB axio-dilaton is given by
\begin{equation}
    \tau_B
    =
    C_{0,B}+i e^{-\phi_B}
    =
    \xi^0
    +
    i\,\frac{R_A M_{p,A}}{\sqrt{\mathcal{V}}}\,.
\end{equation} 

\subsubsection{Black holes under the c-map}\label{sec:runningcmap}

In the previous section we recalled that in four-dimensional $\mathcal{N} = 2$ supergravity compactified on the circle, part of degrees of freedom of the gravity and vector multiplet sector can be mapped via the c-map into degrees of freedom of the hypermultiplet sector. In section \ref{sec:CY} we will see that this map can be turned in a full solution generating techniques which allow us to construct a configuration solving type IIB equations of motion starting with a type IIA BPS black hole. In particular, the c-map can be used to map the dependence of type IIA fields on harmonic functions of the transverse space $\mathbb{R}^3$ to the dependence of type IIB fields on harmonic function of the new transverse space. For a c-map along the time circle we obtain instantons with transverse space $\mathbb{R}^4$; for a c-map along a space circle we obtain strings with transverse space $\mathbb{R}^2$. As a result, non-trivial profiles for the four-dimensional metric, gauge fields and vector multiplets will induce a profile for the three-dimensional scalar fields: the radius $R_A$, the dual axions $(\xi^I,\tilde\xi_I)$ and K\"ahler moduli. 

Let us summarize how the solution generating technique acts on the D0--D4--D4--D4 black hole keeping track only of the c-map part, i.e. leaving the harmonic functions implicit. The radion $R_A$ arising from the reduction of the type IIA metric is mapped to the four-dimensional type IIB dilaton. For black-hole described by static IWP metrics as \eqref{eq:4dMetricBH}, the warp factor implies that the radion acquires a nontrivial radial profile, irrespective of which of the four spacetime directions one T-dualizes along: in four-dimensional Einstein frame, $R_A\propto e^{U}$ for the (Euclidean) time direction, and $R_A\propto e^{-U}$ for a spatial direction along which the solution has been smeared. Consequently, the four-dimensional type IIB dilaton necessarily acquires a radial profile, $e^{-2\varphi_{4,B}}\propto R_A^2$, and can no longer be kept constant. Thus, although on the type IIA side the dilaton stays fixed along the attractor flow of the solutions which explore infinite-distance limits in K\"ahler moduli space, on the type IIB side the corresponding limit necessarily involves the four-dimensional dilaton.

The same logic applies to the axions. Upon reduction of \eqref{eq:GaugeFields} on Euclidean time, $\xi^0$ and $\tilde\xi_i$ are the Wick-rotated electrostatic potentials, $\propto i\,H_0^{-1}$ and $i\,(H^i)^{-1}$. These become the type IIB axions $C_0$ and $d^i$, whose profiles are therefore imaginary, and whose values at the core are shifted by an amount set by the asymptotic string coupling. For a spatial direction $y$ along which the black hole has been smeared, the electric potential has no component along $y$, but the magnetic D4 potentials do, $A^i_y\propto p^i\theta$, and so does the dual potential of the D0 field, $\tilde A_{0,y}\propto q_0\theta$, with $\theta$ the angle around the smeared black hole. The corresponding axions $c^i$ and $c^0$ therefore wind around the T-dual string, which is thus an axionic string with monodromies given by the black-hole charges. Finally, the $B$-field axions $b^i$ are left untouched by the duality, and are set to $b^i_h=0$ in both cases.

\subsection{Instantons}\label{sec:instantonstorus}

We now T-dualize the black holes of section \ref{sec:BHprobes} along the Euclidean time circle. In section \ref{sec:constructionDm1ED3} we construct the resulting D$(-1)$--ED3--ED3--ED3 configuration directly in ten dimensions and reduce it on the orbifold, recovering the hypermultiplet profile anticipated by the c-map in section \ref{sec:runningcmap}. We compute the on-shell action, show that it relates to the ADM mass and entropy of the dual black hole. In section \ref{sec:instUVprobes} we ask whether these Euclidean saddles can probe UV physics as their dual black holes do. We distinguish instantons dual to finite-size and minimal black holes, from those dual to small black holes. Finally, in section \ref{sec:instantonbounds} we compare our instantons with recent bounds on instantons in the literature.

\subsubsection{Instantons dual to black hole probes}\label{sec:constructionDm1ED3}
We start by constructing the instanton T-dual to the D0--D4--D4--D4 black hole of section \ref{sec:BHprobes}.

\paragraph{The D$(-1)$--ED3--ED3--ED3 instanton.}
We T-dualize the ten-dimensional type IIA configuration along the Euclidean time direction, which yields a D$(-1)$--ED3--ED3--ED3 configuration. After reducing type IIB on the same $T^6/(\mathbb{Z}_2\times\mathbb{Z}_2)$ and Wick-rotating to Euclidean signature, we obtain the action
\begin{equation}
S^{(E)}_4 = \frac{M_p^2}{2}\int d^4x\,\sqrt{g_E}\Bigg[ -R_E
+ \frac{1}{2\Phi^2}(\partial \Phi)^2
+ \sum_{i=1}^3\frac{1}{2(t^i)^2}(\partial t^i)^2 + \frac{1}{2\cdot 3!}\sum_{I=0}^3f_I|F_3^I|^2
\Bigg],
\label{eq:euclidean-action-ST}
\end{equation}
where the gauge kinetic functions are
\begin{equation}
    f_0=\frac{\Phi}{t^1t^2t^3}\,,\quad f_1=\Phi\frac{t^2t^3}{t^1}\,,\quad f_2=\Phi\frac{t^1t^3}{t^2}\,,\quad f_3=\Phi\frac{t^1t^2}{t^3}\,.
\end{equation}
where $\Phi$ and $t^i$ have the same interpretation as in the previous section. The solution corresponding to a D$(-1)$--ED3--ED3--ED3 consists of a flat Euclidean Einstein-frame metric
\begin{equation}
d s^2_{E,4} = d r^2 + r^2d\Omega_3^2\,,
\label{eq:flat-E-metric}
\end{equation}
with field strengths 
\begin{equation}
    F^I_3 = \pm \star_4d H^I \,, \qquad I=0,1,2,3\,,
\end{equation}
and the moduli
\begin{equation}
\Phi = \frac{1}{\sqrt{H_0H^1H^2H^3}},
\quad
t^1 = \sqrt{\frac{H_0H^1}{H^2H^3}},
\quad
t^2 = \sqrt{\frac{H_0H^2}{H^1H^3}},
\quad
t^3 = \sqrt{\frac{H_0H^3}{H^1H^2}}.
\label{eq:ST-solution}
\end{equation}
The functions $H$ are harmonic on $\mathbb{R}^4$
\begin{equation}
    H_0(r)=h_0+\frac{q_0 l_p^2}{r^2}\,, \qquad  H^i(r)=h^i+\frac{p^i l_p^2}{r^2} \,, \qquad i=1,2,3\,.
\end{equation}
We have a core at $r=0$ where the dilaton explodes $\varphi_4 = +\infty$, but there are no curvature singularities. The residues of the harmonic functions encode the D-brane charges. The $h$ constants are fixed by the asymptotic normalization of the K\"ahler moduli $(t^i)_\infty$ and the dilaton  $\Phi_\infty= (g_{4,s}^\infty)^{-2} = (h_0h^1h^2h^3)^{-1/2}$.

Notice that the solution can be easily obtained from the D0--D4--D4--D4 configuration. A c-map along the Euclidean time circle yields instantons smeared along the dual circle, with harmonic functions $H \propto 1/r$ on $\mathbb{R}^3$. The localized instanton is then obtained unsmearing the harmonic functions ($H \propto 1/r^2$). The K\"ahler moduli are thus the same functions of the four harmonic functions as in the black-hole solution, while the exponential of the dilaton is controlled by their product.

\paragraph{Instanton action, black hole entropy and the ADM mass.}

An interesting quantity to compute is the instanton action. The choice of duality frame for the action determines which boundary terms are necessary to add to \eqref{eq:euclidean-action-ST}. The frame we are using is the so-called flux frame \cite{Witten:2026twr}, in which the D$(-1)$ and the ED3s are magnetically charged under the 2-form gauge fields dual to the axions $\chi_I=(C_0, d^i)$. This means that the instanton D$(-1)$--ED3--ED3--ED3 has only magnetic charges carried by the 3-form field strengths $F_3^I$. Then, no boundary term is necessary to implement the ensemble. The GHY term, with the flat space subtraction, vanishes on a flat configuration, thus the on-shell action can be evaluated just replacing the solution we found in \eqref{eq:euclidean-action-ST}. We obtain
\begin{equation}
S_{\rm inst} =2\pi^2 \Big( \frac{q_0 }{h_0}+\sum_{i=1}^3 \frac{p^i }{h^i}\Big) =  2\pi^2 \left(\frac{q_0}{g^\infty_{4,s}}\frac{1}{\sqrt{( t^1 t^2 t^3)_\infty}}+\sum_{i=1}^3 \frac{p^i}{g^\infty_{4,s}}\sqrt{\frac{(t^j t^k)_\infty}{(t^i)_\infty}}\right)\,, \label{eq:instaction2}
\end{equation}
where we expressed the action in terms of the moduli at infinity. The familiar action $2\pi N/g_s^\infty$ of $N$ D-instantons is recovered for $q_0=N/\pi$, and similarly for the ED3s; this fixes the integer normalization of the charges in the instanton frame $q_I=N_I/\pi$.\footnote{This normalization differs from the one of the black-hole frame, $q_I=N_I/(2\sqrt\pi)$, because the harmonic functions of the localized instanton on $\mathbb{R}^4$ and of the black hole on $\mathbb{R}^3$ are normalized differently. The relations between the two frames given below hold up to such order-one factors.} 

We now compare the instanton action with the ADM mass and the entropy of the dual black hole. The instanton action is formally related to the mass \eqref{eq:4dMetricBHADM}. Writing $h_I = \Phi_\infty^{-1/2}\hat h_I$ with $\hat h_0\hat h^1\hat h^2\hat h^3=1$ and identifying the charges $q_I$ and the $\hat h_I$ with those of the black hole, we have
\begin{equation}\label{eq:actionADMmass}
    S_{\rm inst}=\frac{\pi}{g^\infty_{4,s}}\,\frac{M_{\rm ADM}}{M_p}\,.
\end{equation}
Using \eqref{eq:gsRA} and ignoring the order-one factors that come from comparing the localized instanton with the smeared one and the conventions for the units, the instanton action reads $S_{\rm inst} \sim  R_AM_{\rm ADM}$ and matches that of the Euclidean worldline of the dual BPS particle wrapping the circle, as expected from T-duality. While the instanton action is linear in the charges, the entropy of the dual black hole is not. The two are related by an inequality: using $h_0h^1h^2h^3=\Phi_\infty^{-2}$, the inequality of arithmetic and geometric means and matching the charges $q_I$ we get
\begin{equation}\label{eq:actionentropybound}
S_{\rm inst} =\;  2\pi^2\sum_I\frac{q_I}{h_I} \; \geq\;8\pi^2 \,\Phi_\infty^{1/2}\,(q_0p^1p^2p^3)^{1/4}= \,\frac{\sqrt{8\pi^2}}{g_{4,s}^\infty}\,\sqrt{{S_{BH}}}\,,
\end{equation}
where $S_{BH}$ is the two-derivative entropy \eqref{eq:firstentropy}. The inequality is saturated if and only if all four ratios $q_I/h_I$ are equal, i.e.\ when the harmonic functions are proportional to one another and the K\"ahler moduli sit at their attractor values everywhere. This is the instanton dual to the double-extremal black hole and equation \eqref{eq:actionentropybound} reduces to the well known relation $S_{BH} = \pi M_{\rm ADM}^2 G_N $.\footnote{Often this is written with $G_N = 1$, i.e. in units of the unreduced Planck mass $m_p$.\label{m_pVSM_p}} Exploiting the fact that the two-derivative black hole entropy provides an upper bound to the species length \cite{Calderon-Infante:2025pls}, equation \eqref{eq:actionentropybound} can be written as
\begin{equation} \label{eq:UpperBoundSS}
g_{4,s}^\infty \, S_{\rm inst} \big \rvert_{\rm IIB}\gtrsim \sqrt{S_{BH}} \big|_{min}\simeq  M_p/\Lambda_{\rm sp}^{\rm IIA}\,.
\end{equation}
The labels $\rm IIA/B$ must be interpreted in the following sense. The species length $M_p/\Lambda_{\rm sp}^{\rm IIA}$ of type IIA, evaluated at the point $z^i_h \in \mathcal{M}_{VM}^{\rm IIA}$ corresponding to the attractor value of a minimal BPS black hole with integer charges $N_I$, is upper bounded (as a function of $N_I$) by $g_{4,s}^\infty \, S_{\rm inst}$. The latter is evaluated at the point in type $\rm IIB$ moduli space corresponding to the asymptotic value of the scalars in the T-dual instanton profile.

A first consequence is that, although the instanton action depends on the asymptotic Kähler moduli $(t^i)_\infty$, the lower bound in \eqref{eq:actionentropybound} does not: it is fixed by the charges and by the asymptotic four-dimensional dilaton $g_{4,s}^\infty$ alone, and it is saturated only when $(t^i)_\infty$ coincide with the attractor values. Along any trajectory in the hypermultiplet moduli space at fixed four-dimensional dilaton the action of an instanton T-dual to a finite-size black hole therefore cannot drop below $\sim\sqrt{S_{BH}}/g_{4,s}^\infty$, however the Kähler moduli are varied. Note that the lower bound in \eqref{eq:actionentropybound} comes from the central charge of the dual BPS state at the end of the attractor flow, which is related to the area term of the entropy of the dual black hole. Indeed, only instantons whose dual black hole has vanishing \emph{classical} entropy, i.e.\ small black holes, for which the bound is trivial, can have an action that goes to zero along such limits. This explains why certain instantons can become unsuppressed along infinite distance limits. For instance, in \cite{Marchesano:2019ifh,Baume:2019sry} they study infinite-distance limits of the type IIB hypermultiplet moduli space at fixed four-dimensional dilaton, where the Kähler moduli grow. These limits do not correspond to the trajectories we obtain T-dualizing type IIA black holes, but they receive the same type of instanton corrections (see section \ref{sec:quantumCorrections} for more details). D$(-1)$ and D$(-1)$--ED1 instantons become unsuppressed and their resummation modifies the classical metric. This is consistent with our bound: the D$(-1)$ and D$(-1)$--ED1 instantons driving the corrections are dual to small black holes. 

Another way to see this is in the T-dual type IIA frame, where the instanton is the Euclidean worldline of the black hole wrapping $S^1_A$, with action $S_{\rm inst} \sim R_A M_{\rm ADM}$. Then, \eqref{eq:UpperBoundSS} becomes a lower bound on the mass of the black hole in terms of the species scale at its horizon, $M_{\rm ADM} \geq \sqrt{8S_{BH}}\, M_p \gtrsim M_p^2/\Lambda_{\rm sp}^{\rm IIA}(z_h)$, where the first inequality is saturated by the double-extremal solution. As emphasised in \cite{Marchesano:2019ifh}, an instanton contributes significantly to the hypermultiplet metric only if the T-dual state lies below the Kaluza--Klein scale of the circle, $M_{\rm ADM} \ll 1/R_A$, in which case its effect can be understood as a one-loop contribution of a three-dimensional state. For instantons T-dual to finite-size black holes this would require $1/R_A \gg M_p^2/\Lambda_{\rm sp}^{\rm IIA}(z_h) \geq M_p$, i.e.\ a Kaluza--Klein scale above the four-dimensional Planck scale. The three-dimensional towers whose one-loop effects generate the corrections of \cite{Marchesano:2019ifh,Baume:2019sry} therefore consist only of states T-dual to small black holes.

\paragraph{Imaginary axions.} So far, we have described the solution in the flux frame. To make contact with the discussion of section \ref{sec:runningcmap}, it is instructive to dualize back to the axion frame. The Hodge dual of the 3-forms $F_3^I$ are the RR axions $\chi_I$. In Euclidean signature, the duality relation acquires a factor of $i$, $d\chi_I=\pm i f_I\star_4F_3^I$, so that on the solution
\begin{equation}\label{eq:instantonaxions}
\chi_I(r)=\chi_{I,\infty}\pm i\left(\frac{1}{h_I}-\frac{1}{H_I(r)}\right)\,.
\end{equation}
The upper (lower) sign corresponds to the instanton (anti-instanton). The axions are therefore imaginary, as for the ten-dimensional D-instanton, and they are the T-dual images of the electrostatic potentials of the black hole. Their saxionic partners are
\begin{equation}\label{eq:instantonsaxions}
s_0\equiv\sqrt{\frac{\Phi}{t^1t^2t^3}}=\frac{1}{H_0}\,,\qquad s_i\equiv\sqrt{\Phi\,\frac{t^jt^k}{t^i}}=\frac{1}{H^i}\,,
\end{equation}
i.e.\ $s_I=\sqrt{f_I}$. Here $s_0$ is the inverse ten-dimensional string coupling, and the $s_i$ are the volumes of the four-cycles $T^2_j\times T^2_k$ wrapped by the ED3s, measured in ten-dimensional Einstein-frame units. Towards the core, the saxions $s_I$ go to zero, the axions approach $\chi_I\to\chi_{I,\infty}\pm i\,s_{I,\infty}$, and the K\"ahler moduli $t^i$ are attracted to \eqref{eq:attractedvalue}. The remaining axions $c^0$, $c^i$ and $\sigma=b^0$ are constant, and $b^i\equiv0$.

\paragraph{Complexified instanton action.} The Euclidean action \eqref{eq:instaction2} is the real part of the complexified action that weighs the instanton contribution to the two-derivative effective action of the hypermultiplets, i.e.\ to the quaternionic-K\"ahler metric on $\mathcal M_{HM}$ \cite{Alexandrov:2008gh,Marchesano:2019ifh}. Besides its real action, the instanton couples to the asymptotic values of the RR axions. In the flux frame this coupling is the term $i\,\chi_{I,\infty}\oint_{S^3_\infty}F_3^I\propto i\,q_I\chi_{I,\infty}$, which is purely imaginary: it does not change the real part of the on-shell action, but it fixes the phase of the instanton contribution. Defining $T_I\equiv s_I\mp i\,\chi_I$, the complexified action is
\begin{equation}\label{eq:complexaction}
S_{\rm inst}=2\pi^2\sum_I q_I\left(s_{I,\infty}\mp i\,\chi_{I,\infty}\right)=2\pi\sum_I N_I\,T_{I,\infty}\,,
\end{equation}
so that $e^{-S_{\rm inst}}=\prod_I e^{-2\pi N_I s_{I,\infty}}\,e^{\pm 2\pi i N_I\chi_{I,\infty}}$. The phase is invariant under the discrete shifts $\chi_I\to\chi_I+1$ exactly when the $N_I$ are integers, which is another way to see the charge quantization $q_I=N_I/\pi$. While the real part controls the size of the instanton effects, the imaginary part determines their dependence on the RR axions, breaking their continuous shift symmetry to a discrete one. In the T-dual frame it is the Aharonov--Bohm phase picked up by the BPS particle winding $S^1_A$ in the presence of the Wilson lines of the four-dimensional gauge fields and of their magnetic duals.

On the solution, the holomorphic combinations $T_I(r)=h_I^{-1}\mp i\,\chi_{I,\infty}$ are constant, while their conjugates $\bar T_I(r)$ flow. This is the familiar BPS condition of D-instantons. Since the metric on each factor of the moduli space is $dT_Id\bar T_I/(2s_I^2)$, the profile is null, and since all $H_I$ are linear in $\lambda=l_p^2/r^2$ it is an affinely parametrized null geodesic in the complexified moduli space. Accordingly, $\text{Re} \,S_{\rm inst}=2\pi^2\sum_Iq_Is_{I,\infty}$ can be read either from the saxions at infinity or from the imaginary excursion of the axions between infinity and the core, $\chi_I(0)-\chi_{I,\infty}=\pm i\,s_{I,\infty}$.

\subsubsection{Instantons as UV probes?}\label{sec:instUVprobes}

A few comments are in order. First and foremost, unlike black holes, instantons are not solitonic objects in Lorentzian spacetime but Euclidean saddles, obtained by Wick-rotating Lorentzian objects. Thus, although the scalar fields have a radial profile, this should not be treated on the same footing as the scalar profile of a Lorentzian black hole. The crucial difference is that the axions become imaginary, so that the scalar profiles explore an analytically continued hypermultiplet target space. Equivalently, we can absorb the factors of $i$ by flipping the sign of the axion kinetic terms, and describe the Euclidean system in terms of real scalars with an indefinite target-space metric. Although the image of the scalar profiles is not a trajectory in the target space of the Lorentzian theory, we can still study what happens as we move in the complexified target space.

As we saw in the previous section, the warping of the black-hole solution induces a nontrivial profile for the four-dimensional dilaton on the type IIB side: from \eqref{eq:ST-solution}, the dilaton runs to strong coupling at the core of the instanton ($\Phi\to0$ as $r\to0$), where the EFT is expected to break down. Moreover, if some of the charges are large, then the moduli $t^i$ can reach large values at the core \eqref{eq:attractedvalue}. However, the BPS D-instantons we obtain as T-duals of BPS black holes follow a light-like trajectory in the complexified target space, so the core sits at zero distance from any point along the profile. Despite the scalars blowing up, we therefore cannot use the standard Swampland Distance Conjecture \cite{Ooguri:2006in} to relate the possible appearance of light towers to an infinite-distance limit. We will briefly come back to this point in section \ref{sec:instantonbounds}, when discussing the imaginary distance conjecture. Note that the instanton solution may receive large corrections near its core \cite{Martucci:2024trp}. Although curvature corrections vanish identically for a flat metric, higher-derivative terms involving gradients of the scalar fields need not vanish and may contribute to the stress-energy tensor. These contributions can backreact on the spacetime metric, generating curvature near the core. If a tower of states also becomes light, the lowering of the species scale can render curvature corrections unsuppressed. Instanton effects provide another potential source of large corrections. Even when backreaction on the spacetime metric and higher-derivative terms are neglected, instanton corrections modify the hypermultiplet metric and can thereby deflect the scalar-field trajectories from their classical behaviour near the core \cite{Marchesano:2019ifh,Baume:2019sry, Kaufmann:2026mha, Kaufmann:2026fli, Kaufmann:2026tsy} (see section \ref{sec:quantumCorrections} for more details).

The question is then whether there is an observable in the instanton solution that can help us understand the physics at the core. We would need an observable that can be followed along the trajectory in the complexified moduli space. The obvious candidate is the Euclidean on-shell action, but it is completely fixed by the charges and the asymptotic moduli and is insensitive to the core, see \eqref{eq:actionADMmass}. Another possible observable is a charge defined by integrating over a sphere of cutoff radius $r_{\rm UV}$, similar to the cylinder prescription common in the string literature (see section \ref{sec:EFTstringtorus}). However, unlike strings, instanton solutions are strongly coupled at the core, instead of weakly coupled. So one has control over the solution from asymptotic infinity until a finite spacelike distance $r_{\rm UV}$ from the core at which strong coupling effects take over. However, there is no simple universal candidate for this cutoff, which has to be determined case by case.  

There is therefore no strict sense in which instantons can be used as classical probes of UV physics in the same way as Lorentzian objects. They probe it in a different way, as saddles of the Euclidean path integral. An instanton of charge $\gamma$ schematically contributes $\delta C_\gamma=\mu_\gamma e^{-S_\gamma}$ to a coupling $C$ of the Lorentzian effective action, where $S_\gamma$ is its complexified classical action \eqref{eq:complexaction}. The prefactor $\mu_\gamma$, called the instanton measure, accounts for quantum fluctuations around the saddle, integration over its collective coordinates, and the multiplicity of its internal states, with the operator insertions required to absorb fermionic zero modes. Both the action and the measure of an instanton can be related to thermodynamic quantities of the dual black hole. As we saw above, the instanton action is set by the ADM mass of the black hole \eqref{eq:actionADMmass}. The measure of the instanton corrections to the type IIB hypermultiplet effective action can be related to the microscopic degeneracy, and hence the entropy, of the T-dual black hole \cite{Gunaydin:2005mx,Alexandrov:2008gh,Pioline:2009ia}. Indeed, under T-duality, the D-instanton is identified with the Euclidean worldline of the dual black hole wrapping the circle $S^1_A$ and the multiplicity entering the measure with the $\mathcal N=2$ BPS index $\Omega(\gamma)$, namely the second helicity supertrace, of the dual black hole. More precisely, for primitive charges $\gamma$, the instanton measure is directly proportional to $\Omega(\gamma)$. For non-primitive charges, the measure is proportional to a dressed index implementing multi-cover contributions \cite{Pioline:2009ia}. This provides a precise relation between the instanton corrections to the hypermultiplet effective action and the entropy and mass of the T-dual black hole. Their relevance, however, depends crucially on which family of black holes the instanton is dual to. 

\paragraph{Instantons dual to finite-size and minimal black holes.} These are bound states such as D$(-1)$ and ED3 instantons carrying all four charges. As for their dual black holes, the K\"ahler moduli are attracted to finite, large values at the core, $t^i\to(t^i)_h$ as in \eqref{eq:attractedvalue}, while the dilaton runs to strong coupling and the axions approach $\chi_{I,\infty}\pm is_{I,\infty}$. Their action is bounded from below by \eqref{eq:actionentropybound}, $S_{\rm inst}\gtrsim\sqrt{S_{BH}}/g^\infty_{4,s}$, so they only become relevant at strong four-dimensional coupling: along any limit at fixed $\varphi_4$ they remain suppressed, however the K\"ahler moduli are varied. Their measure, on the other hand, can be exponentially large. In a large-charge regime dominated by a regular single-centred black hole, and assuming no exponential cancellations in the index, one expects $\log|\Omega(\gamma)|\sim S_{BH}(\gamma)$ \cite{Pioline:2009ia}, so that the horizon entropy is reflected in the exponential growth of the multiplicity entering the instanton measure. For large black holes, with all charges scaled uniformly, the Euclidean solution is under control, and the instanton contributes with action \eqref{eq:instaction2} and a degeneracy $\mu\sim e^{S_{BH}}$. Since $S_{BH}$ grows quadratically in the charges while the action only grows linearly, the degeneracy eventually overwhelms the exponential suppression, and the sum over these instantons is not convergent \cite{Pioline:2009ia}. Instantons dual to minimal black holes, $q_0\gg p^i\sim\mathcal{O}(1)$, deserve a separate comment. On the black-hole side, the K\"ahler moduli are attracted to large values $t^i\sim\sqrt{q_0}$, so that the horizon is large in Planck units but not in units of the species scale at the horizon, $\Lambda_{\rm sp}^{\rm IIA}(z_h)$. As a consequence, the two-derivative entropy of these black holes receives corrections suppressed only by the species scale, which are not parametrically small (see section \ref{sec:blackholestorus}). On the instanton side, the K\"ahler moduli reach the same large values at the core. However, the classical Euclidean profile is presumably substantially modified by quantum corrections: the core explores a large-volume, strong-coupling region of the hypermultiplet moduli space, where the classical metric is itself heavily corrected by instantons and its infinite-distance limits may even disappear \cite{Marchesano:2019ifh,Baume:2019sry}. Since the multiplicity entering the measure is the microscopic BPS index of the dual black hole \cite{Pioline:2009ia}, rather than the exponential of its two-derivative entropy, it should receive the same species-scale suppressed corrections as the entropy of the minimal black holes. The degeneracy of these instantons is therefore presumably not captured by the two-derivative entropy, but by its corrected version. This does not change their role in the instanton sum: along the family $q_0\to\infty$ the action grows linearly in $q_0$, while the logarithm of the measure, including these corrections, is expected to grow only like $\sqrt{q_0}$. Hence their contributions $\mu\,e^{-S_{\rm inst}}$ decrease with the charge such that the sum over this family converges. This is unlike those of the instantons dual to large black holes, for which the degeneracy eventually overwhelms the exponential suppression.

\paragraph{Instantons dual to small black holes.} These are elementary instantons, such as a pure D$(-1)$ instanton, dual to the D0 small black hole, or bound states with fewer than four charges. The harmonic functions of the missing charges are constant and, unlike in the finite-size case, the K\"ahler moduli are not attracted to finite values. For a pure D$(-1)$ charge, for instance, $\Phi=(H_0h^1h^2h^3)^{-1/2}\to0$ and $t^i\propto H_0^{1/2}\to\infty$ at the core. In ten dimensions, this is simply the growth of the string-frame internal volume in the D-instanton background, $e^{\phi/2}\propto H_0^{1/2}$, at fixed Einstein-frame volume. Compared to the finite-size case, the only difference is that all moduli, and not just the dilaton, run off at the core; for charges of order one, they are single Euclidean D-branes with no macroscopic size, so the classical profile cannot be trusted near the core. Their action, $S_{\rm inst}=2\pi^2q_0/g_{s}^\infty$ for the D$(-1)$, is nevertheless reliable, being fixed by the BPS condition and the asymptotic data, and so is their measure, which is of order one and charge independent: it is proportional to $-\chi_E(X)$ for pure D$(-1)$ instantons and the genus-zero Gopakumar--Vafa invariants $n^{(0)}_k$ for D$(-1)$--ED1 bound states. This is consistent with the vanishing classical entropy of the dual small black holes.

Unlike the previous family, these instantons can become unsuppressed at fixed four-dimensional dilaton. At fixed $\varphi_4$, the real part of the complexified action \eqref{eq:complexaction} depends on the K\"ahler moduli exactly as the mass of the dual black hole depends on the type IIA vector-multiplet moduli. Every tower of particles (i.e.\ small black holes) becoming light at an infinite-distance point of $\mathcal{M}^{\rm IIA}_{VM}$ therefore gives rise to a tower of instantons, further labelled by their multi-covering (winding) number around the circle, whose actions vanish at the same rate. By \eqref{eq:actionentropybound}, instantons dual to finite-size black holes cannot join them at fixed $\varphi_4$.

Once $S_{\rm inst}\lesssim1$, these instantons are no longer exponentially suppressed, and their corrections to the hypermultiplet moduli-space metric become of order one, to the point that the infinite distance may be removed altogether \cite{Marchesano:2019ifh,Baume:2019sry}. They also have a simple effect on the scales of the theory. The ratio between the Planck and the string scale is set by the contact potential of the hypermultiplet moduli space, which classically gives $M_p^2\simeq4\pi e^{-2\varphi_4}M_s^2$. Along the large-volume trajectory at fixed $\varphi_4$, the unsuppressed instantons are D$(-1)$ and D$(-1)$--ED1 bound states, T-dual to D0--D2 small black holes. Each of them corrects the contact potential by an order-one amount, so that $M_p^2/M_s^2$ receives a contribution proportional to the number $N_{\rm sp}$ of unsuppressed instantons. Keeping $M_p$ fixed then requires
\begin{equation}\label{eq:MWspecies}
M_s\sim\frac{M_p}{\sqrt{N_{\rm sp}}}\,,
\end{equation}
which takes the form of a species scale, with the string scale as the cutoff and $N_{\rm sp}$ counting unsuppressed instantons instead of light particles \cite{Marchesano:2019ifh}. This is natural from the T-dual type IIA perspective: $S_{\rm inst}\ll1$ is the condition $M_{\rm ADM}\ll1/R_A$, and the correction is reproduced by integrating out at one loop the corresponding three-dimensional states below the Kaluza--Klein scale, in the spirit of the Emergence Proposal. Instantons therefore do contribute to a species scale on the hypermultiplet side, but only the elementary ones T-dual to small black holes, and $N_{\rm sp}$ counts instanton species rather than black-hole microstates. We will come back to this briefly when discussing the emergence proposal in the next section.

\subsubsection{Bounds on instantons}\label{sec:instantonbounds}

We now further characterize the instantons dual to small, minimal and finite-size black holes commenting on their role in fulfilling or saturating several instanton bounds.

\paragraph{Axion weak gravity conjecture.} 
The axionic weak gravity conjecture \cite{ArkaniHamed:2006dz} proposes that, for every axion direction, there exists an instanton whose action and decay constant satisfy $f_qS_{\rm inst}\lesssim M_p$. Our instantons satisfy this bound. From \eqref{eq:euclidean-action-ST}, the metric $dT_Id\bar T_I/(2s_I^2)$ on each factor of the moduli space gives the axion kinetic terms $\frac{M_p^2}{4s_I^2}(\partial\chi_I)^2$. Since $\chi_I\simeq\chi_I+1$, the decay constants are $f_{\chi_I}=M_p/(2\sqrt2\,\pi s_I)$. An instanton with charges $N_I$ couples to the axion direction $\sum_IN_I\chi_I$, with decay constant $f_q^{-2}=\sum_IN_I^2f_{\chi_I}^{-2}$, and has action $S_{\rm inst}=2\pi\sum_IN_Is_{I,\infty}$ by \eqref{eq:complexaction}. Hence
\begin{equation}\label{eq:axionWGC}
\frac{(f_qS_{\rm inst})^2}{M_p^2}=\frac12\,\frac{\big(\sum_IN_Is_{I,\infty}\big)^2}{\sum_I\big(N_Is_{I,\infty}\big)^2}\;\leq\;\frac n2\,,
\end{equation}
where $n\leq4$ is the number of non-vanishing charges, and we used the Cauchy--Schwarz inequality. All our instantons therefore satisfy $f_qS_{\rm inst}\leq\sqrt2\,M_p$. As in \eqref{eq:actionentropybound}, the bound is saturated if and only if all four ratios
$q_I/h_I=N_Is_{I,\infty}/\pi$ are equal, i.e.\ by the instanton dual to the double-extremal black hole. Instantons dual to small black holes, with $n<4$, obey the stronger bound $f_qS_{\rm inst}\leq\sqrt{n/2}\,M_p$, since only $n$ terms contribute to the sums in \eqref{eq:axionWGC}. Within their own $n$-dimensional subsector they saturate $\sqrt{n/2}\,M_p$ when their $n$ ratios $q_I/h_I$ are equal; a pure D$(-1)$ or a pure ED3, for instance, has $f_qS_{\rm inst}=M_p/\sqrt2$, independently of the charge and of the moduli.

\paragraph{Quantum-gravity cutoff.} When an axion comes from a $p$-form on a $p$-cycle, its instanton is a Euclidean brane on that cycle and its axion string is the electric-magnetic dual brane, with tension $\mathcal T\sim2\pi S_{\rm inst}f^2$ \cite{Reece:2025thc}. Since the string core lies at infinite distance, the QG scale must lie below the string scale \cite{Reece:2025thc,Reece:2026hmp},
\begin{equation}
    \Lambda_{\rm sp}\leq 2\pi\sqrt{S_{\rm inst}}\,f \,.\label{eq:QGcutoff}
\end{equation}
For our elementary instantons the dual strings are a D7-brane on $T^6/(\mathbb{Z}_2\times\mathbb{Z}_2)$ for the D$(-1)$ and a D3-brane on $T^2_i$ for the ED3 on $T^2_j\times T^2_k$. The decay constants and actions of the previous paragraph give $2\pi\sqrt{S_{\rm inst}}f_{\chi_I}=\sqrt{\pi/s_{I,\infty}}\,M_p$, i.e.\ $\Lambda_{\rm sp}\leq\sqrt{\pi g_s}\,M_p$ for the D$(-1)$, as in \cite{Reece:2025thc}, and $\Lambda_{\rm sp}\leq\sqrt{\pi g_s/(t^jt^k)}\,M_p$ for the ED3. Both are satisfied by the string scale, $M_s/M_p\sim g_s/\sqrt{\mathcal V}$, at weak coupling and large volume. Bound states only give weaker bounds: their dual strings are BPS bound states, whose tension is the sum of the tensions of the constituents weighted by the charges, $\mathcal T=\sum_IN_I\mathcal T_I$ (see \eqref{eq:EFTstringtension} in section \ref{sec:EFTstringtorus}), so that $\sqrt{2\pi\mathcal T}$ is larger than for any of the fundamental constituents and grows with the charges.

\paragraph{Emergence.}  For the $\frac12$-BPS $R^4$ coupling of type IIA string theory on $T^2$ and $T^3$, \cite{Artime:2026kfq} proposed, and checked along infinite-distance limits, that the one-loop Schwinger integral over the towers below the species scale reproduces the full contribution of precisely those instantons whose action lies, parametrically along the limit, in the window
\begin{equation}
    M_{\rm light}/\Lambda_{\rm sp}\lesssim S\lesssim\Lambda_{\rm sp}/M_{\rm light}\,, \label{eq:emergence}
\end{equation}
with $M_{\rm light}$ the mass of the lightest tower. These are the instantons whose action is a ratio of light masses, e.g.\ $S_{\rm ED0}\sim m_{\rm D0}/M_{\rm KK}=m_{\rm D0}R$ for a light particle wrapping a circle whose KK modes are also light; in emergent string limits $\Lambda_{\rm sp}\sim M_{\rm light}$ and only worldsheet instantons, with $S\sim\mathcal O(1)$, emerge. Our $S_{\rm inst} \sim R_AM_{\rm ADM}$ has this structure, and lies in the window \eqref{eq:emergence} when both the dual particle and the KK modes on $S^1_A$ lie below the three-dimensional species scale. This can happen for instantons dual to small black holes in the light towers (cf.\ section \ref{sec:instUVprobes}), but never for those dual to finite-size black holes, with $M_{\rm ADM}>M_p>\Lambda_{\rm sp}$. Although our setup only has eight supercharges, this suggests that the former are emergent while the latter are not.

\paragraph{Imaginary distance bound and EFT breakdown.}

\cite{Maldacena:2026jqd} and \cite{DiUbaldo:2026rly} proposed the \emph{imaginary distance bound}: in a UV-complete theory of quantum gravity, some effect must invalidate the low-energy description at or before the moduli reach an imaginary distance\footnote{In \cite{DiUbaldo:2026rly} they argued that otherwise the positivity of the gravitational path integral is violated.}
\begin{equation}
\tau_{\rm IDB}=\frac{\pi}{2}\sqrt{\frac{2(D-1)}{D-2}}\;\overset{D=4}{=}\;\frac{\sqrt3\,\pi}{2} \,,
\end{equation}
from the real slice of moduli space, with kinetic terms normalized as in \eqref{eq:euclidean-action-ST}. The effect is typically an instanton whose action vanishes when continued to complex moduli. This bound does not apply to our D-instantons, despite the fact that the axions develop an imaginary profile. The Euclidean wormholes of \cite{Maldacena:2026jqd} describe timelike geodesics once the axions are Wick-rotated. Our null trajectories lie on the boundary of this timelike region, and the bound as formulated does not constrain them.

Nevertheless, as discussed in section \ref{sec:instUVprobes}, the core of our instantons sits at strong coupling and, for large charges, at large values of the K\"ahler moduli \eqref{eq:attractedvalue}, where the classical description should break down. This breakdown, however, does not fall within the standard scope of either the imaginary distance bound or the Swampland Distance Conjecture, since the core lies at zero distance along a null trajectory. In Appendix \ref{sec:appendixB} we will explore how the theory could break down along classical light-like trajectories in the complexified moduli space.

\paragraph{EFT instantons.} In four-dimensional $\mathcal N=1$ theories, EFT instantons are BPS instantons whose charges lie in the dual saxionic cone $\mathcal P$, so that their backreacted profile remains within the EFT down to the UV cutoff \cite{Lanza:2021udy,Martucci:2024trp}. On the truncation of the hypermultiplet moduli space spanned by the four RR axions $\chi_I$ and their saxions $s_I$, our instantons are precisely of this type, with the harmonic functions $H_I$ playing the role of the dual saxions, and generalize the D$(-1)$ solution of \cite{Gibbons:1995vg}. Since the saxionic cone is the positive orthant, so is $\mathcal P$, and all our instantons are EFT instantons, if $\mathcal P$ is taken to include its boundary as in \cite{Martucci:2024trp}: the bound states dual to finite-size black holes lie in its interior, where all BPS string tensions grow towards the core of the instanton \cite{Martucci:2024trp}, while those dual to small black holes lie on its infinite-distance boundary. The arguments of \cite{Licciardello:2026roi} suggest that, in a saxionic sector of homogeneity degree $k$, the imaginary distance bound can be realized by towers of BPS EFT instantons only if $k\leq3$. Our truncation has $k=4$, but its kinetic potential factorizes, so that turning on at most three of the four charges defines consistent subsectors with $k'\leq3$. These are precisely the instantons dual to \emph{small} black holes. The bound states dual to finite-size black holes, which carry all four charges, belong instead to the $k=4$ sector, where, according to \cite{Licciardello:2026roi}, the bound could only be realized by non-BPS instantons.

\paragraph{Summary.} The instantons T-dual to black holes have an action set by the ADM mass of the dual black hole and a measure set by its BPS index, while their classical cores, at strong coupling and, for large charges, at large volume, should not be trusted. On the black-hole side, small black holes are the species themselves, i.e.\ the BPS particles of the light towers, while minimal black holes are macroscopic objects of species-scale size, whose entropy $S_{BH}\sim(M_p/\Lambda_{\rm sp})^2\sim N_{\rm sp}$ counts the number of species of the theory. This is reflected in the instantons. Those dual to small black holes have an order-one, charge-independent measure: they are T-dual to the light three-dimensional states, become unsuppressed along infinite-distance limits at fixed $\varphi_4$, and their number sets the cutoff \eqref{eq:MWspecies}. Those dual to minimal black holes remain suppressed, but carry a measure $\log|\Omega|\sim S_{BH}\sim N_{\rm sp}$, i.e.\ the same degeneracy as their dual black hole.

\subsection{ Strings}\label{sec:EFTstringtorus}

We consider now the configurations obtained performing a T-duality along a space-like circle. They are string-like, codimension two defects.

\subsubsection{Strings dual to black hole probes}
We start by constructing the string T-dual to the D0--D4--D4--D4 black hole of section \ref{sec:BHprobes}.

\paragraph{The D1--D5--D5--D5 string.} We T-dualize the ten-dimensional type IIA configuration along a spatial direction. This yields a four-charge D1--D5--D5--D5 configuration, in which the D1 is electrically charged and the D5s are magnetically charged. Compactifying on $T^6/(\mathbb{Z}_2\times\mathbb{Z}_2)$ as in the previous two cases, we obtain the action
\begin{equation}
\begin{aligned}
    S=\frac{M_p^2}{2}\int d^4x \sqrt{-g}\Big(&R- \frac{1}{2\Phi^2}(\partial \Phi)^2 - \sum_{i=1}^3\frac{1}{2(t^i)^2}(\partial t^i)^2 -\frac{1}{2\cdot3!}f_0|F_3^0|^2-\frac{1}{2}\sum_{i=1}^3f_i |F_1^i|^2\Big)\,,
    \end{aligned}
\end{equation}
with the gauge kinetic functions
\begin{equation}
    f_0=\Phi t^1t^2t^3\,,\quad f_1=\frac{t^2t^3}{\Phi t^1}\,,\quad f_2=\frac{t^1t^3}{\Phi t^2}\,,\quad f_3=\frac{t^1t^2}{\Phi t^3}\,.
\end{equation}
where $\Phi$ and $t^i$ have the same interpretation as for the previous sections. The 1-form field strengths are those of the RR axions $c^i$, the periods of $C_2$ over the two-tori $T^2_i$, i.e. $F_1^i=dc^i$. The 3-form is the field strength of the four-dimensional RR 2-form which is dual to the $c^0$ axion of section \ref{sec:cmap} via $dc^0=\pm f_0\star_4F_3^0$. The solution corresponding to a D1--D5--D5--D5 is given by the Einstein-frame metric
\begin{equation}
    ds^2_{4}=(-dt^2+dx^2)+H_0H^1H^2H^3(dr^2+r^2d\theta^2)\,,
\end{equation}
the dilaton and K\"ahler moduli profiles 
\begin{equation}
    \Phi =\sqrt{H_0H^1H^2H^3}\,, \quad t^1=\sqrt{\frac{H_0H^1}{H^2H^3}}\,,\quad t^2=\sqrt{\frac{H_0H^2}{H^1H^3}}\,,\quad t^3=\sqrt{\frac{H_0H^3}{H^1H^2}}\,,
    \label{moduli} 
\end{equation}
and the 3-form and 1-form field strength
\begin{equation}
    F_3^0=d(H_0^{-1})\wedge dt\wedge dx \,, \qquad F_1^i=r\,\partial_rH^i\,d\theta=-p^i\,d\theta \,.
\end{equation}
The functions $H$ are harmonic on $\mathbb{R}^2$ and have a logarithmic profile
\begin{equation}
    H_0=h_0-q_0\log(r/r_0) \,, \qquad  H^i=h^i-p^i\log(r/r_0) \,.
\end{equation}
The constants $h_I=(h_0,h^i)$ are fixed in terms of the asymptotic values of the scalar fields via \eqref{moduli}. $q_I=(q_0,p^i)$ are related with the integer D-brane charges of the configurations and $r_0$ is an integration constant which we leave general for now but it can be used to fix the units of $r$. Notice that this solution can be obtained performing a c-map along a space-like direction, provided that we first smear the  black hole along it, in order to have an isometry direction. Then the harmonic functions only depend on the two remaining transverse directions, and become logarithmic.

Let us comment on the spacetime profile of the defect. To do this, let us write the transverse metric as $ds_\perp^2 = e^{-2U}\left(dr^2+r^2d\theta^2\right)$, with
$e^{-2U} \equiv H_0 H^1 H^2 H^3$. Since the harmonic functions are logarithmic there is no event horizon. Near the core of the string, $r\to 0$, all harmonic functions grow, so that $\Phi=e^{-U}\to\infty$ and the four-dimensional string coupling $g_{4,s}=\Phi^{-1/2}$ goes to zero: the theory is weakly coupled and the supergravity description is under control. However we have a curvature singularity and curvature corrections are not suppressed. Moving away from the defect, the harmonic functions decrease, and some of the moduli eventually become of order one at a finite radius $r^*$. In a generic Calabi--Yau compactification, the solution can no longer be trusted beyond this point. If we keep moving in the radial direction the harmonic functions eventually vanish, providing another curvature singularity which signal the complete breakdown of the solution. Thus, we cannot reach the asymptotically flat region. In this sense, these solutions are \emph{confined}: they cannot exist as solitonic objects in asymptotically flat spacetime.

\paragraph{Tension, black hole ADM mass and axions.} The absence of an asymptotic region implies that we do not only lack a horizon surrounding the singular core; we are also not able to extend spacetime integrals of a conserved current till the asymptotic region. This means that we can not define charges and tensions as for branes of higher codimension. One of the quantities people investigated is the string tension $\mathcal{T}$. It can be equivalently computed as the energy density per unit-length contained in a space-like cylinder enclosing the string or as the deficit angle of a circumference on the transverse plane. Following the second prescription, let us introduce the proper circumference and the proper radius,
\begin{equation}
    C(r) \equiv 2\pi r\sqrt{e^{-2U(r)}}\,, \qquad
    \rho(r) \equiv \int^r dr'\,\sqrt{e^{-2U(r')}}\,.
\end{equation}
The local deficit angle $\delta(r)$ is
\begin{equation}
     \delta(r) \equiv 2\pi - dC/d\rho = \pi \sum_I \frac{q_I}{H_I} \,, \label{eq:deficit_angle}
\end{equation}
and it determines the tension of the string via
\begin{equation}\label{eq:EFTstringtension}
    \mathcal{T}(r) = |\delta(r)|M_p^2 = \pi M_p^2 e^{\varphi_4} \left(\frac{q_0}{\sqrt{t^1 t^2 t^3}}+\sum_i p^i\frac{\sqrt{t^1 t^2 t^3}}{t^i}\right)\,,
\end{equation}
where  in the last step we expressed the harmonic functions in terms of the moduli in order to compare it with the familiar tensions of wrapped branes. The two terms scale as the tension of a D1-brane and of a D5-branes wrapping the four-cycles $T^2_j\times T^2_k$:
\begin{equation}
    \frac{\mathcal{T}_{\rm D1}}{M_p^2}=\frac{g_s}{\mathcal{V}}\,, \qquad \frac{\mathcal{T}_{\rm D5_i}}{M_p^2}=\frac{g_s\,t^j t^k}{\mathcal{V}} \,.
\end{equation}
The tension of the string is thus the sum of the tensions of its constituents, as expected for a BPS bound state.

Let us analyze the features of the tension \eqref{eq:EFTstringtension}. At the core it vanishes
\begin{equation}
    \lim_{r\to 0}\mathcal{T}(r)=0\,.
\end{equation}
This is not unlike a D7-brane in ten-dimensional type IIB, whose core tension also goes to zero. As discussed above, the core is the locus of a curvature singularity and the place where $\Phi$ blows up toward the weak coupling limit. There, towers of states become light, as per the Swampland Distance Conjecture, and we lose control of the EFT. It was therefore proposed in \cite{Lanza:2021udy} that we should stop trusting the solution at a distance $r_{UV}$ from the core. This $r_{UV}$ can be identified with the radius at which the proper distance probed is at the order of the UV cutoff scale $C(r_{UV}) \sim \rho(r_{UV}) \sim \Lambda_{UV}^{-1}$. One can then define the \emph{core tension} of the string as the tension evaluated at such cutoff
\begin{equation}
    \mathcal{T}_{core}= \mathcal{T}(r_{UV}) \,.
\end{equation}
The core tension is the quantity that should know about the dual black hole entropy and the species scale, but it can not be computed with EFT data alone.

Away from the core, one reaches a point at finite distance at which the first of the harmonic functions vanishes: there, the tension diverges, $\mathcal{T}\to\infty$. Before that, as discussed above, some moduli become of order one at $r^*$, thus entering the bulk of moduli space, where quantum corrections are unsuppressed. The exact value of $r^*$ depends on the solution at hand. We should not trust the solution in this regime either. The string solutions we discuss are therefore under control in the region in between $r_{UV}$ and $r^*$. 

Away from the special point $r_{UV}$,  like the instanton action, the tension is a sum over the charges rather than a product. It is more natural to match it with the ADM mass of the dual black hole, rather than to its entropy. Indeed, writing $h_I = \Phi(r_0)^{1/2}\hat h_I$ with $\hat h_0 \hat h^1 \hat h^2 \hat h^3=1$, where $\Phi(r_0)=g_{4,s}(r_0)^{-2}$, and identifying the charges $q_I$ and the $\hat h_I$ with those of the black hole, \eqref{eq:4dMetricBHADM} gives
\begin{equation}\label{eq:TensionADMmass}
    \frac{\mathcal{T}(r_0)}{M_p^2} = \frac{g_{4,s}(r_0)}{2}\,\frac{ M_{\rm ADM}}{ M_p}\,.
\end{equation}
Although there is no asymptotic infinity, we can measure the tension at the finite distance point $r_0$ and recover the ADM mass of the dual black hole. This perhaps motivates identifying $r_0$ with $r^*$ which identifies the largest radial distance away from the string where the solution is trustable. In this way $r^*$ plays the role of asymptotic infinity in measuring the ``ADM tension''. 

Finally, let us consider the axion duality frame. We have the scalar profiles
\begin{equation}
    c^i=-p^i\theta+{\rm const} \,, \qquad c^0=\pm q_0\theta+{\rm const} \,,
\end{equation}
with all the other axions, $C_0$, $d^i$, $b^i$ and $\sigma=b^0$, which are constant, and $b^i$ takes its attractor value $b^i_h=0$. The string is therefore an axionic string: going around it, the axions shift by the charges, $(c^0,c^i)\to(c^0\pm q_0,\,c^i-p^i)$ in units of $2\pi$, as anticipated in section \ref{sec:runningcmap}. Their saxionic partners are the harmonic functions themselves,
\begin{equation}\label{eq:EFTsaxions}
s^0\equiv H_0=e^{-\phi}\mathcal{V}\,,\qquad s^i\equiv H^i=e^{-\phi}t^i\,,
\end{equation}
which run logarithmically as for the EFT strings of \cite{Lanza:2021udy}. In terms of the K\"ahler potential $K=-\sum_I\log s^I$ of this truncation, the tension \eqref{eq:EFTstringtension}  takes the form $\mathcal{T}=2\pi M_p^2\sum_Iq_I\ell_I$ with $\ell_I=-\frac12\partial K/\partial s^I$, matching the EFT-string tension formula of \cite{Lanza:2021udy} with their charges $e^I=2\pi q_I$.

\subsubsection{Strings probing the boundary of moduli space}\label{sec:stringprobestorus}

Strings are Lorentzian configurations, thus we can map the scalar field profile into a trajectory of the moduli space. As we explained, we can not start from the asymptotic region of the spacetime, so we consider trajectories from the reference radial IR cutoff $r = r_0$ to the core $r = 0$. The point in moduli space we are setting at the IR cutoff can be fixed by a choice of the constant part of the harmonics. At the core, the dilaton is always driven to weak coupling and the K\"ahler moduli are attracted to the same values \eqref{eq:attractedvalue} they had at the horizon or at the singularity of the T-dual black hole. The absence of a horizon implies that the strings T-dual to finite-size and small black holes always explore infinite distance in moduli space. One might wonder if we could explore limits with large K\"ahler moduli in the same way we did with families of finite-size black holes. As in section \ref{sec:blackholestorus}, we can consider families of scalar field profiles which start at the same point of moduli space. Different choices of D-brane charges provide then different profiles as a function of the radial coordinate. However, a finite rescaling of the charges does not modify the infinite distance limit we approach following the attractor flow. The only way out is then to stop before reaching the core. A natural prescription is then to stop when we probe distances at the cutoff energy of the theory, i.e. at $r_{UV}$. In the following we analyze in details the differences among the T-duals of the two families we considered in section \ref{sec:BHprobes}. Then we make concrete the proposal for the limits with infinite K\"ahler moduli in setups in which the UV cutoff scale is under control.

\paragraph{Strings dual to black hole attractors.}\label{sec:EFTminimal}
Let us now focus on the strings dual to finite-size black holes, and in particular to minimal black holes. For the D1--D5--D5--D5 string, as $r\to0$ volumes approach the constant values and the only \textit{running} quantity is the four-dimensional dilaton
\begin{equation}\label{eq:scalingNearTheCore}
\begin{aligned}
 &  t^1 \xrightarrow{r\to0} \Big(\frac{q_0p^1}{p^2p^3} \Big)^{1/2}\,, \qquad t^2 \xrightarrow{r\to0} \Big(\frac{q_0p^2}{p^1p^3} \Big)^{1/2}\,,\qquad t^3\xrightarrow{r\to0} \Big(\frac{q_0p^3}{p^1p^2} \Big)^{1/2}\,, \\[2mm] 
 &\mathcal{V} \xrightarrow{r\to0}\frac{q_0^{3/2}}{(p^1p^2p^3)^{1/2}}\,,\quad \qquad  g_{4,s} \sim \left|\frac{(q_0p^1p^2p^3)^{-1/4}}{\log(r/r_0)}\right|\to0 \,.
    \end{aligned}
\end{equation} 
As an infinite distance limit, we are describing a weak-coupling limit with large constant volumes. The fundamental string is then the lightest tensionless string. At the core its tension is
\begin{equation}
   \frac{\mathcal{T}_{\rm F1}}{M_p^2}\sim(g_{4,s})^2\sim \frac{1}{\log^2(r/r_0)}\xrightarrow{r\to0}0\,,
\end{equation}
while all D-brane and solitonic brane strings, including the source string, and particle-like states obtained wrapping D3 branes on cycles scale with lower powers of the string coupling constant. The KK modes of the two cycles scale as $M_{\rm KK} \sim g_{4,s} M_p $, thus we have an equi-dimensional emergent-string limit, in which the fundamental string is the lightest string.

The emergent F1 limit is not special of the system we analyzed, but it is a universal feature of all strings T-dual to finite-size black holes. The finiteness of all volumes at the core has been directly inherited from the black hole attractor mechanism. Therefore, despite the fact that the dual black holes were exploring different infinite-distance limits in IIA vector-multiplet moduli space, all of the dual strings explore the same limit in type IIB hypermultiplet moduli space. The latter is affected by instanton corrections, but we will argue in section \ref{sec:quantumCorrections} that the F1 limit survives these corrections.   

Finally, observe that near the core we have
\begin{equation}\label{eq:dilatonSbh}
e^{-2\varphi_4}\sim {S_{BH}} \log^2(r/r_0)\,,\qquad r\to0\,,
\end{equation}
after we identified the D-brane charges of the string with those of the T-dual system. Consequently, at a fixed radius the entropy of the dual black hole determines how light the F1 tower is
\begin{equation}\label{eq:TF1entropy}
    \frac{\mathcal{T}_{\rm F1}}{M_p^2}\sim\big(g_{4,s}\big)^2=\Phi^{-1}\sim\frac{1}{S_{BH}\,\log^2(r/r_0)}\,.
\end{equation}

\paragraph{String probes and charge rescaling.}\label{sec:BHlimitRescalingCharges}

It is natural to ask what happens if, at some finite distance from the core, the charges are rescaled. By itself the radial coordinate is not physical, but we can select a value $r_{UV}$ as the radius at which our proper radius is probing the UV cutoff scale. Once $r_{UV}$ is fixed we can then verify whether we can evade the universal F1-limit identified in the previous section. We will now present two examples showing that this is indeed possible. To do so, we plug the scalar profiles into \eqref{eq:EFTstringtension} at the fixed radius $r_{UV}$,
\begin{equation}
    \frac{\mathcal{T}(r_{UV})}{M_p^2}=\pi\sum_I\frac{q_I}{H_I(r_{UV})}\,,\qquad H_I(r_{UV})=h_I+q_I\log(r_0/r_{UV})\,,
\end{equation}
so that each unit of charge contributes the tension $\pi M_p^2/H_I(r_{UV})$ of the corresponding constituent, and blow up some of the charges. For the strings to be dual to black holes in the large-volume regime, all charges must be large, including those denoted $\mathcal{O}(1)$, which are kept fixed along the limit. Thus, also the volumes denoted by $1$ mean that they are constant and large. As a charge $q_I$ grows, its contribution to $\mathcal T(r_{UV})$ saturates to $\pi M_p^2/\log(r_0/r_{UV})$, while the tension of the corresponding constituent decreases as $1/q_I$, and the moduli at $r_{UV}$ move as follows. 
\begin{itemize}
    \item $q_0\to \infty,\,\,p^i\sim \mathcal{O}(1)$: the moduli at $r_{UV}$ scale as
    \begin{equation}
        g_s\sim q_0^{1/2}, \,\,\,
         t^i \sim q_0^{1/2},\,\,\, \mathcal{V} \sim q_0^{3/2}.
    \end{equation}
    The lightest objects are the D1-branes, $\mathcal{T}_{\rm D1}/M_p^2=\pi/H_0\sim 1/q_0$, and the KK modes, $M^2_{KK}/M_p^2\sim 1/q_0$. We are in a D1-limit, i.e.\ an equi-dimensional emergent-string limit with the D1 as the critical string, S-dual to a weak-coupling F1 limit at fixed volumes.
\item  $q_0=p^1=\lambda \to \infty,\,\,p^2 \sim p^3\sim \mathcal{O}(1)$: the moduli at $r_{UV}$ scale as
    \begin{equation}
        g_s\sim 1, \,\,\,
         t^1 \sim \lambda,\,\,\, t^2\sim t^3 \sim 1,\,\,\, \mathcal{V} \sim \lambda.
    \end{equation}
    The D1 and the D5 on $T^2_2\times T^2_3$, with $\mathcal T/M_p^2\sim1/\lambda$, and the fundamental string, $\mathcal{T}_{\rm F1}/M_p^2\sim1/\lambda$, are heavier than the KK modes on $T^2_1$, $M^2_{KK}/M_p^2\sim 1/\lambda^2$. This is a decompactification of $T^2_1$ to six dimensions at fixed ten-dimensional coupling. The same holds for permutations of $ p^1, p^2, p^3$. 
    \end{itemize}
Thus, we see that if one were able to stop at some finite distance from the core and start rescaling charges, one can classically probe other limits. In a Calabi--Yau compactification, the corresponding limits can be modified by quantum corrections to the hypermultiplet metric, which we discuss in section \ref{sec:quantumCorrections}.

Note that these D1/D5 string limits are obtained in the limiting case where the charges $q_0$ or $\lambda$ are strictly infinite. For any large but finite charge representative of these solutions, the dilaton will take over at the core. Then the infinite distance probed will always be the F1 string limit. There, we observe that for $\rho(r_{UV}) \sim l_s$ we get $r_{UV} \sim {l_s}/{S_{BH}}$, then the terms $\log(r_{UV}/r_0)$ are not independent of the charges we are rescaling. Despite the presence of the logarithm will never dominate on the polynomial scaling, it can drive apart objects with the same polynomial scaling. The K\"ahler moduli $t^i$ are not affected in the cases analyzed because at leading order the logarithmic contribution cancels. From \eqref{eq:TF1entropy} we can easily read that instead $g_s$ is suppressed by a factor $\log(S_{BH})$. This means that objects whose tensions or mass squared have the highest power of $g_s$ will be the lightest. In the cases previously identified as equi-dimensional emergent-string limits, this implies that the KK scale would be always logarithmically lighter (or equal for an F1-limit) to the emergent-string string scale.

\paragraph{Strings dual to small black holes.}\label{sec:EFTsmall}
Recall from section \ref{sec:blackholestorus} that small black holes are solutions whose charges are not sufficient to support a finite-size horizon. In the two-derivative theory their horizon area vanishes, they sit at a curvature singularity and the scalars are dragged to infinite distance at their core. The same properties are shared by the T-dual strings except that the four-dimensional dilaton is not constant but always runs to weak coupling at their core.

Near the core, each charged harmonic function grows like $L\equiv\log(r_0/r)\to\infty$, while the others remain constant. Since $\Phi=\sqrt{H_0H^1H^2H^3}\to\infty$ as soon as one charge is non-vanishing, the four-dimensional dilaton always runs to weak coupling, $g_{4,s}\sim L^{-n/4}$, with $n$ the number of non-vanishing charges. The saxions $s^I$ dual to the non-vanishing charges grow linearly in $L$, and the tension \eqref{eq:EFTstringtension} vanishes at the core as $\mathcal{T}\simeq n\pi M_p^2/L$. All these strings are weakly coupled at the core, which lies at infinite distance in moduli space, and become tensionless there. Note that the ten-dimensional coupling need not go to zero; for the D1 string it grows. The different bound states can produce different classical limits.

Within the D0--D4--D4--D4 system, small black holes are obtained by switching off some of the charges, i.e.\ by setting the corresponding harmonic functions to constants, so that $q_0p^1p^2p^3=0$. T-dualizing along a spatial direction as before, the D0 becomes a D1 string and each D4 becomes a D5 string wrapping the corresponding four-cycle, so that the dual strings are the D1, D5, D1--D5, D5--D5, D1--D5--D5 and D5--D5--D5 strings. Only the axions dual to the non-vanishing charges wind around the string. Unlike in the finite-size case, the K\"ahler moduli are no longer attracted to finite values, and the $B$-field axions are no longer fixed at the core. We find the behaviour summarized in table \ref{tab:EFTsmall}:
\begin{table}[t]
    \centering
    \small
    \renewcommand{\arraystretch}{1.3}
    \setlength{\tabcolsep}{4pt}
    \begin{tabular}{l|c|c|c|c|l|c}
      String & $g_{4,s}$ & $g_s$ & $(t^1,t^2,t^3)$ & lightest, $\mathcal{T}/M_p^2$ & limit & $w$ \\ \hline
      D1 & $L^{-1/4}$ & $L^{1/2}$ & $(L^{1/2},L^{1/2},L^{1/2})$ & KK, D1: $L^{-1}$ & emergent D1 & 1 \\
      D5 & $L^{-1/4}$ & $L^{-1/2}$ & $(L^{1/2},L^{-1/2},L^{-1/2})$ & KK, wind., D5: $L^{-1}$ & emergent D5 & 1 \\
      D1--D5 & $L^{-1/2}$ & $1$ & $(L,1,1)$ & KK $T^2_1$: $L^{-2}$ & decomp.\ to 6d & 2 \\
      D5--D5 & $L^{-1/2}$ & $L^{-1}$ & $(1,1,L^{-1})$ & wind.\ $T^2_3$: $L^{-2}$ & decomp.\ to 6d & 2 \\
      D1--D5--D5 & $L^{-3/4}$ & $L^{-1/2}$ & $(L^{1/2},L^{1/2},L^{-1/2})$ & KK, wind.: $L^{-2}$ & decomp.\ to 10d & 2 \\
      D5--D5--D5 & $L^{-3/4}$ & $L^{-3/2}$ & $(L^{-1/2},L^{-1/2},L^{-1/2})$ & wind.: $L^{-2}$ & decomp.\ to 10d & 2 \\ \hline
      D1--D5--D5--D5 & $L^{-1}$ & $L^{-1}$ & $(1,1,1)$ & KK, F1: $L^{-2}$ & emergent F1 & 2
    \end{tabular}
 \caption{Classical behaviour at the core, $L=\log(r_0/r)\to\infty$, of the strings dual to the small black holes of the D0--D4--D4--D4 system on $T^6/(\mathbb{Z}_2\times \mathbb{Z}_2)$. The non-vanishing charges are $q_0$, $p^1$, $(q_0,p^1)$, $(p^1,p^2)$, $(q_0,p^1,p^2)$ and $(p^1,p^2,p^3)$, respectively; other choices follow by permuting the two-tori. The last line is the finite-size case of section \ref{sec:EFTminimal}. For strings, $\mathcal{T}$ stands for the tension; ``wind.'' denotes winding modes; $w$ is the scaling weight \eqref{eq:scalingweight}, computed with the tension of the string $\mathcal T/M_p^2\simeq n\pi/L$, with $n$ the number of non-vanishing charges. Every $1$ in this table should be interpreted as a large but finite number, as to stay in the large volume approximation.}
  
    \label{tab:EFTsmall}
\end{table}

Whenever some two-tori shrink in table \ref{tab:EFTsmall}, the limit is read in the frame obtained by double T-duality along them, $z^i\to-1/z^i$, which maps $T^6/(\mathbb{Z}_2\times\mathbb{Z}_2)$ to a geometric orbifold \cite{Pradisi:2014fqa} (see also \eqref{eq:double-T-duality}) and in which all volumes are large; in a generic Calabi--Yau, the corresponding regimes would instead be sensitive to small-volume corrections.

\begin{itemize}
    \item \textbf{Emergent-string limits:} D1 and D5. For the D1 string, the volumes and the ten-dimensional coupling grow at the same rate, so that, in the S-dual frame, the volumes $t^i/g_s$ remain fixed while the coupling goes to zero as $L^{-1/2}$. This is a weak-coupling limit in which the D1 is the emergent critical string. Its tension scales like the KK scale, $\mathcal{T}_{\rm D1}/M_p^2\sim M_{KK}^2/M_p^2\sim L^{-1}$, so the limit is equi-dimensional. The D5 string wrapping $T^2_2\times T^2_3$ is mapped to the D1 by T-duality along the two shrinking two-tori. In both cases, the string itself becomes the critical string. This is the same kind of limit as the D1 limit of \cite{Baume:2019sry}, which in Calabi--Yau compactifications only arises after quantum corrections are taken into account. On $T^6/(\mathbb{Z}_2\times\mathbb{Z}_2)$, it is realized classically.
    \item \textbf{Decompactification to six dimensions:} D1--D5 and D5--D5. For the D1--D5 string, the ten-dimensional coupling stays fixed, as do $t^2$ and $t^3$, while $t^1\sim L$ grows. The KK modes on $T^2_1$, with $m^2/M_p^2\sim L^{-2}$, are lighter than the string scale, $\mathcal{T}_{\rm F1}/M_p^2\sim L^{-1}$, and the limit is a decompactification of $T^2_1$ to six-dimensional type IIB on the K3 fibre $T^4/\mathbb{Z}_2$. The D5--D5 string is mapped to a D1--D5 one by T-duality along $T^2_1\times T^2_3$.
    \item \textbf{Decompactification to ten dimensions:} D1--D5--D5 and D5--D5--D5. After T-duality along the shrinking two-tori, all three volumes grow uniformly, $t^i\sim L^{1/2}$, at fixed ten-dimensional coupling. All KK modes become light at the same rate, $m^2/M_p^2\sim L^{-2}$, and the limit is a full decompactification to ten-dimensional type IIB string theory. The leading tower is the KK tower, and the species scale is the ten-dimensional Planck scale, $\Lambda_{\rm sp}^2/M_p^2\sim (M_{KK}/M_p)^{3/2}\sim L^{-3/2}$, which at fixed ten-dimensional coupling is parametrically the string scale.
\end{itemize}
\noindent Comparing with section \ref{sec:blackholestorus}, we see that even if we can produce several limits, T-duality does not preserve the type of limit probed by small black holes. The single-charge black holes probe a decompactification to five dimensions, while their T-dual strings probe emergent-string limits. The two-charge black holes probe emergent-string limits, while their T-dual strings decompactify two dimensions. The three-charge black holes again probe a decompactification to five dimensions, while their T-dual strings decompactify all six. The reason is the same as for finite-size black holes: on the type IIB side, the four-dimensional dilaton necessarily runs, $g_{4,s}\sim L^{-n/4}$, and the nature of the limit is determined by how this running competes with that of the K\"ahler moduli. 

\subsubsection{Relation to EFT strings}\label{sec:stringbounds}

On the truncation spanned by the RR axions $c^I$ and their saxions $s^I$, our strings are EFT strings in the sense of \cite{Lanza:2021udy}: the axions wind with monodromies given by the charges, the saxions $s^I=H_I$ flow linearly along the charge vector, which lies in the cone of EFT string charges, and the tension \eqref{eq:EFTstringtension} vanishes at the core, which lies at infinite distance. Bound states are part of the framework of \cite{Lanza:2021udy}, although its examples mostly focus on elementary flows. Along bound-state flows only the ratios of charges enter the ten-dimensional moduli, so that some of the K\"ahler moduli can reach \emph{classically} finite values at the core, and for the strings dual to finite-size black holes all of them do, leaving only the four-dimensional dilaton running (see section \ref{sec:EFTminimal}).

There is, however, an important difference. In four-dimensional $\mathcal N=1$ theories, the EFT condition on the charges guarantees that the instantons breaking the axionic shift symmetries are suppressed along the flow. Here, these are the ED1- and ED5-instantons, electric-magnetic dual to the D1- and D5-strings, whose actions are linear in $s^I$. However, the $\mathcal N=2$ hypermultiplet sector also contains D$(-1)$-, ED3-, world-sheet and NS5-instantons, which can become unsuppressed along EFT-string flows. The EFT condition alone therefore does not guarantee that the flows are under control, and we come back to this in section \ref{sec:quantumCorrections}.

\paragraph{Integral scaling relation.} Along such flows, \cite{Lanza:2021udy} conjectured that the mass scale of the leading tower behaves as
\begin{equation}\label{eq:scalingweight}
\frac{m_*^2}{M_p^2}\sim\Big(\frac{\mathcal T}{M_p^2}\Big)^{w}\,,\qquad w\in\mathbb Z_{>0}\,,
\end{equation}
with $w=1$ when the EFT string itself becomes the emergent critical string. Near the core $\mathcal T/M_p^2\simeq n\pi/\log(r_0/r)$, with $n$ the number of non-vanishing charges, and table \ref{tab:EFTsmall} gives $w=1$ for the D1 and D5 strings, which are themselves the emergent critical strings, and $w=2$ for all the others. In particular, for the strings dual to finite-size black holes the tower is that of the fundamental string, and \eqref{eq:TF1entropy} gives $m_*^2/M_p^2\sim\mathcal T_{\rm F1}/M_p^2\sim(\mathcal T/M_p^2)^2/S_{BH}$: the scaling weight is $w=2$ because the emergent critical string is the fundamental string rather than the EFT string, and the coefficient is set by the entropy of the dual black hole. These weights also satisfy the convexity $w_{e_1+e_2}\leq w_{e_1}+w_{e_2}$ observed in \cite{Lanza:2021udy}.

These weights fit the bottom-up analysis of \cite{Etheredge:2026jun}. For K\"ahler potentials $K=-\log P(s)$ with $P$ a single monomial, such as ours $K=-\sum_I\log s^I$, the $\alpha$-vector of a bound state is a weighted average of those of its constituents, and the scaling weight of a given tower along a bound-state flow is the sum of its weights along the elementary flows. Our strings dual to finite-size black holes realize this in a non-trivial way: the F1 tower, with $\mathcal T_{\rm F1}/M_p^2\propto(s^0s^1s^2s^3)^{-1/2}$, is subleading along each elementary D1 or D5 flow, with half-integer weight $w=\frac12$, and these add up to the integer $w=2$ along the bound-state flow. This also accounts for the convexity above, since the leading weight of a bound state is a maximum over towers of such sums. As in the examples of \cite{Etheredge:2026jun}, $w=2$ then signals an emergent string limit in which the emergent string, here the F1, is lighter than the EFT string sourcing the flow. Whether such limits survive quantum corrections is an open question there \cite{Etheredge:2026jun}, which we will answer in section \ref{sec:quantumCorrections}.

\paragraph{QG cutoff scale.} The tension of the string also sets an upper bound on the cutoff. As for the strings electric-magnetic dual to our instantons in section \ref{sec:instantonbounds}, a fundamental string whose core lies at infinite distance cannot be lighter than the QG cutoff, $\Lambda_{\rm sp}^2\lesssim2\pi\mathcal T$, and by electric-magnetic co-scaling $\mathcal T\sim2\pi S_{\rm inst}f^2$, with $S_{\rm inst}$ and $f$ the action and decay constant of the electric-magnetic dual instanton. This is the bound \eqref{eq:QGcutoff} seen from the string side, and \eqref{eq:scalingweight} sharpens it along the string flow. We thus obtain a square of dualities: electric-magnetic duality relates the D$(-1)$--ED3 instantons to D7--D3 strings, and the D1--D5 strings to ED5--ED1 instantons, while double T-duality along one of the $T^2_i$, see \eqref{eq:double-T-duality}, relates the instantons, and the strings, of the two pairs,
\begin{equation*}
\begin{tikzcd}[column sep=large]
\text{D}(-1)\text{--ED3--ED3--ED3} \arrow[r,leftrightarrow,"\text{e-m}"] \arrow[d,leftrightarrow,"T^2_i"'] & \text{D7--D3--D3--D3} \arrow[d,leftrightarrow,"T^2_i"]\\
\text{ED5--ED1--ED1--ED1} \arrow[r,leftrightarrow,"\text{e-m}"'] & \text{D1--D5--D5--D5}
\end{tikzcd}
\end{equation*}

\section{General case: Calabi--Yau black holes and the c-map}\label{sec:CY}
In this section we extend the construction of section \ref{sec:torus} to type II compactifications on a generic Calabi--Yau threefold. We first review BPS black holes of type IIA in the two-derivative, large-volume approximation, built from the stabilization equations and the attractor mechanism. We then derive explicitly the dictionary between type IIA and type IIB realizing the c-map and T-duality, along the Euclidean time circle and along a spatial circle, which maps these black holes into D-instantons and strings in the type IIB hypermultiplet sector, respectively. We study the properties of both, and in particular the infinite-distance limits classically probed by the strings near their core. As for the toroidal orbifold, strings dual to finite-size black holes always probe an emergent F1 limit, since the attractor mechanism fixes the K\"ahler moduli at the core, whereas those dual to small black holes can explore different limits.

\subsection{Type II on a CY$_3$} \label{sec:conventions}

We now set up the four-dimensional $\mathcal N=2$ effective theories of type IIA and type IIB on a Calabi--Yau threefold, and recall how the c-map and T-duality relate the IIA vector multiplet sector to the IIB hypermultiplet sector, generalizing section \ref{sec:cmap}. We then fix our conventions for $\mathcal N=2$ supergravity, which we use throughout the rest of this section.

\paragraph{c-map and T-duality.} CY compactifications of type II string theories are described at low energies by a four-dimensional, $\mathcal{N} =2$ ungauged supergravity with $n_V$ vector multiplets and $n_H$ hypermultiplets. The bosonic fields of the theory are the metric, $n_V+1$ gauge fields $A^I$, including the graviphoton, $n_V$ complex scalars $z^i$, and $4n_H$ real scalars $q^u$. In the classical, two derivative approximation, the scalar manifold locally factorizes as $\mathcal{M}_{4d}=\mathcal{M}_{VM}\times\mathcal{M}_{HM}$, where the two factors have special K\"ahler and quaternionic-K\"ahler geometry, respectively. As we explained in section \ref{sec:cmap}, the c-map is realized in the three-dimensional theory obtained compactifying one further space-like dimension. Compactifying on a circle the vector and gravity multiplets we produce new vectors and scalars. The latter corresponds to the KK radius $R$ and the Wilson lines $\xi^I$. Dualizing the $n_V+2$ three-dimensional vectors gives the scalars $(\tilde{\xi}_I,\sigma)$, where $I=0,\ldots,n_V$. Together with $z^i$ and $R$ and $\xi^I$, these form $n_V+1$ twisted hypermultiplets of three-dimensional $\mathcal{N}=4$ supergravity. The classical spatial c-map is the embedding of $\mathcal{M}_{VM}$ into $\mathcal{M}_{twisted}$.

For type IIA and type IIB compactified on the same Calabi--Yau threefold $X$, the four-dimensional multiplet numbers satisfy $n_V^{\rm IIA}=n_H^{\rm IIB}-1=h^{1,1}(X)$ and $n_V^{\rm IIB}=n_H^{\rm IIA}-1=h^{2,1}(X)$. Further compactification on the dual spatial circles $S_A^1$ and $S_B^1$ gives equivalent three-dimensional theories. T-duality is realized exchanging their ordinary and twisted hypermultiplet sectors, i.e. the c-map of their vector multiplet sectors.
\begin{equation}
\label{eq:cmap_Tduality}
\mathcal{M}_{ VM}^{\rm IIA/B} \xrightarrow{\rm c-map} \mathcal{M}_{twisted}^{\rm IIA/B} \xleftrightarrow{\rm T-duality} \mathcal{M}^{\rm IIB/A}_{HM} \,.
\end{equation}
For instance, performing a c-map from type IIA to type IIB, the complexified K\"ahler moduli become the IIB hypermultiplet K\"ahler coordinates. The scalars $(\xi^I,\tilde{\xi}_I)$ encode the IIB RR axions, while $R_A$ and $\sigma$ encode the four-dimensional dilaton and the NSNS axion, with the standard axionic redefinitions~\cite{Alexandrov:2009qq}. In particular, $\xi^0=C_{0,B}$ and $g_{4,B}^{-1}=R_A M_{p,A}$, as in \eqref{eq:gsRA}. Notice that one can also perform a c-map along the time circle, provided that we perform first a Wick rotation to Euclidean signature. Indeed, if we apply the temporal c-map directly in a Lorentzian type $\rm IIA/B$ we obtain type $\rm IIB^*/A^*$ after the time-like T-duality, which have flipped sign for RR kinetic terms. Performing a Wick rotation first, T-duality maps Euclidean type $\rm IIA/B$ to Euclidean type $\rm IIB/A$.

\paragraph{Conventions and identities in $\mathcal{N}=2$ supergravity.}
We now briefly review our conventions for four-dimensional, $\mathcal{N}=2$ sugra realized as type IIA on a threefold $X$. Given that we are interested in black hole solutions with constant hypermultiplets we truncate them in the following. The bosonic action for the supergravity multiplet and the $n_V$ vector multiplets is then
\begin{equation} \label{eq:actionNonDemocratic}
    S^{\rm IIA} = \int d^4x \sqrt{g} \left\{ R- 2 g_{i\bar{j}} \partial_\mu z^i \partial^\mu \bar{z}^{\bar{j}} + \frac{1}{4}{\rm Im}\mathcal{N}_{IJ} F^{I}_{\mu\nu} F^{J \mu\nu} - \frac{1}{4} {\rm Re} \mathcal{N}_{IJ} F^I_{\mu\nu} (\star F^J)^{\mu\nu}\right\} \,.
\end{equation}
This action is written with mostly plus signature, we dropped the overall $M^2_p/2$ factor, the Ricci is negative on the sphere, ${\rm Im}\mathcal{N}_{IJ} < 0$, $ g_{i\bar{j}} > 0$ and the hodge star is defined exactly as in \cite{Behrndt:1997ny}.\footnote{i.e. we have the Levi-Civita symbol normalized with $\epsilon^{0123} = -\epsilon_{0123} = +1$ and it enters in the Hodge star operator via the relation $$\star \omega = \frac{\sqrt{g}}{k! (d-k)!} dx^{\mu_1} \wedge \cdots \wedge dx^{\mu_{d-k}} \epsilon_{\mu_1 \dots \mu_{d-k} \nu_1 \dots \nu_k} \omega^{\nu_1 \dots \nu_k} \,. $$}
The non-trivial kinetic matrices of the scalar and gauge sector are fully determined by a holomorphic function $\mathcal{F}(X^I)$ called prepotential 
\begin{equation}
    g_{i\bar j}=\partial_i \partial_{\bar{j}} K \,, \qquad  \mathcal{N}_{IJ} = \overline{\mathcal{F}}_{IJ} + 2i \frac{(\text{Im}\, \mathcal{F})_{IA} X^A (\text{Im}\, \mathcal{F})_{JB} X^B}{X^A (\text{Im}\, \mathcal{F})_{AB} X^B}\,,
\end{equation}
where \(\mathcal{F}_{IJ} = \partial_{X^I} \partial_{X^J} \mathcal{F}\) and $K$ is the K\"ahler potential
\begin{equation} \label{eq:kahlerpotential}
K = - \log i\left( \bar{X}^I \mathcal{F}_{I}- X^I \bar{\mathcal{F}}_I  \right)\, .
\end{equation}
The prepotential is homogeneous of degree two in $X^I$, i.e. $\mathcal{F} = \frac{1}{2} X^I \mathcal{F}_I$, where $\mathcal{F}_I = \partial_{X^I} \mathcal{F}$. The $X^I$, $I = 0,1,\dots n_V$ are homogeneous (projective) coordinates on a K\"ahler manifold and they are related to the lowest component of the vector multiplets (K\"ahler moduli) by $z^i = b^i + i t^i =  X^i/X^0$, $i = 1, \dots n_V$. $F^I$ are the field strengths of the  gauge fields $A^I$.

Let $J$ be the dimensionless K\"ahler 2-form of the CY. Introducing a basis $\{\omega_i\}$ of $H^{1,1}(X,\mathbb R)$, where $i=1,\ldots,h^{1,1}(X)$, normalized such that $\mathcal{K}_{ijk}=\int_X\omega_i\wedge\omega_j\wedge\omega_k = - 6 D_{ijk}$, we have the expansion $J=t^i\omega_i$. Using that $J^k$ is a calibration form, the string-units volume of fixed nonzero effective curves $C$, divisors $D$ and the threefold $X$ are
\begin{equation}
 \mathcal V= \frac{1}{6}\int J^3 = \frac16\mathcal{K}_{ijk}t^it^jt^k,\qquad v_D=\frac12\int_DJ^2,\qquad
 v_C=\int_CJ .
 \label{eq:intclass-volumes}
\end{equation}

Finally, it is useful to recast the quantities just introduced into symplectic vectors. We can introduce the symplectic section  $\Omega^I$, the holomorphic symplectic section $V^I$ and the symplectic matrix $\Omega_{IJ}$ 
\begin{equation}
 \Omega^I = \begin{pmatrix}
     X^I \\ \mathcal{F}_I
 \end{pmatrix}\,, \qquad
 V^I = \begin{pmatrix}
     L^I \\ M_I
 \end{pmatrix} =  e^{K/2} \Omega^I  \,, \qquad
    \Omega_{IJ} = \begin{pmatrix}
0 & \mathbf{1} \\
\mathbf{-1} & 0
\end{pmatrix}\,.
\end{equation}
They satisfy \cite{Ceresole:1995ca}
\begin{equation} \label{eq:specialGeom}
i \bar{V}^IV^J\Omega_{IJ} = 1 \,, \qquad U^I_i = D_i V^I \,, \qquad D_{\bar{j}} V^I = 0 \,, \qquad g_{i\bar{j}} = i \Omega_{IJ} U^I_i \bar{U}^J_{\bar{j}}\,,
\end{equation}
where $D_i = \partial_i + \frac{1}{2}K_i$ is the K\"ahler-covariant derivative with $K_i = \partial/\partial z^i K$. If we indicate the magnetic duals to the $F^I$ with $G_I$ 
\begin{equation}
    G_I ={\rm Re}\mathcal{N}_{IJ} F^J +{\rm Im}\mathcal{N}_{IJ} \star F^J \,,
\end{equation}
we can introduce the symplectic doublet $(F^I, G_I)$ and build a democratic (pseudo) action.\footnote{Both gauge fields and their duals appear in the action in a manifestly duality covariant way. The equations of motion of the democratic action are equivalent to those of the original action provided that an external constraint halves the amount of degrees of freedom. The necessity of such constraint is the reason for the name \emph{pseudo} action. See \cite{Fernandez-Melgarejo:2023kwk} for more details.} Writing the gauge sectors with forms we have
\begin{equation} \label{eq:actionDemocratic}
    S^{\rm IIA} = \int d^4x \sqrt{g} \left\{ R- 2 g_{i\bar{j}} \partial_\mu z^i \partial^\mu \bar{z}^{\bar{j}} + \frac{1}{8}\mathcal{M}_{IJ} F^I_{\mu\nu} F^J{}^{\mu\nu}\right\} \,,
\end{equation}
where with an abuse of notation we used $F^I$ also to indicate the doublet $(F^I, G_I)$ and we introduced the symplectic kinetic matrix $\mathcal{M}$ 
\begin{equation}
    (\mathcal M_{IJ}) =
\begin{pmatrix}
\mathcal I + \mathcal R\, \mathcal I^{-1}\, \mathcal R 
& -\,\mathcal R\, \mathcal I^{-1} \\[6pt]
-\,\mathcal I^{-1}\, \mathcal R 
& \mathcal I^{-1}
\end{pmatrix} \,,  \qquad 
\mathcal{I} ={\rm Im}\mathcal{N}_{IJ} \,, \qquad \mathcal{R} ={\rm Re}\mathcal{N}_{IJ} \,.
\end{equation}
To avoid the doubling of degrees of freedom we must enforce the twisted self-duality constraint $ \star F^I= - \Omega^{IJ} \mathcal{M}_{JK} F^K$ which can be equivalently written as $G_I^-=\bar{\mathcal{N}}_{IJ}F^{J,\, -}$, where the minus indicates the anti self-dual parts.

\subsection{Black holes}\label{sec:explicitBHs}

In this section we are going to review BPS black holes in type IIA compactified on a Calabi--Yau threefold in the two-derivative, large volume approximation. We focus on configurations which arise from the back-reaction of wrapped D-branes. In particular, we extend the AdS$_2 \times S^2 $ near horizon geometry that arises as a solution of the well-known attractor mechanism to a full black hole solution, i.e. a spacetime configuration interpolating  AdS$_2 \times S^2 $ and asymptotic Minkowski space with non-trivial scalar field profiles. We follow the method of \cite{Bergshoeff:1996gg, Behrndt:1997ny, Meessen:2006tu} based on the solution of the  so-called \textit{stabilization equations}  (see \cite{Ortin:2015hya} for a comprehensive review). We provide the explicit solutions in the large volume approximation, i.e. considering a generic cubic prepotential. We do so adapting the approach originally used in \cite{Shmakova:1996nz} to solve the attractor equations for a generic Calabi--Yau black hole (see \cite{Castellano:2025yur} for more details on the procedure). 

\paragraph{D-brane BPS Calabi--Yau black holes.}
In \cite{Behrndt:1997ny, Meessen:2006tu} it is shown that one can build a static, supersymmetric configuration considering the metric
\begin{equation} \label{eq:ansatzMetric}
ds^2 = - e^{2U} dt^2 + e^{-2U} dx^m dx^m \,,
\end{equation}
where $x^m$ are Cartesian coordinates on $\mathbb{R}^3$ and the gauge fields
\begin{equation} \label{eq:ansatzGaugeFields}
     \begin{pmatrix}
        F^I \\ G_I 
    \end{pmatrix}  = - 2 \sqrt{2} \, d({\rm Re} \,\Omega^I e^{2U}) \wedge dt - 2 \sqrt{2} \, e^{2U} \star (dt \wedge d \,{\rm Im}\,\Omega^I) \,.
\end{equation}
The configuration is a supersymmetric solution provided that the warping factor is related to the K\"ahler potential via  $U = K/2$, the K\"ahler connection  vanishes $K_\mu = 0$\footnote{Assuming a generic IWP metric, we obtain that the gravitino Killing spinor equation (KSE) has nontrivial solutions if $\partial_{[m}Q_{n]} = 0$. Then $Q_m = 0$ iff the configuration is static.} and we can solve the stabilization equations \footnote{Notice that there is a sign with respect to \cite{Behrndt:1997ch}. We are picking these conventions to profit from the results of \cite{Castellano:2025yur}.}
\begin{equation}   \label{eq:stabilization}
    2 \,{\rm Im}\, \mathcal{F}_I  = H_I \,, \qquad 2 \,{\rm Im}\, {X}^I  = \tilde{H}^I \,,
\end{equation}
with $ H_I$ and $\tilde{H}^I$ harmonic functions in $\mathbb{R}^3$. For single-center BH solutions they will have the general form
\begin{equation} \label{eq:harmonics}
    H_I = a_I + \frac{b_I}{r} \,, \qquad \tilde{H}^I = \tilde{a}^I + \frac{\tilde{b}^I}{r} \,,
\end{equation}
with $a_I,b_I, \tilde{a}^I, \tilde{b}^I$ some constants. The constants can be expressed in terms of physical quantities in the following way. The $b_I$, $\tilde{b}^I$ can be related via the expressions of $F^I$ and $G_I$ to the brane charges, usually indicated with $q_I$ and $p^I$ in the literature
\begin{equation} 
    (\text{D0},\text{D2},\text{D4},\text{D6}):\quad  (q_0, q_i, p^i, p^0)\,.
\end{equation}
$2 n_V$ of the $a_I$, $\tilde{a}^I$ can be fixed  via the asymptotic values of the $n_V$ (complex) scalar fields $z^i_\infty$. The vanishing of the K\"ahler connection imposes the algebraic constraint $b^I \Omega_{I J} a^J = 0$, where $a^I = (a_I, \tilde{a}^I)$, $b^I = (b_I, \tilde{b}^I)$ and $\Omega$ is the symplectic matrix with rank $2n_V + 2$. Finally, one integration constant is fixed by asymptotic flatness.\footnote{Notice that the integration constants must also satisfy proper inequalities to ensure that the ADM mass of the BH is positive.}

\paragraph{Stabilization equations in the large volume approximation.} \label{ss:cubicPrepotential}

The prepotential for type IIA on a CY in the 2-derivative, large volume approximation has the simple cubic form
\begin{equation}\label{eq:prepotentialCY}
    \mathcal{F} = \frac{D_{ijk} X^i X^j X^k}{X^0} \,.
\end{equation}
Our strategy to solve the stabilization equations is to use the same trick suggested in \cite{Behrndt:1997ny}. Formally the stabilization equations \eqref{eq:stabilization} are quite similar to the usual attractor mechanism equations
\begin{equation}
    2 \,{\rm Im}\, C\mathcal{F}_I  = q_I \,, \qquad 2 \,{\rm Im}\, C {X}^I  = p^I \,,
\end{equation}
where $C$ is the so-called compensator which makes the symplectic section neutral under K\"ahler transformations, and $p^I$ and $q_I$ are the D-brane charges we introduced previously. We can then make use of all the computations performed in the context of the standard attractor mechanism and extract solutions to the stabilization equations \eqref{eq:stabilization} performing the replacement
\begin{equation}
    p^I \rightarrow \tilde{H}^I \,, \qquad q_I \rightarrow  H_I \,.
\end{equation}
Following the steps performed in \cite{Castellano:2025ljk}, we obtain that 
\begin{subequations}
    \begin{align} \label{eq:firstAttractorEq}
    H_i = 2 \,{\rm Im}F_i  \quad  & \Rightarrow  \quad y^i = \frac{\sqrt{3}}{|X^0|} (\tilde{H}^0{\rm Re}X^i -{\rm Re}X^0 \tilde{H}^i)  \\[2mm]
    H_0 = 2 \,{\rm Im}F_0  \quad &  \Rightarrow \quad |X^0|^2 = \frac{(\tilde{H}^0)^2}{4}\frac{(\Delta_i y^i)^2}{(\Delta_i y^i)^2 - \frac{27}{4} D^2} \,, \label{eq:X0norm}
\end{align}
\end{subequations}
where $y^i$ is defined as the solution (if it exists) of $D_{ijk} y^j y^k = \Delta_i$ with
\begin{equation}
    \Delta_i = - \tilde{H}^0 H_i + 3 D_{ijk} \tilde{H}^j \tilde{H}^k \,, \qquad
    D = \tilde{H}^0 \tilde{H}^I H_I - 2 D_{ijk} \tilde{H}^i\tilde{H}^j\tilde{H}^k \,.
\end{equation}
From \eqref{eq:X0norm} also follow
\begin{equation} \label{eq:X0relations}
   {\rm Re}X^0 = \frac{\tilde{H}^0 3 \sqrt{3} D}{2 T} \,, \qquad |X^0| = \frac{\tilde{H}^0 \Delta_i y^i}{T} \,,
\end{equation}
with $ T = \sqrt{4 (\Delta_i y^i)^2 - 27 D^2 }$. Replacing \eqref{eq:X0relations} in \eqref{eq:firstAttractorEq} we get an expression for ${\rm Re} X^i$ 
\begin{equation}
   {\rm Re}X^i = \frac{1}{2T}\left(3 \sqrt{3}D \tilde{H}^i + \frac{2}{\sqrt{3}} \Delta_j y^j y^i \right) \,,
\end{equation}
and then $X^I$ is given by
\begin{subequations}
\begin{align}
    X^i & = \frac{1}{2T}\left(3 \sqrt{3}D \tilde{H}^i + \frac{2}{\sqrt{3}} \Delta_j y^j y^i \right) + \frac{i}{2}\tilde{H}^i \,, \\[2mm]
    X^0 & = \frac{\tilde{H}^0}{2}\left(\frac{3 \sqrt{3} D}{T}+i\right) \,.
\end{align}
\end{subequations}
Let's focus now on the warping factor. We can then write
\begin{equation}
    e^{-2U} = e^{-K} = 2 ({\rm Im}X^I {\rm Re}\mathcal{F}_I - {\rm Re}X^I \text{Im}\mathcal{F}_I) = 4 \left({\rm Im}\mathcal{F} - \frac{1}{2} H_I{\rm Re}X^I \right) \,,
\end{equation}
where we used the stabilization equations and the homogeneity properties of the prepotential. Comparing with the expression for the BPS black hole entropy \cite{Ooguri:2004zv} (see \cite{Castellano:2025yur} for the expression in our conventions) we realize that 
\begin{equation}
    e^{-K} = \frac{S_{BH}}{\pi} [p^I = \tilde{H}^I, q_I = H_I] \,.
\end{equation}
We can then evaluate the warping factor borrowing the result of \cite{Shmakova:1996nz}
\begin{equation}
    e^{-2U} =  \frac{1}{3 \tilde{H}^0} \sqrt{\frac{4}{3}(\Delta_i y^i)^2 - 9 D^2} \,.
\end{equation}
With a little bit of algebra it is not difficult to show that the K\"ahler moduli have the explicit expression in terms of $y^i$, $H_I$, $\tilde{H}^I$
\begin{equation} \label{eq:moduliKaehler}
    z^i = \frac{X^i}{X^0} = \frac{\tilde{H}^i D + \frac{2}{9}(\Delta_j y^j) y^i + i \tilde{H}^i \tilde{H}^0 e^{-2U} }{\tilde{H}^0 D + i (\tilde{H}^0)^2 e^{-2U} } \,.
\end{equation}
We recover the solution of section \ref{sec:BHprobes} if we select $D_{ijk} = -1/6 $ with $(i,j,k)$ a permutation of $(1,2,3)$.

\subsection{Instantons}
We now map these black holes to type IIB D-instantons. As in section \ref{sec:instantonstorus}, we first derive the temporal c-map relating the type IIA vector multiplet sector to the Euclidean type IIB hypermultiplet sector, and then use it to construct the instantons dual to the Calabi--Yau black holes above and study their properties.

\subsubsection{Temporal c-map: Euclidean Type IIB action from Type IIA}

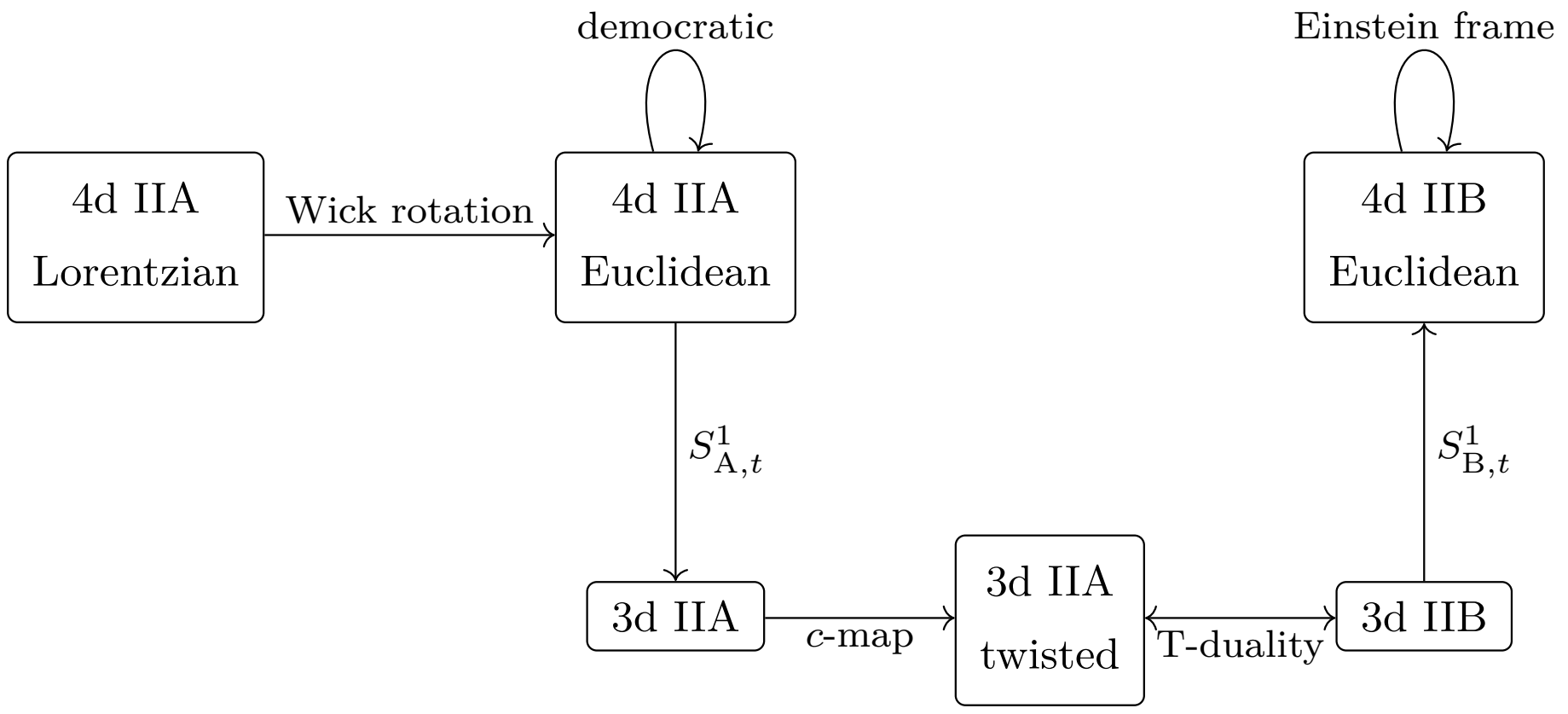
\begin{figure}[t]
\centering
\resizebox{0.8\linewidth}{!}{%
\begin{tikzcd}[
    ampersand replacement=\&,
    row sep=4em,
    column sep=3em,
    cells={nodes={
        font=\footnotesize,
        draw,
        rounded corners=2pt,
        inner xsep=5pt,
        inner ysep=4pt
    }}
]
\begin{gathered}
    \text{4d IIA}\\
    \text{Lorentzian}
\end{gathered}
    \arrow[r, "\text{Wick rotation}"]
\&[2.5em]
\begin{gathered}
    \text{4d IIA}\\
    \text{Euclidean}
\end{gathered}
    \arrow[loop above, "\text{democratic}"]
    \arrow[d, "S^1_{\mathrm{A},t}"]
\&
\&
\begin{gathered}
    \text{4d IIB}\\
    \text{Euclidean}
\end{gathered}
    \arrow[loop above, "\text{Einstein frame}"]
\\
\&
\text{3d IIA}
    \arrow[r, "c\text{-map}"']
\&
\begin{gathered}
    \text{3d IIA}\\
    \text{twisted}
\end{gathered}
    \arrow[r, leftrightarrow, "\text{T-duality}"']
\&
\text{3d IIB}
    \arrow[u, "S^1_{\mathrm{B},t}"']
\end{tikzcd}%
}
\caption{Schematic illustration of the chain of maps, changes of frame and dualities performed to map Lorentzian type IIA to Euclidean type IIB, both compactified on the same threefold.}
\label{fig:duality-chain}
\end{figure}

We are now going to derive the explicit dictionary embedding type IIA gravity and vector multiplet sectors into Euclidean Type IIB gravity and hypermultiplet sectors following \cite{Behrndt:1997ch}. We do so performing a chain of maps, changes of frame and dualities which is summarized in figure \ref{fig:duality-chain} for the ease of the reader. We start considering Lorentzian type IIA and performing a Wick rotation to the Euclidean. As we explained, this avoids eventually landing on type IIB$^*$. We then embed the gauge fields into symplectic doublets switching to the democratic (pseudo) action. This allows us to obtain expressions explicitly invariant under symplectic transformations and simplifies the derivation presented in \cite{Behrndt:1997ch}. Assuming that euclidean time is an isometry of the fields, we dimensionally reduce along $S^1_{A,t}$. We then implement the c-map dualizing all the three-dimensional gauge fields into axions. T-duality is realized embedding the field content into a democratic Euclidean type IIB theory compactified on the dual circle $S^1_{B,t}$. Decompactifying the IIB radius, i.e. the inverse IIA radius, we get four-dimensional Euclidean IIB theory in string frame. We finally go to Einstein frame performing a Weyl rescaling.

Let us implement the chain of transformations. Our starting point is the action \eqref{eq:actionNonDemocratic}. Performing a Wick rotation we obtain
\begin{equation} 
    S_E^{\rm IIA} = -\int d^4x \sqrt{g} \left\{ R- 2 g_{i\bar{j}} \partial_\mu z^i \partial^\mu \bar{z}^{\bar{j}} + \frac{1}{4}{\rm Im}\mathcal{N}_{IJ} F^{I}_{\mu\nu} F^{J \mu\nu} - \frac{i}{4}{\rm Re} \mathcal{N}_{IJ} F^I_{\mu\nu} (\star_E F^J)^{\mu\nu}\right\} \,.
\end{equation}
The democratic action one builds from the euclidean theory is equal to the result of Wick rotating the  Lorentzian democratic action \eqref{eq:actionDemocratic}. One gets (with an abuse of notation, we will keep using the Latin labels $I,J,\dots$ for both doublets and their components. We also suppress space-time indexes)
\begin{equation}
    S_E^{\rm IIA} \supset - \int d^4x \sqrt{g} \, \left[\frac{1}{8} \mathcal{M}_{IJ} F^{I} F^{J} \right] \,,
\end{equation}
where the $\mathcal{M}$ matrix has not changed, but we have now the duality relations
\begin{equation}
    G_I = i \,\star_E{\rm Im}\mathcal{N}_{IJ} \star F^J  +{\rm Re}\mathcal{N}_{IJ} F^J \,, \qquad  \star_E F^I = i \Omega^{IJ} \mathcal{M}_{JK} F^K \,.
\end{equation}
We now follow \cite{Behrndt:1997ch} to implement the c-map. We want to dimensionally reduce the action to three dimensions along the euclidean time. We consider the ansatz for the metric 
\begin{equation}
    ds^2_{\rm IIA} = e^{-2\varphi} (dt + A)^2  + h_{mn}\, dx^m dx^n \,,
\end{equation}
where $h$ and $A$ represent, respectively, the lower dimensional metric and the Kaluza--Klein (KK) vector. $e^{-\varphi}$ instead plays the role of the KK scalar. Assuming that none of these fields depend on the euclidean time we can easily perform a dimensional reduction. The KK reduction of the scalar and gravity sector (see \cite{Gladush:1998xf} for useful relations) provides
\begin{equation}
    \sqrt{g} (R^{(4)} - 2 g_{i\bar{j}} \partial z^i \partial \bar{z}^{\bar{j}}  ) = \sqrt{h} e^{-\varphi} \left[R^{(3)} - \frac{1}{4}e^{-2\varphi} F^2 - 2 g_{i\bar{j}} \partial z^i \partial \bar{z}^{\bar{j}} \right]\,,
\end{equation}
where $F = dA$ and we discarded total derivatives. For the gauge sector, we consider the following decomposition of the four-dimensional gauge potential $A^I$ 
\begin{subequations}
\begin{align}\label{eq:defXi}
   & A^{(3) I}_m = A^I_m -  A^I_t A_m \,, \qquad \sqrt{2}\xi^I = A^I_t \,, \\[2mm]
   & F^{(3) I} \equiv d  A^{(3) I} \,, \qquad \hat{F}^I \equiv F^{(3) I} + \sqrt{2}\xi^I F\,.
\end{align}
\end{subequations}
This ansatz implies the decomposition (see  \cite{Maharana:1992my} for useful relations)
\begin{equation}
    \sqrt{g} \left[\frac{1}{8} \mathcal{M}_{IJ} F^I F^J\right] = \sqrt{h} e^{-\varphi} \left[\frac{1}{8} \mathcal{M}_{IJ} \hat{F}^I \hat{F}^J + \frac{1}{2}e^{2\varphi}\mathcal{M}_{IJ} \partial \xi^I \partial \xi^J\right] \,.
\end{equation}
The full Euclidean 3d theory is
\begin{equation}
    \begin{split}
        S^{3d} = - \int d^3x \sqrt{h}& e^{-\varphi} \bigg[R^{(3)} - \frac{1}{4}e^{-2\varphi} F^2 - 2 g_{i\bar{j}} \partial z^i \partial \bar{z}^{\bar{j}}  \\[2mm]
        & \quad +\frac{1}{8} \mathcal{M}_{IJ} \hat{F}^I \hat{F}^J + \frac{1}{2}e^{2\varphi}\mathcal{M}_{IJ} \partial \xi^I \partial \xi^J \bigg]  \,. 
    \end{split}
\end{equation}
We now want to dualize the three-dimensional gauge vectors $A_m$ and $A^{(3)I}_m$ into scalars. We do so adding a proper Lagrange multiplier. For $A^{(3)I}_m$ we deform the action adding
\begin{equation}
    \Delta S_{\tilde{\xi}} = \int \frac{\sqrt{2}}{2} \tilde{\xi}_I d F^{(3)I} \,.
\end{equation}
The EOMs of $F^{(3)I}$  give us the duality relation
\begin{equation}\label{eq:dualityRelationF3A}
    F^{(3)I} = - \sqrt{2}\xi^I F - \sqrt{2}e^{\varphi} \mathcal{M}^{IJ} \star_{(3)} d \tilde{\xi}_J \,,
\end{equation}
where $\star_{(3)}$ is the hodge star defined with the metric $h_{mn}$. Removing $F^{(3)I}$ we get the terms in the bulk action $S^{3d}{}' = S^{3d} +  \Delta S_{\tilde{\xi}} $
\begin{equation}
    S^{3d}{}'\supset - \int \sqrt{h}e^{-\varphi} \left[-\frac{1}{2}e^{2\varphi} \mathcal{M}^{IJ} \partial \tilde{\xi}_I\partial \tilde{\xi}_J - e^{\varphi} \xi^I \partial_m \tilde{\xi}_I (\star_{(3)} F)^m \right] \,.
\end{equation}
Now we dualize $A_m$. We add the Lagrange multiplier 
\begin{equation}
    \Delta S_{\sigma} = - \int \frac{1}{2}dF (\sigma - \xi^I \tilde{\xi}_I) 
\end{equation}
and the EOMs of $F$ produce the duality constraint
\begin{equation}
    F = - \frac{1}{2} e^{3\varphi} \star_{(3)} \left[d\sigma - d\xi^I \tilde{\xi}_I + \xi^I d \tilde{\xi}_I \right]\,.
\end{equation}
Integrating out $F$, we obtain the theory $S^{3d}{}' = S^{3d} +  \Delta S_{\tilde{\xi}} +  \Delta S_{\sigma} $
\begin{equation}
    \begin{split}\label{eq:action3dFinal}
        S^{3d}{}' = & - \int d^3x \sqrt{h}  e^{-\varphi} \bigg[R^{(3)} + \frac{1}{8}e^{4\varphi} (\partial\sigma - \partial\xi^I \tilde{\xi}_I + \xi^I \partial \tilde{\xi}_I)^2 - 2 g_{i\bar{j}} \partial z^i \partial \bar{z}^{\bar{j}}  \\[2mm]
        & \hspace{2.5cm} -\frac{1}{2} e^{2\varphi} \mathcal{M}^{IJ} \partial \tilde{\xi}_I \partial \tilde{\xi}_J + \frac{1}{2}e^{2\varphi}\mathcal{M}_{IJ} \partial \xi^I \partial \xi^J \bigg]  \,. 
    \end{split}
\end{equation}
Once we unpack the democratic formulation, the action \eqref{eq:action3dFinal} is nothing but the one discussed in \cite{Ferrara:1989ik}. In \cite{Ferrara:1989ik} they show that after the compactification on a spacelike direction and the dualization of the scalar fields, the resulting theory has a quaternionic structure. It can be obtained therefore equivalently compactifying the four-dimensional IIB hypermultiplet sector in euclidean signature. We go to such type IIB theory decompactifying \eqref{eq:action3dFinal} on a circle with inverse radius and no KK vector, i.e. we want to obtain the type IIB metric
\begin{equation}
    ds^2_{\rm IIB} = e^{2\varphi} dt^2 + h_{mn} dx^m dx^n \,.
\end{equation}
It is not difficult to verify that we have the four-dimensional uplift
\begin{equation}
    \begin{split}
        S^{\rm IIB}_{E} = & - \int d^4x \sqrt{g} e^{- 2\varphi}\bigg[R^{(4)} + \frac{1}{8}e^{4\varphi} (\partial\sigma - \partial\xi^I \tilde{\xi}_I + \xi^I \partial \tilde{\xi}_I)^2 - 2 g_{i\bar{j}} \partial z^i \partial \bar{z}^{\bar{j}}  \\[2mm]
        & \hspace{2.5cm} -\frac{1}{2} e^{2\varphi} \mathcal{M}^{IJ} \partial \tilde{\xi}_I \partial \tilde{\xi}_J + \frac{1}{2}e^{2\varphi}\mathcal{M}_{IJ} \partial \xi^I \partial \xi^J + 4 (\partial\varphi)^2\bigg]  \,. 
    \end{split}
\end{equation}
Notice that the dualized scalars $\tilde{\xi}_I$ and $\sigma$ seem to have the wrong sign in the kinetic term. This is however fixed when we go in IIB, Lorentzian signature by the fact that they transform as pseudo-scalars, i.e. with an extra $i$ factor. We can now go to Einstein via the rescaling $g^E_{\mu\nu} = g_{\mu\nu}\,  e^{-2\varphi}$. We end up with 
\begin{equation}
    \begin{split} \label{eq:actionIIBfinal}
        S^{\rm IIB}_{E} = & - \int d^4x \sqrt{g} \bigg[R - 2(\partial\varphi)^2  - 2 g_{i\bar{j}} \partial z^i \partial \bar{z}^{\bar{j}} + \frac{1}{8}e^{4\varphi} (\partial\sigma - \partial\xi^I \tilde{\xi}_I + \xi^I \partial \tilde{\xi}_I)^2   \\[2mm]
        & \hspace{2.5cm} -\frac{1}{2} e^{2\varphi} \mathcal{M}^{IJ} \partial \tilde{\xi}_I \partial \tilde{\xi}_J + \frac{1}{2}e^{2\varphi}\mathcal{M}_{IJ} \partial \xi^I \partial \xi^J \bigg]  \,. 
    \end{split}
\end{equation}
where the Einstein frame metric now is given by 
\begin{equation}
    ds^2_{\rm IIB,\, E} = dt^2 + e^{-2\varphi} h_{mn} dx^m dx^n \,.
\end{equation}

\subsubsection{Instantons from CY black holes}

Having the Euclidean type IIB action \eqref{eq:actionIIBfinal}, we now construct its instanton solutions. We take over the black hole ansatz, with the harmonic functions now localized in four-dimensional Euclidean space, and show that it solves the equations of motion.

\paragraph{The ansatz.}
A solution of the theory \eqref{eq:actionIIBfinal} can be obtained using exactly the same ansatz we used for Calabi--Yau black holes of type IIA (cfr. equations  \eqref{eq:ansatzMetric}, \eqref{eq:ansatzGaugeFields} and \eqref{eq:stabilization}) but leaving implicit the harmonic functions $H_I$, $\tilde{H}^I$. Such functions, originally harmonic functions in $\mathbb{R}^3$, must be replaced with harmonic functions in $\mathbb{R}^4$
\begin{equation}  \label{eq:harmonics2}
    H_I = a_I + \frac{b_I}{r^2} \,, \qquad \tilde{H}^I = \tilde{a}^I + \frac{\tilde{b}^I}{r^2} \,.
\end{equation}
Physically, we can see this as localizing the brane sources in euclidean time. Then, the instantons dual to static black holes satisfy
\begin{equation} \label{eq:staticConditions}
    A = 0 = d\sigma - d\xi^I \tilde{\xi}_I + \xi^I d \tilde{\xi}_I \,, \qquad ds^2 = \delta_{\mu \nu} dx^\mu dx^\nu \,, \qquad  e^{2\varphi} = e^{-K} \,,
\end{equation}
where $\varphi$ is the type IIB four-dimensional dilaton and $K$ is the K\"ahler potential defined in section \ref{sec:conventions}. The expression of the axions $\xi^I$ in terms of the type IIA ansatz can be easily obtained plugging the ansatz \eqref{eq:ansatzGaugeFields} in the definition
\eqref{eq:defXi} and recalling that \eqref{eq:defXi} is involving the Euclidean components of the gauge field.\footnote{\eqref{eq:ansatzGaugeFields} is the Lorentzian ansatz}  We obtain 
\begin{equation}\label{eq:dictionary2}
    \xi^I = 2 i \, e^{-2\varphi} \, {\rm Re}\Omega^I \,, \qquad e^{2\varphi} \mathcal{M}_{IJ}  d \xi^J = - 2i \, \Omega_{IJ} \, d{\rm Im}\Omega^J \,.
\end{equation}
where $\Omega^I$ is the symplectic section defined in section \ref{sec:conventions}. For the axions $\tilde{\xi}_I$ we replace the ansatz \eqref{eq:ansatzGaugeFields} inside \eqref{eq:dualityRelationF3A} recalling that it holds \eqref{eq:staticConditions}. After some algebra we get
\begin{equation} \label{eq:dictionary1}
    \tilde{\xi}_I = 2 \, e^{-2\varphi}\,\Omega_{IJ} \, {\rm Re}\Omega^J \,, \qquad e^{2\varphi} \mathcal{M}^{IJ}  d \tilde\xi_J = 2 \, d{\rm Im}\Omega^I \,,
\end{equation}
Finally obtaining the dependence of the K\"ahler moduli $z^i$ and of the dilaton on the harmonic functions requires to solve the stabilization equations. For the case of a cubic prepotential it was solved explicitly in section \ref{ss:cubicPrepotential}. 

\paragraph{Solving the EOMs.}
Let us now verify that the equations of motion (EOMs) of \eqref{eq:actionIIBfinal} are satisfied. We focus on the equations of the axions, the dilaton and on the trace of Einstein equations, which are crucial to show the appearance of $\mathbb{R}^4$ harmonics and to evaluate the Euclidean on-shell action. A complete analysis of the full set of equations of the type IIB action we obtain after performing a T-duality and c-map is left for section \ref{sec:CYStrings}. The EOMs of $\tilde{\xi}_I$ and $\xi^I$ are 
\begin{equation}
    \mathbb{E}_{\xi^I,\, \tilde \xi_I}: \qquad d \star (e^{2\varphi} \mathcal{M}^{IJ} d \tilde{\xi}_J) = 0 \,, \qquad  d \star (e^{2\varphi} \mathcal{M}_{IJ} d {\xi}^J) = 0 \,. 
\end{equation}
Using \eqref{eq:dictionary1} and \eqref{eq:dictionary2} it is easy to verify that they are satisfied provided that $2{\rm Im}\Omega^I$ is a harmonic function in $\mathbb{R}^4$. Let us now verify that the trace of the Einstein equations is vanishing on-shell. For an Euclidean metric, all what we have to verify is the vanishing of the trace of the stress energy tensor, which can be written as
\begin{equation} \label{eq:Trace0}
    \text{Tr} \, \mathbb{E}_{g^{\mu\nu}} : \quad- 2(\partial\varphi)^2   -\frac{1}{2} e^{2\varphi} \mathcal{M}^{IJ} \partial \tilde{\xi}_I \partial \tilde{\xi}_J + \frac{1}{2}e^{2\varphi}\mathcal{M}_{IJ} \partial \xi^I \partial \xi^J = 2 g_{i\bar{j}} \partial z^i \partial \bar{z}^{\bar{j}} \,.
\end{equation}
It's not difficult to show that the 2nd and 3rd terms of the LHS of \eqref{eq:Trace0} are equal. Replacing our ansatz we get
\begin{align}
     &  - 2(\partial\varphi)^2   = -\frac{1}{2}(\partial K)^2  \,, \label{eq:Trace1}\\[2mm]
     &  - e^{2\varphi} \mathcal{M}^{IJ} \partial \tilde{\xi}_I \partial \tilde{\xi}_J  = -4 \left[\partial_\mu (e^{ K}{\rm Re}\mathcal{F}_I) \partial^\mu{\rm Im}X^I - \partial_\mu (e^{ K} {\rm Re}{X}^I ) \partial^\mu{\rm Im}\mathcal F_I\right] \label{eq:Trace2} \,.
\end{align}
In order to massage the RHS of \eqref{eq:Trace0} we start observing that
\begin{equation}
    \partial_\mu V^I  = (\partial_\mu z^i D_i - i Q_\mu )V^I = \partial_\mu z^i D_i V^I \,,   \qquad Q_\mu = - \frac{i}{2}(K_i \partial_\mu z^i - K_{\bar{j}}\partial_\mu \bar{z}^{\bar{j}}) \,,
\end{equation}
where we used the special geometry identities \eqref{eq:specialGeom} and the assumption that we work with static BPS CY BHs, which implies the vanishing of the K\"ahler connection $Q_\mu = 0$. This implies that the kinetic term of the K\"ahler moduli is $g_{i\bar{j}} \partial z^i \partial \bar{z}^{\bar{j}}  = i \Omega_{IJ} \partial_\mu V^I \partial^\mu \bar{V}^J$. With a little bit of algebra one can then show
\begin{equation} \label{eq:Trace3}
     - 2 g_{i\bar{j}} \partial z^i \partial \bar{z}^{\bar{j}}  =  \frac{1}{2}(\partial K)^2 + 4 \left[\partial_\mu (e^{ K}{\rm Re}\mathcal{F}_I) \partial^\mu{\rm Im}X^I - \partial_\mu (e^{ K} {\rm Re}{X}^I ) \partial^\mu{\rm Im}\mathcal F_I\right] \,.
\end{equation}
Combining \eqref{eq:Trace1}, \eqref{eq:Trace2} and \eqref{eq:Trace3} we prove \eqref{eq:Trace0}. We finally consider the dilaton EOM. It is 
\begin{equation}
   \mathbb{E}_\varphi: \quad 4 d  \star d  \varphi - e^{2\varphi} \mathcal{M}^{IJ} d\tilde{\xi}_I \wedge \star d\tilde{\xi}_J + e^{2\varphi} \mathcal{M}_{IJ} d{\xi}^I \wedge \star d{\xi}^J = 0 \,.
\end{equation}
Using our ansatz we can write 
\begin{equation} \label{eq:dilatonV2}
   d  \star \left[ 2 \,d K  +8 \, e^{ K} \left({\rm Re}\mathcal{F}_I\, d\,{\rm Im}X^I - {\rm Re}{X}^I d\,{\rm Im}\mathcal F_I\right) \right] = 0 \,.
\end{equation}
Let us massage the first term of \eqref{eq:dilatonV2}. One can consider the identity $\Omega_{IJ} V^I U^J_i = 0 $ that when contracted with $\partial_\mu z^i$ and using $Q_\mu = 0$ gives $\Omega_{IJ} V^I \partial_\mu V^J = 0 $ which implies $\mathcal F_I \partial_\mu X^I - X^I \partial_\mu \mathcal F_I = 0$ or equivalently ${\rm Im}X^I \partial_\mu{\rm Re}\mathcal F_I - \text{Im}\mathcal F_I \partial_\mu{\rm Re}X^I = {\rm Re}\mathcal F_I \partial_\mu{\rm Im}X^I -{\rm Re}X^I \partial_\mu{\rm Im}\mathcal F_I$. We can then write  (we use that $\text{Im}\Omega^I$ is harmonic)
\begin{align}
    & \partial K = -4 e^K \left[\partial( {\rm Im}X^I){\rm Re}\mathcal F_I - \partial(\text{Im}\mathcal F_I){\rm Re}X^I \right] \,, \\[2mm]
    & \partial^2 K = -4  \left[\partial({\rm Im}X^I) \partial(e^K{\rm Re}\mathcal F_I ) - \partial(\text{Im}\mathcal F_I) \partial( e^K{\rm Re}X^I  )\right] \,. \label{eq:indentityDilaton}
\end{align}
With the identity \eqref{eq:indentityDilaton} it is straightforward to verify that the dilaton EOM is satisfied.

\subsubsection{Properties of the instantons T-dual to CY black holes}

Fixing the integration constants of the $H$ functions in agreement with the conventions used in section \ref{sec:instantonstorus} and selecting units $G_N = 1$ we have
\begin{equation} 
    H_I = h_I + \frac{q_I}{r^2} \,, \qquad \tilde H^I = \tilde h^I + \frac{p^I}{r^2}  \,. 
\end{equation}
The core is located at $r =  0$. The charges $q_I$ and $p^I$ are related to the D-brane charges according to 
\begin{equation}
    (\text{D}(-1),\, \text{ED1}, \, \text{ED3}, \, \text{ED5}):\quad  (q_0, q_i, p^i, p^0)\,.
\end{equation}
In the following we will assume that  $q_I$ has the same sign of the $h_I$ and $p^I$ has the same sign of $\tilde h^I$.  

\paragraph{Euclidean on-shell action.}

We are interested now in the evaluation of the Euclidean on-shell action. Given that the theory \eqref{eq:actionIIBfinal} contains only the Ricci and the kinetic terms of the scalar fields, it vanishes on-shell because it is proportional to the trace of the Einstein equations. The value of the instanton action is therefore entirely determined by the boundary terms. The boundary terms can be reconstructed imposing that the boundary conditions we select for the variational problem (namely, which charges and fields we keep fixed) imply the vanishing of the fluctuations. 
The instanton solutions we consider have flat metric. Then the Gibbons--Hawking--York (GHY) boundary term is not necessary because the extrinsic curvature contribution is cancelled exactly by the regularization procedure which subtract the asymptotic flat space contribution (see for instance the discussion in \cite{Massai:2023cis}). The configuration is electrically charged with respect to $\tilde{\xi}_I$ and $\xi^I$. The boundary terms which allow us to fix such electric charges are then\footnote{Given that we are using a democratic formulation one might be worried about magnetic charges. However, boundary terms are not necessary for them. Being topological in nature, they are automatically fixed at the level of the path integral measure.}
\begin{equation}
     S^{\rm IIB}_{E, {\rm bound}} =  \int d^4x \sqrt{g} \,  D_\mu \left[- e^{2\varphi} \mathcal{M}^{IJ} \tilde{\xi}_I \partial^\mu \tilde\xi_J+ e^{2\varphi} \mathcal{M}_{IJ} {\xi}^I \partial^\mu \xi^J\right]\,.
\end{equation}
Unpacking the doublets and using the dilaton equation \eqref{eq:dilatonV2} we finally get the compact form for all the instantons induced by static CY black holes 
\begin{equation}
S_{\rm inst}= -\frac{2}{\kappa_4^2}\int d^4x \sqrt{g} \, D^2 \varphi =  -\frac{1}{\kappa_4^2} \int d \star d K \,,
\end{equation}
where we restored the overall $1/2\kappa_4^2$ factor. The expression can be further simplified performing the integral explicitly. Using that $Q_\mu = 0$ and the harmonics \eqref{eq:harmonics2}, the derivative of the K\"ahler takes the form
\begin{equation} \label{eq:IdentityPartialK}
    \partial K =  \frac{4}{r^3}e^{K} \left[p^I \mathcal F_I - q_I X^I\right] =  \frac{4}{r^3}e^{K/2} Z  \,,
\end{equation}
where we identified the combination $Z = e^{K/2} \left[p^I \mathcal F_I - q_I X^I\right]$ which formally, i.e. as a function of the harmonics, is the type IIA central charge with the odd D-brane charges $q_I$, $p^I$ identified with those of even D-branes. Observe that in static configurations $\mathrm{Re} Z = Z$. Assuming that $Z$ is never vanishing along the T-dual black hole attractor flow, restoring units and identifying $Z_\infty = M_{\rm ADM}/m_p > 0 $ we obtain\footnote{Notice that the combination $e^{-\varphi_\infty} |Z_\infty| $ is invariant under the rescaling $H\rightarrow \lambda H$ and we don't have to worry about the normalization of the constant part of the harmonics.} 
\begin{equation} \label{eq:InstantonAction}
    S_{\rm inst}= \pi e^{-\varphi_\infty}  \frac{M_{\rm ADM}}{m_p} =  \pi e^{-\varphi_\infty}  |Z_\infty| \,,
\end{equation}
We thus generalize \eqref{eq:actionADMmass}.

\paragraph{Warping factor and attractor flow.}

The warping factor of the IWP metric \eqref{eq:ansatzMetric} of the initial IIA  black hole satisfies an attractor flow equations of the form
\begin{equation} \label{eq:attractorFlowEquationWarping}
    \partial_\rho{U} = - e^{U} |Z| \,, \qquad \rho = 1/r \,.
\end{equation}
where $Z$ is the type IIA central charge. Replacing our solution $K = 2 U$ we get $\partial_\rho (e^{-K/2}) = |Z|$ which implies that as long as we pick a flow along which the central charge is not vanishing\footnote{These are the \textit{good flows} one uses to build BHs.} the warping factor $e^{-K}$ is monotonic and increases when approaching the core of the BH. If we replace the harmonic functions of $\mathbb{R}^3$ with those of  $\mathbb{R}^4$ it is not difficult to check that the warping factor should now satisfy the differential equation (cf. \eqref{eq:IdentityPartialK})
\begin{equation}
    \partial_\rho{U} = - 2 \rho \, e^{U} |Z| \,. 
\end{equation}
We conclude that along the radial profile of the instanton solution $e^{-K}$ is still monotonic with $e^{\varphi} \ge e^{\varphi_\infty}$ and it is divergent at the core. This confirms that all D-instantons configurations have their core at strong four-dimensional coupling.

\paragraph{Minimal size of the string-frame wormhole.}
With the explicit spacetime profile for the Calabi--Yau D-instantons we built we can test the proposal of \cite{Behrndt:1997ch}, where they proposed a relation among the black hole entropy and a geometric quantity of the string-frame metric. In string frame the metric has a warping factor controlled by $ e^{2 \varphi} = e^{-K}$
\begin{equation}
    ds^2 = e^{2 \varphi} dx^m dx^m \,.
\end{equation}
In \cite{Behrndt:1997ch} they analyzed instantons T-dual to axion-free, double extremal black holes, observing that the geometry is the one of a wormhole, as in the case of a single D-brane charge \cite{Gibbons:1995vg}. They noticed that the value of the central charge at the core $|Z_h|$, which controls the black hole entropy in the IIA side, also controls a geometric quantity for the double extremal wormhole: the minimal size of the wormhole throat. Our goal is to show that for generic CY D-instantons we still have a relation between the minimal area of the string-frame wormhole and the entropy only of the T-dual black hole. However, this relation is an inequality similar to the one relating the on-shell action and the entropy of the dual black hole (cf. equation \eqref{eq:actionentropybound}). The proper radius of the 3-sphere is controlled by $R = e^{\varphi}r$. The condition $\partial_r R = 0$ has not simple solution $r_{\rm min}$, but we can easily find the implicit expression
\begin{equation}
    R_{\rm min}^2 = 2 e^{\varphi} |Z| \bigg \rvert_{r = r_{\rm min}} \,.
\end{equation}
Using the monotonicity of $e^{-K}$ and the central charge $|Z|$ along the attractor flow we can establish the lower bound
\begin{equation}
     R^2_{\rm min} \ge 2 e^{\varphi_\infty} |Z_h| \equiv R_0^2\,.
\end{equation}
This is precisely the lower bound we anticipated. We can then wonder how good it is. In the double extremal case we can solve the minimization condition exactly and it is simple to verify that $R_{\rm min}^2 = 2 R_0^2$, generalizing the observation of  \cite{Behrndt:1997ch} to generic double extremal black holes. 

\subsection{Strings}\label{sec:CYStrings}
We now turn to the strings obtained by T-dualizing the black holes along a spatial direction. As for the instantons, and as in section \ref{sec:EFTstringtorus}, we first derive the spatial c-map relating the type IIA vector multiplet sector to the Lorentzian type IIB hypermultiplet sector. We then construct the EFT strings dual to the Calabi--Yau black holes and study the infinite-distance limits they probe near their core.

\subsubsection{Spatial c-map: Lorentzian type IIB action from type IIA}

\begin{figure}[t]
\centering
\resizebox{0.5\linewidth}{!}{%
\begin{tikzcd}[
    ampersand replacement=\&,
    row sep=4em,
    column sep=3em,
    cells={nodes={
        font=\footnotesize,
        draw,
        rounded corners=2pt,
        inner xsep=5pt,
        inner ysep=4pt
    }}
]
\begin{gathered}
    \text{4d IIA}\\
    \text{Lorentzian}
\end{gathered}
    \arrow[loop above, "\text{democratic}"]
    \arrow[d, "S^1_{\mathrm{A},y}"]
\&
\&
\begin{gathered}
    \text{4d IIB}\\
    \text{Lorentzian}
\end{gathered}
    \arrow[loop above, "\text{Einstein frame}"]
\\
\text{3d IIA}
    \arrow[r, "c\text{-map}"']
\&
\begin{gathered}
    \text{3d IIA}\\
    \text{twisted}
\end{gathered}
    \arrow[r, leftrightarrow, "\text{T-duality}"']
\&
\text{3d IIB}
    \arrow[u, "S^1_{\mathrm{B},y}"']
\end{tikzcd}%
}
\caption{Schematic illustration of the chain of changes of frame, maps and dualities performed to map Lorentzian type IIA to Lorentzian type IIB,
both compactified on the same threefold.}
\label{fig:duality-chain-2}
\end{figure}
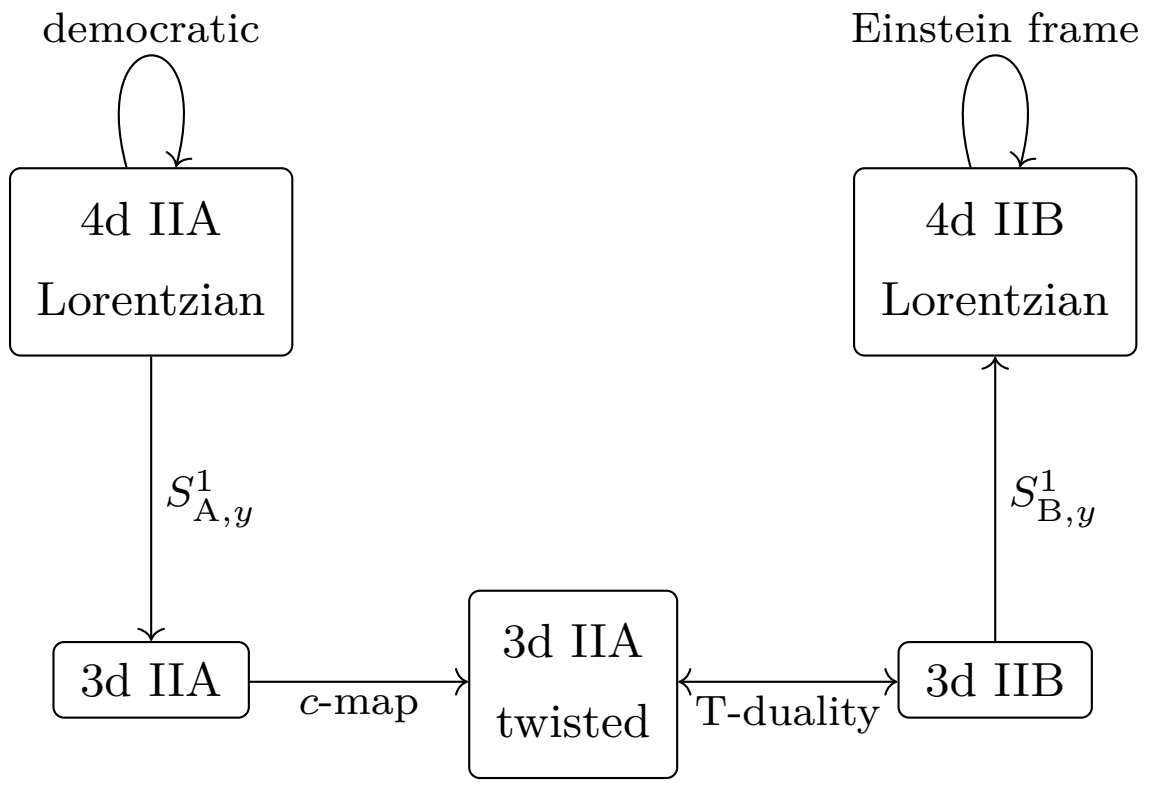

We are now going to derive the explicit dictionary embedding type IIA gravity and vector multiplet sectors into Type IIB gravity and hypermultiplet sectors. We do so performing a chain of maps, changes of frame and dualities whose core is a c-map and T-duality along a space-like direction. As in the previous section, we use the democratic type IIA action. We then assume that one of the four-dimensional space-like directions $y$ is an isometry and we dimensionally reduce along $S^1_{A,y}$. We implement c-map and T-duality dualizing all the three-dimensional gauge fields into axions and embedding the field content into a democratic Lorentzian type IIB theory compactified on the dual circle $S^1_{B,y}$. We obtain a four-dimensional, Lorentzian, type IIB action decompactifying the dual circle and we go to Einstein frame performing a Weyl rescaling. The steps are summarized in the figure \ref{fig:duality-chain-2}.

We consider the democratic action \eqref{eq:actionDemocratic} and we pick the ansatz 
\begin{equation} \label{eq:metricSpacelikeReduction}
    ds^2_{\rm IIA} = e^{-2\varphi} (dy + A)^2  + h_{mn}\, dx^m dx^n \,,
\end{equation}
where $y$ is a space-like isometry coordinate and $ h_{mn}$ a three-dimensional Lorentzian metric. The reduced three-dimensional Lorentzian action is 
\begin{equation}
    \begin{split}
        S^{3d} =  \int d^3x \sqrt{h}& e^{-\varphi} \bigg[R^{(3)} - \frac{1}{4}e^{-2\varphi} F^2 + 2 e^\varphi D^m (e^{-\varphi}\partial_m \varphi) - 2 g_{i\bar{j}} \partial z^i \partial \bar{z}^{\bar{j}}  \\[2mm]
        & \quad +\frac{1}{8} \mathcal{M}_{IJ} \hat{F}^I \hat{F}^J + \frac{1}{2}e^{2\varphi}\mathcal{M}_{IJ} \partial \xi^I \partial \xi^J \bigg]  \,,
    \end{split}
\end{equation}
where we introduced the field strengths
\begin{equation}
    F = dA\,, \qquad F^{(3) I} = d  A^{(3) I} \,, \qquad \hat{F}^I \equiv F^{(3) I} + \sqrt{2}\xi^I F\,,
\end{equation}
and decomposed the four-dimensional gauge potential $A^I$ as
\begin{equation}
    A^{(3) I}_m = A^I_m -  A^I_y A_m \,, \qquad \sqrt{2}\xi^I = A^I_y.
\end{equation}
Adding the Lagrange multipliers 
\begin{equation}
    \Delta S = \int \frac{\sqrt{2}}{2} \tilde\xi_I d F^{(3) I} + \frac{1}{2}d\left(\sigma - \xi^I \tilde \xi_I\right) \wedge F \,,
\end{equation}
we can dualize the three-dimensional vectors into scalars 
\begin{equation}
    F^{(3) I} + \sqrt{2} \xi^IF=  \sqrt{2} e^\varphi \mathcal{M}^{IJ}\star_{(3)} d\tilde \xi_I  \,, \qquad  F = \frac{1}{2} \star_{(3)}\left(d\sigma - d\xi^I \tilde{\xi}_I + \xi^I d \tilde{\xi}_I \right)\,,
\end{equation}
and we obtain the dualized three-dimensional action
\begin{equation}
    \begin{split}
        S^{3d}{}' = &  \int d^3x \sqrt{h}  e^{-\varphi} \bigg[R^{(3)} -\frac{1}{8}e^{4\varphi} (\partial\sigma - \partial\xi^I \tilde{\xi}_I + \xi^I \partial \tilde{\xi}_I)^2 - 2 g_{i\bar{j}} \partial z^i \partial \bar{z}^{\bar{j}}  \\[2mm]
        & \hspace{2.5cm} +\frac{1}{2} e^{2\varphi} \mathcal{M}^{IJ} \partial \tilde{\xi}_I \partial \tilde{\xi}_J + \frac{1}{2}e^{2\varphi}\mathcal{M}_{IJ} \partial \xi^I \partial \xi^J \bigg]  \,,
    \end{split}
\end{equation}
where we dropped boundary terms. Notice that with respect to \eqref{eq:action3dFinal} we have all the canonical signs of kinetic terms in Lorentzian signature. We now decompactify on a circle with inverse radius, i.e. we consider the four-dimensional metric
\begin{equation}
    ds^2_{\rm IIB} = e^{2\varphi} dy^2 + h_{mn} dx^m dx^n \,.
\end{equation}
It is not difficult to verify that we have the four-dimensional uplift
\begin{equation}
    \begin{split}
        S^{\rm IIB}_{{\rm bulk}} = &  \int d^4x \sqrt{g} e^{- 2\varphi}\bigg[R^{(4)} - \frac{1}{8}e^{4\varphi} (\partial\sigma - \partial\xi^I \tilde{\xi}_I + \xi^I \partial \tilde{\xi}_I)^2 - 2 g_{i\bar{j}} \partial z^i \partial \bar{z}^{\bar{j}}  \\[2mm]
        & \hspace{2.5cm} +\frac{1}{2} e^{2\varphi} \mathcal{M}^{IJ} \partial \tilde{\xi}_I \partial \tilde{\xi}_J + \frac{1}{2}e^{2\varphi}\mathcal{M}_{IJ} \partial \xi^I \partial \xi^J + 4 (\partial\varphi)^2\bigg]  \,. 
    \end{split}
\end{equation}
We go to Einstein via the rescaling $g^E_{\mu\nu} = g_{\mu\nu}\,  e^{-2\varphi}$
\begin{equation} \label{eq:LorentzianTypeIIB}
    \begin{split}
        S^{\rm IIB}_{E, {\rm bullk}} = &  \int d^4x \sqrt{g} \bigg[R - 2(\partial\varphi)^2  - 2 g_{i\bar{j}} \partial z^i \partial \bar{z}^{\bar{j}} - \frac{1}{8}e^{4\varphi} (\partial\sigma - \partial\xi^I \tilde{\xi}_I + \xi^I \partial \tilde{\xi}_I)^2   \\[2mm]
        & \hspace{2.5cm} +\frac{1}{2} e^{2\varphi} \mathcal{M}^{IJ} \partial \tilde{\xi}_I \partial \tilde{\xi}_J + \frac{1}{2}e^{2\varphi}\mathcal{M}_{IJ} \partial \xi^I \partial \xi^J \bigg]  \,,
    \end{split}
\end{equation}
where the metric now is given by 
\begin{equation}
    ds^2 = dy^2 + e^{-2\varphi} h_{mn} dx^m dx^n \,.
\end{equation}

\subsubsection{Strings from CY black holes}

With the Lorentzian type IIB action at hand, we now construct its string solutions. As for the instantons, we take over the black hole ansatz, now with harmonic functions on the plane transverse to the string, and show that it solves the equations of motion.

\paragraph{The ansatz.}
As before, we inherit the structure of our ansatz as a function of the harmonic functions directly from type IIA. Starting from the type IIA metric \eqref{eq:ansatzMetric} we identify $e^{-2\varphi} = e^{-K}$ and we obtain in type IIB Einstein frame  
\begin{equation}  \label{eq:metricEFTString}
ds^2 = -  dt^2 + dy^2 + e^{-2 K} \left( dx^m dx^m \right) \,,
\end{equation}
where $x^m$ are cartesian coordinates in $\mathbb{R}^2$ parametrizing the space transverse to the EFT string. Static black holes have a trivial KK vector $A$ (cfr. \eqref{eq:metricSpacelikeReduction}) which implies
\begin{equation} \label{eq:NoKKCondition}
    A = 0 = d\sigma - d\xi^I \tilde{\xi}_I + \xi^I d \tilde{\xi}_I \,.
\end{equation}
For the rest of the axions we can not obtain expressions as simple as \eqref{eq:dictionary1} and \eqref{eq:dictionary2}. Still, we have a compact expression for their field strengths
\begin{equation} \label{eq:AxionsEFTString}
    d\xi^I = 2\star_{(2)} d  \, {\rm Im} \Omega^I \,, \qquad d \tilde\xi_I = - 2\, \Omega_{IJ} \star_{(2)} d \, {\rm Im} \Omega^J \,,
\end{equation}
where $\star_{(2)}$ is the hodge star in the transverse $\mathbb{R}^2$ with euclidean metric, which implies 
\begin{equation} \label{eq:RelationXi}
    \tilde{\xi}_I = -\Omega_{IJ} \xi^J \,.
\end{equation}
Some useful relations are also
\begin{equation} 
    e^K \mathcal{M}_{IJ} \, d\xi^J = 2 \,\Omega_{IJ} \star_{(2)} d \left({\rm Re} \Omega^J e^K \right) \,, \qquad  e^K \mathcal{M}^{IJ} \, d\tilde \xi_J =  2\,\star_{(2)} \,  d \left({\rm Re} \Omega^I e^K \right)\,.   \label{eq:IdentityUsefulforEOM}
\end{equation}
The IIA stabilization equations \eqref{eq:stabilization} relates ${\rm Im} \Omega^I$ with a collection of harmonic functions
\begin{equation}
      2 \,{\rm Im}\, \mathcal{F}_I  = H_I \,, \qquad 2 \,{\rm Im}\, {X}^I  = \tilde{H}^I \,.
\end{equation}
This ansatz produces a solution of the IIB equations of motion representing a string provided that the $H$, $\tilde H$ are harmonics of the new transverse space $\mathbb{R}^2$. This can be realized  smearing the black hole harmonic functions along the new isometry direction and can be understood as the fact that such direction must be longitudinal to the string and to all D-branes forming the bound state. Using cylindrical coordinates we have the general structure
\begin{equation}   
    H_I = a_I + b_I \log{(r)} \,, \qquad \tilde H^I = \tilde a^I + \tilde b^I \log{(r)} \,. 
\end{equation}
Finally, the dependence of the $z^i$ and of the warping factor on the harmonic functions requires to solve the stabilization equations. For the case of a cubic prepotential it was solved explicitly in section \ref{ss:cubicPrepotential}.

\paragraph{Solving the EOMs.}

Let us verify that the ansatz produced solves the type IIB EOMs. We start determining the equations produced by the action \eqref{eq:LorentzianTypeIIB}. The Einstein equation can be written in trace-reversed form as 
\begin{equation}
\begin{split}
\operatorname{Tr}  \mathbb{E}_{g^{\mu\nu}}: \qquad R_{\mu\nu}={}&
2\,\partial_\mu\varphi\,\partial_\nu\varphi
+2g_{i\bar j}\,
\partial_{(\mu}z^i\partial_{\nu)}\bar z^{\bar j}
+\frac18 e^{4\varphi}\Theta_\mu\Theta_\nu \\
& -\frac12 e^{2\varphi}\mathcal M^{IJ}
\partial_\mu\tilde\xi_I\partial_\nu\tilde\xi_J
-\frac12 e^{2\varphi}\mathcal M_{IJ}
\partial_\mu\xi^I\partial_\nu\xi^J \,,\label{eq:EOMeinstein}
\end{split}
\end{equation}
where we introduced the one-form
\begin{equation}
\Theta_\mu = 
\partial_\mu\sigma
-\tilde\xi_I\,\partial_\mu\xi^I
+\xi^I\,\partial_\mu\tilde\xi_I \,.
\end{equation}
The equations for the dilaton  and special-K\"ahler scalars are (together with their complex conjugates)
\begin{align}
& \mathbb{E}_{z^i}: \quad  \nabla^2 z^i
+\Gamma^i{}_{jk}\,
\partial_\mu z^j\partial^\mu z^k
= -\frac14 e^{2\varphi}g^{i\bar j}
\left(
\partial_{\bar j}\mathcal M^{IJ}\,
\partial_\mu\tilde\xi_I\partial^\mu\tilde\xi_J
+
\partial_{\bar j}\mathcal M_{IJ}\,
\partial_\mu\xi^I\partial^\mu\xi^J
\right) \,, \label{eq:EOMkaehlerScalars}\\[2mm]
& \mathbb{E}_{\varphi}: \quad \nabla^2\varphi =
\frac18 e^{4\varphi}\Theta_\mu\Theta^\mu
-\frac14 e^{2\varphi}
\left(
\mathcal M^{IJ}\partial_\mu\tilde\xi_I\partial^\mu\tilde\xi_J
+
\mathcal M_{IJ}\partial_\mu\xi^I\partial^\mu\xi^J
\right) \label{eq:EOMdilaton} \,,
\end{align}
where $\nabla^2\equiv\nabla_\mu\nabla^\mu$ and $\nabla_\mu$ is the Levi-Civita connection for the space-time metric and $\Gamma^i{}_{jk}$ are the Christoffel symbols of the special-K\"ahler metric $g_{i\bar j}$. The equation for the axions $\sigma$, $\xi^I$ and $\tilde \xi_I$  can be written equivalently as
\begin{align}
&\mathbb{E}_{\sigma}: \qquad  \nabla_\mu \left(e^{4\varphi}\Theta^\mu\right)=0 \,, \label{eq:EOMTheta}\\[2mm]
&\mathbb{E}_{\xi^I}: \qquad  \nabla_\mu\left(
e^{2\varphi} \mathcal M_{IJ}\partial^\mu\xi^J
\right)
+\frac12 e^{4\varphi}\Theta^\mu\partial_\mu\tilde\xi_I
=0 \,, \label{eq:EOMxi} \\[2mm]
&\mathbb{E}_{\tilde{\xi}_I}: \qquad  \nabla_\mu \left(
e^{2\varphi} \mathcal M^{IJ}\partial^\mu\tilde\xi_J
\right)
-\frac12 e^{4\varphi}\Theta^\mu\partial_\mu\xi^I
=0 \,.\label{eq:EOMxiTilde} 
\end{align}

Let us start checking the axion equations. In our static ansatz $\Theta_\mu = 0$ and \eqref{eq:EOMTheta} is trivially satisfied. \eqref{eq:EOMxi}  and \eqref{eq:EOMxiTilde} reduce to 
\begin{equation}
    d \star_{(4)} e^K \mathcal{M}^{IJ} d\tilde\xi_J = 0 \,, \qquad d \star_{(4)} e^K \mathcal{M}_{IJ} d\xi^J = 0 \,,
\end{equation}
where $\star_{(4)}$ is the hodge star operator for the metric \eqref{eq:metricEFTString}. Replacing \eqref{eq:IdentityUsefulforEOM} we see that both the equations have the structure $d \star_{(4)} \star_{(2)} d \omega $, with $d \omega$ and $\star_{(2)}$ a 1-form and the hodge star operator for the transverse space $\mathbb{R}^2$. It's not difficult to check that $\star_{(4)} \star_{(2)} d \omega  = dt \wedge dy \wedge d\omega $, so the equations are satisfied.

We now consider the dilaton equation \eqref{eq:EOMdilaton}. The first term of the RHS vanishes. The second and the third term are equal. This can be easily checked noticing that \eqref{eq:AxionsEFTString} implies \eqref{eq:RelationXi} and that $\mathcal{M}^{IJ} = \Omega^{IK}\Omega^{JL}\mathcal{M}_{KL}$. Combining \eqref{eq:AxionsEFTString} and \eqref{eq:IdentityUsefulforEOM} one obtains
\begin{equation}\label{eq:RelationOccuringInTheEOM}
    \partial_m \partial_m K + 4 \, \Omega_{IJ} \, \partial_m{\rm Im}\Omega^I  \, \partial_m (e^K {\rm Re}\Omega^J) = 0 \,,
\end{equation}
which is nothing but identity \eqref{eq:indentityDilaton}.

Let us analyze the Einstein equations \eqref{eq:EOMeinstein}. The third term of the RHS vanishes. The first term is just $\partial_\mu K \partial_\nu K/2$. The fourth and fifth term are equal. Following the same steps that lead us to obtain equation \eqref{eq:Trace3} we can show that
\begin{equation}
     2 g_{i\bar{j}} \partial_\mu z^i \partial_\nu \bar{z}^{\bar{j}}  =  -\frac{1}{2}\partial_\mu K \partial_\nu K - 4 \left[\partial_{(\mu} (e^{ K}{\rm Re}\mathcal{F}_I) \partial_{\nu)}{\rm Im}X^I - \partial_{(\mu} (e^{ K} {\rm Re}{X}^I ) \partial_{\nu)}{\rm Im}\mathcal F_I\right] \,, \label{eq:DummyIdentity1}
\end{equation}
Replacing our ansatz in the fourth term we get
\begin{equation}
    - e^{2\varphi}\mathcal M^{IJ}
\partial_\mu\tilde\xi_I\partial_\nu\tilde\xi_J = 4 \Omega_{IJ} \, \partial_\mu{\rm Im}\Omega^I  \, \partial_\nu (e^K {\rm Re}\Omega^J) - 4 \, \eta^\perp_{\mu \nu} \, \Omega_{IJ} \, \partial_m{\rm Im}\Omega^I  \, \partial_m (e^K {\rm Re}\Omega^J)\,, \label{eq:DummyIdentity2}
\end{equation}
where $\eta^\perp_{\mu \nu}$ acts as projector on the transverse space $\mathbb{R}^2$; it is a symmetric tensor  such that $\eta^\perp_{t\mu} = \eta^\perp_{y\mu} = 0 $, $\eta^\perp_{mn} = \delta_{mn}$. Combining \eqref{eq:DummyIdentity1} and \eqref{eq:DummyIdentity2} we reduce \eqref{eq:EOMeinstein} to
\begin{equation}
    R_{\mu\nu} = - 4 \, \eta^\perp_{\mu \nu} \, \Omega_{IJ} \, \partial_m{\rm Im}\Omega^I  \, \partial_m (e^K {\rm Re}\Omega^J) \,.
\end{equation}
Given that the metric \eqref{eq:metricEFTString} describes a direct product of two-dimensional flat and conformally-flat manifolds, the Christoffel symbols with mixed indexes vanish and the only non-trivial entries to check are those with indexes in the transverse space $R_{mn}$. The Ricci tensor of a conformally flat metric with coordinates \eqref{eq:metricEFTString} satisfies $R_{mn} = \delta_{mn} \partial_{l} \partial_l K $. The Einstein equations are then satisfied because we end up again with the relation \eqref{eq:RelationOccuringInTheEOM}.

Finally, we consider the EOM for the K\"ahler scalars \eqref{eq:EOMkaehlerScalars}. The two terms of the RHS are equal. Using the coordinates of \eqref{eq:metricEFTString} and the fact that all derivatives in $t$ and $y$ vanish we get 
\begin{equation}
\partial_m \partial_m  z^i+\Gamma^i{}_{jk}\partial_m z^j\partial_m z^k
+2e^K g^{i\bar j}\partial_{\bar j}\mathcal M_{IJ}\partial_m{\rm Im}\Omega^I \partial_m{\rm Im}\Omega^J=0 \,.
\end{equation}
We will now show that the relation is always satisfied because of he identities
\begin{align}
\partial_m \partial_m  z^i+\Gamma^i{}_{jk}\partial_m z^j\partial_m z^k
& =
K_m\partial_mz^i
-i g^{i\bar j}\bar C_{\bar j\bar k\bar p}
\partial_m\bar z^{\bar k}\partial_m\bar z^{\bar p}\,, \label{eq:UsefulId1} \\[2mm]
    2e^K g^{i\bar j}\partial_{\bar j} \mathcal{ M}_{IJ}\partial_m{\rm Im}\Omega^I \partial_m{\rm Im}\Omega^J
& =
-K_m\partial_mz^i
+i g^{i\bar j}\bar C_{\bar j\bar k\bar p}
\partial_m \bar z^{\bar k}\partial_m \bar z^{\bar p} \,, \label{eq:UsefulId2} 
\end{align}
where $C_{ijk}$ is the completely symmetric, covariantly holomorphic Yukawa tensor of special K\"ahler geometry
\begin{equation}
C_{ijk}=e^K \mathcal F_{IJK}\,\partial_iX^I\partial_jX^J\partial_kX^K \,,
\qquad
\mathcal F_{IJK}\equiv \partial_I \partial_J\partial_K \mathcal{F} \,.
\end{equation}
We start using that $\text{Im}\,\Omega^I$ is harmonic to write
\begin{equation}
0 = \partial_m \partial_m \text{Im}\Omega^I
= \partial_m \left[ e^{-K/2}\left(
\partial_m{\rm Im}V^I -\frac12K_m({\rm Im}V^I) \right) \right]\,.
\end{equation}
Exploiting the fact that for a static solution the K\"ahler connection vanishes and $\partial_m V = U_i\partial_m z^i$ and using the standard identities of local special K\"ahler geometry
\begin{equation} \label{eq:setIDPesce}
D_iU_j=iC_{ijk}g^{k\bar k}\bar U_{\bar k}\,,
\qquad
D_{\bar i}U_j=g_{j\bar i}V \,,
\end{equation}
one finds \eqref{eq:UsefulId1}.
The second relation follows from using 
\begin{equation}
\mathcal M_{IJ}V^J= - i \Omega_{IJ} V^J \,,
\qquad
\mathcal M_{IJ} U_i^J = i \Omega_{IJ} U_i^J \,.
\end{equation}
together with \eqref{eq:specialGeom} and \eqref{eq:setIDPesce}.

\subsubsection{Properties of the strings T-dual to CY black holes}

We start fixing the integration constants of the $H$ functions in agreement with the conventions used in the torus compactification section \ref{sec:EFTstringtorus}. We have
\begin{equation}  \label{eq:HfunctionsLogarithmic}
    H_I = h_I - q_I \log{(r/r_0)} \,, \qquad \tilde H^I = \tilde h^I -  p^I \log{(r/r_0)} \,. 
\end{equation}
The core is located at $r =  0$, $r_0$ is a reference IR scale we use to fix the normalization of the $H$. In the following we will assume that $q_I$ and $p^I$ have the same sign of the constant parts $h_I$ and $\tilde h^I$ and they encode the D-brane charges 
\begin{equation}
    (\text{D1},\, \text{D3}, \, \text{D5}, \, \text{D7}):\quad  (q_0, q_i, p^i, p^0)\,.
\end{equation}

\paragraph{Warping factor profile.}

We study the profile of the warping factor along the radial profile of the string. The main takeaway from this analysis is the fact that the fields do not smoothly interpolate between the core and the asymptotic region because of the presence of curvature singularities.

Smearing the attractor flow equation satisfied by the original BH warping \eqref{eq:attractorFlowEquationWarping} and replacing  $K = 2 U$ we get for  $ 0 < r < r_0$
\begin{equation} \label{eq:AttractorEqEftString}
    \partial_\rho (e^{-K/2}) = \frac{1}{\rho}|Z| \,, \qquad \rho = r_0/r \ge 1 \,.
\end{equation}
where the $Z$ is formally the type IIA central charge of a BH carrying charges $p^I$ and $q_I$. Approaching the core of the defect we have a monotonic growth which eventually lead to a divergence. We have indeed the expansion
\begin{equation} \label{eq:scalingWarpingFactor}
     e^{-2 K} \sim |Z_h|^4 \log{(r/r_0)}^4 \,, \qquad \text{for} \quad  r \rightarrow 0 \,,
\end{equation}
with $Z_h$ the central charge evaluated at the attractor value of the T-dual black hole. For $ r > r_0$  the monotonic growth is not anymore guaranteed. In general there will be radii $r_I$, $\tilde{r}^I$ at which the $H$ functions will vanish and flip sign. In simple systems such as the D1--D5--D5--D5 for which the warping factor $e^{-2K}$ can vanish in correspondence of these sign-flip radii, the configuration develops curvature singularities. Indeed the Ricci scalar is
\begin{equation} \label{eq:RicciScalar}
    R = \frac{2}{r}e^{2K}(K' + r K'') = 2 D^\mu D_\mu K\,.
\end{equation}
For $r > \hat{r}_0 \equiv \max{\{r_I \,, \tilde r^I \}}$ again we have a well defined attractor flow satisfying 
\begin{equation}
    \partial_\rho (e^{-K/2}) = -\frac{1}{\rho}|Z| \,, \qquad \rho \le  r_0/\hat{r}_0   \,.
\end{equation}
When studying this geometry we have then to work in the region $0 < r < r_0$. Notice that this is not a restriction because exploring the region with $r > r_0 $ is equivalent to perform an analytic continuation of the original BH harmonic functions to negative values of the radius. 

\begin{figure}[!h]
    \centering
    \begin{subfigure}[t]{0.48\textwidth}
        \centering
        \includegraphics[width=\textwidth]
        {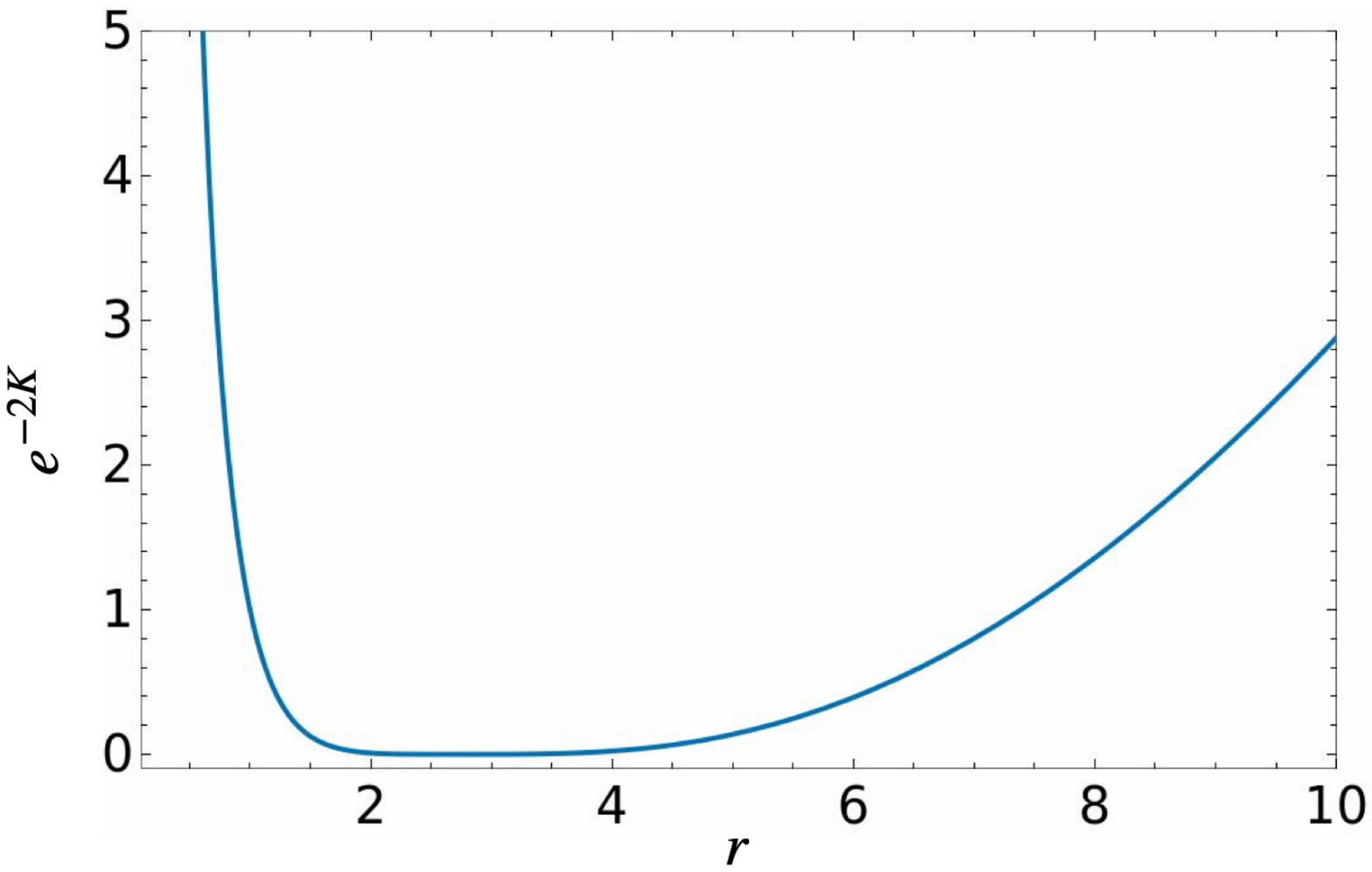}
        \caption{
            Warp factor $e^{-2K}$ for the double-extremal configuration.
        }
        \label{fig:double_extremal_warp}
    \end{subfigure}
    \hfill
    \begin{subfigure}[t]{0.48\textwidth}
        \centering
        \includegraphics[width=\textwidth]
        {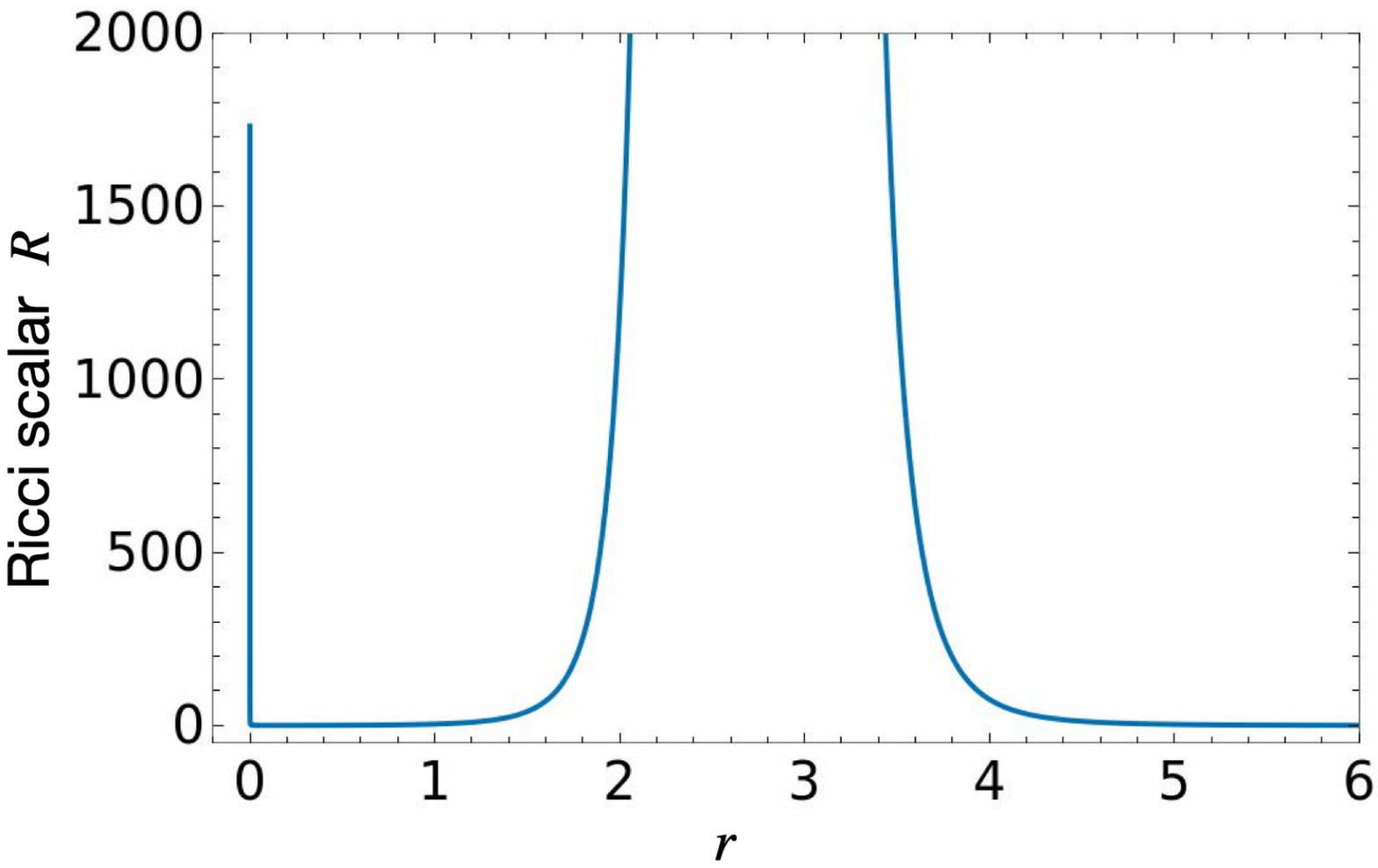}
        \caption{
            Ricci scalar ${R}$ for the double-extremal configuration. 
        }
        \label{fig:double_extremal_ricci}
    \end{subfigure}
    \caption{
        Radial dependence of the warp factor $e^{-2K}$ and the
        Ricci scalar ${R}$ for the double-extremal solution. We set $|Z| = |r_0| = g_{4,s} =  1$ and $\hat{r}_0 = e \sim 2.72\dots$
    }
    \label{fig:double_extremal}
\end{figure}

A simple class of solutions where we can explicitly keep track of the warping factor profile are double extremal BHs, i.e. attractor flows with $|Z| \ne 0$ constant. Solving \eqref{eq:AttractorEqEftString} we obtain
\begin{equation} \label{eq:warpingFromDoubleExtremal}
    e^{-K/2} =  const. - |Z| \log \left(\frac{r}{r_0}\right) \quad  \Rightarrow  \quad  e^{-2K} = \left(\frac{1}{g_{4,s}(r_0)} - |Z| \log \left(\frac{r}{r_0}\right)\right)^4 \,,
\end{equation}
In figure \ref{fig:double_extremal} we plot the warping factor of this simple configuration (notice $r_I =  \tilde{r}^I = \hat{r}_0 = r_0 \, e^{1/(g_{4,s}(r_0)|Z|)} > r_0 $) and the Ricci scalar (notice the singularity at $r = 0 $ and $r = \hat{r}_0$).

\paragraph{Attractor mechanism for the scalar fields.} The scalar fields have a well defined behaviour at $r = 0$. Exploiting the relation among the four-dimensional approaches weak coupling
\begin{equation}
    \varphi = K/2 \sim - \log \log \frac{r_0}{r} + \dots  \,, \qquad e^{\varphi} \rightarrow 0 \,, \qquad \text{for}\qquad r/r_0 \rightarrow 0 \,.
\end{equation}
For the axions $\xi^I$ we can solve \eqref{eq:AxionsEFTString} and obtain (we unpack the doublet)
\begin{equation}
    \xi^I =  \, p^I  \theta \,, \qquad \xi_{I} = \, q_I  \theta \,.
\end{equation}
Observe that the $\xi^I$ are not globally defined. What is globally defined is their field strengths, which are proportional to the harmonic form $d\theta$. The duals $\tilde{\xi}^I_h$ are simply given by \eqref{eq:RelationXi}. Replacing into \eqref{eq:NoKKCondition} we get that the scalar field $\sigma$ satisfies
\begin{equation}
    d\sigma = 0 \,. 
\end{equation}
Finally, let's consider the $z^i$ which were originally the type IIA K\"ahler moduli. For a cubic prepotential (valid in the large volume approximation) the stabilization equations can be solved explicitly. Keeping the harmonic functions implicit one gets \eqref{eq:moduliKaehler}, which we repeat for the ease of the reader.
\begin{equation}
    z^i = \frac{X^i}{X^0} = \frac{\tilde{H}^i D + \frac{2}{9}(\Delta_j y^j) y^i + i \tilde{H}^i \tilde{H}^0 e^{-2U} }{\tilde{H}^0 D + i (\tilde{H}^0)^2 e^{-2U} } \,.
\end{equation}
It's not difficult to verify that replacing \eqref{eq:HfunctionsLogarithmic} and taking the limit $r = 0$ all divergences cancel out. We simply recover the value $z^i$ that the scalars would have taken at the horizon of a BH with charges $q_I$ and $p^I$
\begin{equation}
    z^i \rightarrow z^i_h \equiv z^i [ \tilde{H}^I = p^I ,  H_I = q_I] \,. 
\end{equation}

\paragraph{String tension.} As we said in section \ref{sec:EFTstringtorus}, for a string described by the metric \eqref{eq:metricEFTString} it's quite difficult to define macroscopic quantities which can inform us on the microscopic nature of the back-reacted solution. The reason is that the spacetime of a string obtained from the back-reaction of D-branes has many curvature divergences. We have the naked singularity corresponding to the core of the string, but we also have the other finite distance singularities (recall fig. \ref{fig:double_extremal_ricci}). This means that we do not only lack a horizon surrounding the singular core; we are also not able to extend spacetime integrals of a conserved current till the asymptotic region. One of the quantities people investigated in the context of EFT strings is the so-called EFT string tension $\mathcal{T}$, computed as the energy density per unit-length in space-like cylinder enclosing the string. We now evaluate it for the T-dual of a D-brane configuration.

The energy contained in a cylinder of radius $r$ is 
\begin{equation} \label{eq:diskFormula}
    E(r) = \int_{D(r) \times \mathbb{R}_y} \star_{(4)} T^t  = R_y \int_{D(r)} dr d\theta \, r \,T_{tt}  \, e^{-2K}
\end{equation}
where $\star_{(4)}$ is the Hodge star operator for the string metric \eqref{eq:metricEFTString}, $D(r)$ is the disk of radius $r < r_0$, $T^t = T^t{}_\mu dx^\mu$ is the 1-form built out of the stress energy tensor and $R_y$ is the length of the $y$ direction. The tension can be obtained regularizing the infinite cylinder computing the energy per unit of length. Exploiting the Einstein equations, that $R_{tt} = 0$ for the metric \eqref{eq:metricEFTString} and recalling the expression for the Ricci \eqref{eq:RicciScalar} we get
\begin{equation} \label{eq:EFTstringTension}
    \mathcal{T}(r) = \frac{E(r)}{R_y} =   2\pi M_p^2 \left[r K'\right]_0^r =  4\pi M_p^2 |Z| g_{4,s}\,,
\end{equation}
where we used the attractor flow equation \eqref{eq:AttractorEqEftString} in the second step and both $Z$ and $g_{4,s} = e^{\varphi} = e^{K/2}$ are running quantities evaluated at $r$.

We now comment on the properties of the string tension. In the D1--D5 setup the tension is given by the sum of the tensions of its component (cfr.  \eqref{eq:EFTstringtension}). This additivity is not manifest in the general case \eqref{eq:EFTstringTension}. $\mathcal{T}(r)$  still vanishes at $r = 0$. This can be easily understood, in \eqref{eq:diskFormula} we are integrating  over a disk of zero size. Expanding \eqref{eq:scalingWarpingFactor} close to the core the information about the bound state of branes we used has been washed out because the central charge dropped out of the leading term
\begin{equation}
    \mathcal{T}(r) \sim  4\pi M_p^2 \frac{1}{ \log{(r_0/r)}} + \dots
\end{equation}
In fact, this behaviour of the tension near the core will be a common property of all of our strings, regardless of if they are dual to a finite or zero-sized black hole. Therefore, at $r\to 0$, the tension knows nothing about the microscopic components of the string. We finally test the approach of \cite{Lanza:2021udy} to determine the \textit{core tension}: we introduce  a UV cutoff as the smallest scale that could be probed by the EFT and then we evaluate the tension at such scale. As we did in section \ref{sec:EFTstringtorus}, we can do so imposing that the proper circumference of the circle of radius $r_{UV}$ in the transverse plane to the string (or the proper distance from the core to $r_{UV}$) is of the order of the species scale $\Lambda_{\rm sp}$. We implement this in the simple case of a double extremal string. Using \eqref{eq:warpingFromDoubleExtremal} we can see that we always obtain at leading order a condition of the form $r \, (\log r)^2 \sim c \, \Lambda_{\rm sp}^{-1} $ where $c$ is a constant. For instance, using $\oint \sqrt{g_{\theta \theta}(r_{UV})}\;d\theta = \Lambda_{\rm sp}^{-1}$ we get
\begin{equation}
      w^2 e^w = x^2  \,, \qquad e^w = \frac{r}{r_0 \exp(1/|Z| g_{4,s}(r_0))}  \,, \quad x^2 = \frac{\Lambda_{\rm sp}^{-1}}{|Z|^2 r_0 \exp(1/|Z| g_{4,s}(r_0))} \,.
\end{equation}
This equation is a generalized Lambert equation. Assuming $|Z| g_{4,s}(r_0) \ge 1$ and $r/r_0 < 1/e$ we can restrict to solutions with $w < -2 $ and the function is invertible in the range $(0,4/e^2]$. For small $x$ we get at leading order $w \sim 2 \log x + \mathcal{O}\left(\log\log(x)\right)$. We can then determine
\begin{equation}
    r_{UV} \sim \frac{\Lambda_{\rm sp}^{-1}}{|Z|^2} \,,
\end{equation}
and get an estimate of the core tension of a string T-dual to a double extremal black-hole which knows about the central charge of the black hole. Assuming a large central charge $|Z| \gg r_0 \Lambda_{\rm sp}$ we get
\begin{equation}
    \mathcal{T}_{\rm core} = \mathcal{T}(r_{UV}) \sim  4\pi M_p^2 \frac{1}{ \log{|Z|^2}} \,.
\end{equation}

\subsubsection{Classical infinite distance limits}\label{sec:CYStringsLimits}

We now study the infinite distance limits probed by the scalar fields attractor flow. We keep track of whether a configuration is the T-dual of a finite-size black hole or of a small black hole. The picture that emerged in section \ref{sec:EFTstringtorus} is confirmed by the analysis of the general CY solutions. A regular, horizon-full black hole solving the attractor equations will generate a T-dual bound state of D-branes whose classical attractor flow probes an F1 limit. A horizonless, small black hole will not fix the values of all of the K\"ahler moduli at the core of the string and the associated classical attractor flow can explore different types of limits.

\paragraph{F1 attractor from regular BHs.} Let us focus on the strings T-dual to finite-size black holes. The four-dimensional string coupling constant vanishes at the core. The volumes of the CY cycles in string units maintain a finite value, due to the attractor mechanism fixing the value of the K\"ahler moduli at the string core. Therefore, we have a type-IIB hypermultiplet trajectory for which the axions are constant and the K\"ahler saxions approach a non-degenerate finite-distance point at large volume $t^i\to t^i_h = \text{const.}$, dictated by the charges. Let us analyze the states which can become light along this trajectory. Charged BPS particle towers are generated by odd-cycles wrappings and their mass depend on vector-multiplet complex-structure moduli rather than on hypermultiplets. Four-dimensional strings states are produced by wrapping even-cycles. Using the standard string-frame tensions $\mathcal{T}_{\rm F1}\sim M_s^2$, $\mathcal{T}_{\rm Dp}\sim M_s^{p+1}/g_{s}$, and $\mathcal{T}_{\rm NS5}\sim M_s^6/g_{s}^2$ \cite{Polchinski:1995mt,Callan:1991ky}, and noting that every wrapped cycle entering the four-dimensional string tensions has large but fixed volume in string units, one obtains\footnote{Recall that at fixed finite string-frame Calabi--Yau volume $\mathcal V$, the classical relation $ e^{-2\varphi}= g_{s}^{-2}\mathcal V$ implies $g_{s}\sim g_{4,s} = e^{\varphi}$, while $M_s^2/M_p^2\sim e^{2\varphi}$.}
\begin{equation}\label{eq:nicescalings}
 \frac{\mathcal{T}_{\rm F1}}{M_p^2}\sim e^{2\varphi}\,,\qquad
 \frac{\mathcal{T}_{\rm D1}}{M_p^2}\sim
 \frac{\mathcal{T}_{{\rm D3}|C_2}}{M_p^2}\sim
 \frac{\mathcal{T}_{{\rm D5}|D_4}}{M_p^2}\sim
 \frac{\mathcal{T}_{{\rm D7}|X}}{M_p^2}\sim e^{\varphi}\,,\qquad
 \frac{\mathcal{T}_{{\rm NS5}|D_4}}{M_p^2}\sim O(1)\,,\end{equation}
where we have dropped all multiplicative constants induced by $t^i_h = \text{const.}$. This approximation is justified near the core, where the exponential dependence on $\varphi$ kicks in. Hence the F1 becomes tensionless parametrically faster than all D-brane strings, while the wrapped NS5 does not become tensionless. In this limit, all characteristic internal radii are fixed to be large in string units and therefore $M_{\rm KK}\sim M_s$; consequently 
\begin{equation}
 \frac{M_{\rm KK}}{M_p}\sim\frac{M_s}{M_p}\sim e^{\varphi} \,.
 \end{equation}
Thus the KK and F1 scales vanish with the same power. According to the Emergent String Conjecture classification the attractor flow trajectory describes an equi-dimensional weak-coupling limit with an emergent type-IIB fundamental string. We refer to this limit as F1 limit.

Observe that the string we produced via the c-map \eqref{eq:EFTstringTension} is tensionless along the infinite distance limit. However, the tension has the same scaling of a D-brane string and it is always parametrically heavier than a fundamental string. This should not come as a surprise, since the object we are describing is a four-dimensional string arising from the backreaction of a bound state of wrapped D-branes. It is true independently of the details of branes involved, as long as the brane boundstate produces a well-defined attractor mechanism and the saxions are non-degenerate. More generally, the fact that the tensions of such strings are always higher (or at) the species scale is at the root of the bounds on the QG scale discussed in sections \ref{sec:instantonbounds} and \ref{sec:stringbounds}.

\paragraph{Charge rescalings.}
Note that the analysis above assumed that the K\"ahler saxions are attracted to a large, but \emph{finite} value $t^i\to t^i_h=\text{const.}$, dictated by the charges of the string through the attractor mechanism, so that the exponential dependence on $\varphi$ in \eqref{eq:nicescalings} is the leading effect near the core. As in section \ref{sec:BHlimitRescalingCharges}, we can instead stop at the radius $r_{UV}$ at which the string probes the UV cutoff, and send (some of) the charges to be large. The moduli at $r_{UV}$ can then run to increasingly large values as the charges get larger. In the limit where the charges become strictly infinite, these moduli go to infinity and one can probe another infinite distance limit than the F1 string limit discussed above, before even reaching the core. Then \eqref{eq:nicescalings} no longer applies: other towers can take the lead and fall below the string scale. Recall that on the orbifold in section \ref{sec:stringprobestorus}, rescaling $q_0$, or $q_0=p^1$, reproduced the D1 and D1--D5 strings of table \ref{tab:EFTsmall}. In a Calabi--Yau, rescaling the charges of a single type of brane is similarly expected to reproduce the flows of the strings dual to small black holes studied below, while rescaling several types of charges, possibly at different rates, can in principle give more general infinite-distance limits. We do not attempt to classify them here, since the orbifold examples already illustrate the possible outcomes, and leave this for future work. As for the flows below, quantum corrections can modify these limits, as we discuss in section \ref{sec:quantumCorrections}.

\paragraph{Attractors from small black holes.} We now study the T-duals of small type IIA black holes built with a single type of D$2p$-branes wrapping $2p$-cycles. Their c-map duals along an external space-like direction are strings obtained by wrapping D$(2p+1)$-branes on the same $2p$-cycles: D1-branes, D3-branes on holomorphic curves, D5-branes on effective divisors, or D7-branes on the whole Calabi--Yau $X$. We derive the naive classical solutions obtained by solving the stabilization equations \eqref{eq:stabilization} with the cubic prepotential \eqref{eq:prepotentialCY} at two derivatives. We neglect all $\alpha'$, world-sheet instanton, string-loop, D-instanton and NS5-instanton corrections. The resulting trajectories are classical, and many of them leave the regime in which these corrections are suppressed, so that the corrections may obstruct or deflect them. We discuss them in section \ref{sec:quantumCorrections}.

We want to determine the attractor flow generated by the string and evaluate the mass of probe objects along it. Since all the branes sit at the same point of the transverse plane, the codimension-two solution is parametrized by a single harmonic function $H(r)=h+Q\log(r_0/r)$, with $r_0$ a reference scale and $h\geq0$, $Q>0$ constants such that $H>0$ for $r_0>r>0$. The constants $h$ and $Q$ will depend on the case at hand. We are interested in the leading behaviour of the fields at the core $r\to 0$. The K\"ahler moduli are $t^i={\rm Im}\,z^i$, with $z^i=X^i/X^0$ given by \eqref{eq:moduliKaehler}, the four-dimensional dilaton is $ e^{-2\varphi}=e^{-K}=8|X^0|^2\mathcal V\,,
    \label{eq:intclass-gs}$ and the ten-dimensional string coupling is $g_s\sim|X^0|^{-1}$.

On $T^6/(\mathbb Z_2\times\mathbb Z_2)$, the effect of the sources is captured by the \textit{harmonic function rule}: each stack of branes shrinks the two-tori it wraps by a factor $H^{-1/2}$, and expands the transverse ones by a factor $H^{1/2}$ (cf.\ table \ref{tab:EFTsmall}). On a general Calabi--Yau, the wrapped curves or divisors do not split $X$ into wrapped and transverse directions, and the role of this rule is played by the leading behaviour of the K\"ahler form $J=t^i\omega_i$ at the core. In all the flows below, $J$ is dominated at the core by a fixed class $J_+\in H^{1,1}(X,\mathbb R)$, multiplied by a power of $H$. Along the flows we consider, $J$ stays in the K\"ahler cone, as required for \eqref{eq:prepotentialCY} to apply. This is a restriction on the sources and on the constants: the flow must start in the K\"ahler cone, the total class of the divisors wrapped by the D5-branes must be nef, and that of the curves wrapped by the D3-branes must be movable (see appendix \ref{app:singleDbranes}). It follows that $J_+$ is also nef: $J_+\cdot C\geq0$ for every effective curve $C$. The volumes of the curves with $J_+\cdot C>0$ then scale with the same power of $H$ as $J$, while the curves with $J_+\cdot C=0$ are parametrically smaller, with volumes set by the subleading terms in $J$.

For the D1 and D7 sources, $J$ is exactly proportional to a class fixed by the constants in the harmonic functions, which must then lie in the interior of the K\"ahler cone, and all curves grow (D1) or shrink (D7) at the same rate (see appendix \ref{app:singleDbranes}). The D3 and D5 sources are richer. For a D5 source, the stabilization equations \eqref{eq:stabilization} give $t^i=\tilde H^i/(2X^0)$ with $\tilde H^i=\tilde h^i+p^i\log(r_0/r)$, i.e.\ $J=\big(J_b+\log(r_0/r)\,J_+\big)/(2X^0)$, with $J_b=\tilde h^i\omega_i$ and $J_+=p^i\omega_i$. Hence $J_+$ is the Poincar\'e dual of the total class of the divisors wrapped by the D5-branes. For a D3 source, $X^0$ is constant and \eqref{eq:stabilization} fixes instead the quadratic combination $\frac12\mathcal K_{ijk}t^jt^k=H_i/\tilde H^0$, with $H_i=h_i+q_i\log(r_0/r)$. Since $\frac12\mathcal K_{ijk}t^jt^k=\frac12\int_XJ\wedge J\wedge\omega_i$ and the $q_i$ are the components of the total class of the wrapped curves, this states that, up to normalization, $\frac12J\wedge J$ grows as $\log(r_0/r)$ times its Poincar\'e dual. If $J\simeq H^{1/2}J_+$, then $J_+\wedge J_+$ is proportional to this Poincar\'e dual, i.e.\ $\frac12\mathcal K_{ijk}t_+^jt_+^k\propto q_i$ with $J_+=t_+^i\omega_i$ nef. Otherwise, the flow can have $J\simeq HJ_+$ with $J_+\wedge J_+=0$, and it is $J_+\wedge J_b$, with $J_b$ the next term in $J$, that is proportional to it.

How the volumes scale depends on how degenerate $J_+$ is, which is measured by its numerical dimension
\begin{equation}
 \nu\equiv\nu(J_+)=\max\{k\in\{1,2,3\}:J_+^k\not\equiv0\}\,.
 \label{eq:intclass-nu-rank}
\end{equation}
For instance, on the orbifold, D5-branes wrapping one, two or three of the four-tori $T^2_j\times T^2_k$ give $\nu=1,2,3$, and reproduce the D5, D5--D5 and D5--D5--D5 rows of table \ref{tab:EFTsmall}. With $J_b$ the subleading term in $J$ ($J_b=\tilde h^i\omega_i$ for a D5 source), we further assume $J_+^\nu J_b^{3-\nu}\neq0$, which holds e.g.\ if $J_b$ is ample, so that $\nu$ fixes the leading power of $H$ in $\mathcal V$.

Inserting the expansion of $J$ in \eqref{eq:intclass-volumes}, the leading behaviour of $v_C$, $v_D$ and $\mathcal V$ is set by $J_+\cdot C$, $J_+^2\cdot D$ and $J_+^3$ when these are non-zero. Thus $\nu$ determines the scaling of $\mathcal V$ and, for a D5 source, also that of $g_s\sim|X^0|^{-1}$, since $(X^0)^2\propto\mathcal K_{ijk}\tilde H^i\tilde H^j\tilde H^k$ (see appendix \ref{app:singleDbranes}). The data (source, $\nu$) do not, however, fix the flow uniquely. Curves with $J_+\cdot C=0$ and divisors with $J_+^2\cdot D=0$ are controlled by the subleading terms in $J$, which depend on the constants $h_i$ and $\tilde h^i$, and so does whether they shrink. We state the corresponding assumptions case by case in appendix \ref{app:singleDbranes}.

Since these trajectories are classical, we do not attempt to identify the limits they probe, but only track how the masses and tensions of familiar objects in Planck units behave as we approach the core. We consider four-dimensional strings, i.e.\ F1- and NS5-branes and D$(2p+1)$-branes wrapped on $2p$-cycles, and KK modes. The masses of BPS particles from D3-branes on 3-cycles depend on the complex-structure moduli, which are not sourced by the string. Since a D$(2p+1)$-brane (NS5-brane) wrapped on a $2p$-cycle $\Sigma$ has string-frame tension $\mathcal T\sim M_s^2\,v_\Sigma/g_s$ ($\mathcal T\sim M_s^2\,v_\Sigma/g_s^2$), with $v_\Sigma$ given by \eqref{eq:intclass-volumes}, the tensions in Planck units, $\tau\equiv\mathcal T/M_p^2$, are
\begin{equation}
\begin{aligned}
 \tau_{\rm F1}&\sim g_s^2/\mathcal V,
 &\tau_{\rm D1}&\sim g_s/\mathcal V,
 &\tau_{{\rm D3}[C]}&\sim g_s v_C/\mathcal V,\\
 \tau_{{\rm D5}[D]}&\sim g_s v_D/\mathcal V,
 &\tau_{{\rm D7}[X]}&\sim g_s,
 &\tau_{{\rm NS5}[D]}&\sim v_D/\mathcal V .
\end{aligned}
\label{eq:intclass-tensions}
\end{equation}
For a characteristic radius $\mathcal R_\alpha=M_sR_\alpha$ in string units, the KK modes have mass $M_{{\rm KK},\alpha}\sim M_s/\mathcal R_\alpha$, i.e.
\begin{equation}
 M_{{\rm KK},\alpha}/M_p
 \sim \frac{g_s}{\sqrt{\mathcal V}\,\mathcal R_\alpha}.
 \label{eq:intclass-KK}
\end{equation}
In the following we assume that the shape of each curve is preserved, so that its characteristic radius is $\mathcal R_C\sim v_C^{1/2}$. Different curves can scale with different powers of $H$, and the lightest KK modes come from the curve with the largest volume. If instead some cycle shrinks, $\mathcal R_\alpha\to0$, the large-volume approximation underlying \eqref{eq:prepotentialCY} breaks down. 

In appendix \ref{app:singleDbranes} we construct these flows explicitly for each source and each value of $\nu$. Our goal is not a complete classification but to provide concrete realizations; whether a given Calabi--Yau supports them depends on the details of its geometry. In all cases the four-dimensional dilaton runs to weak coupling, $g_{4,s}=g_s\mathcal V^{-1/2}\to0$, while the ten-dimensional coupling grows for the D1 source, $g_s\sim H^{1/2}$, is constant for D3 sources, and vanishes for D5 and D7 sources, $g_s\sim H^{-\nu/2}$ and $g_s\sim H^{-1}$, respectively. All volumes grow for the D1 source and for D3 sources with $\nu=3$ and $J_+$ ample, while some cycles shrink for D3 sources with $\nu\leq2$ (if $\nu=1$, only for special curves), for all D5 sources and for the D7 source. Where $g_s$ grows, i.e.\ for the D1 source, S-duality, $g_s'=1/g_s$ and $t'^i=t^i/g_s$, may provide a better description. It does not help when cycles shrink, since there $g_s$ is constant or vanishes, and the S-dual frame either has the same shrinking cycles or is strongly coupled.

Unlike for the orbifold of section \ref{sec:torus}, no double T-duality is available in general, and most of these classical flows are therefore not under control near the core: some cycles shrink at fixed or weak coupling, so that neither the large-volume approximation nor S-duality applies. Determining what these flows probe requires the full stringy corrections to the hypermultiplet moduli space, and we leave it for future work. There are, however, a few exceptions. The first is the D1 flow, which in the S-dual frame runs to weak coupling at fixed volumes, $t'^i\propto\tilde h^i$, that can be chosen large. The second are the D3 flows with $\nu=3$ and $J_+$ ample, and with $\nu=1$ when the subleading term in $J$ is positive on all curves with $J_+\cdot C=0$ (see appendix \ref{app:singleDbranes}). Along these D3 flows no cycle shrinks and $g_s$ stays fixed, and can be chosen small. As we will see in section \ref{sec:QCtoD1}, the D1 flow survives quantum corrections: in the S-dual frame, the instanton actions either grow, so that the corresponding corrections are parametrically suppressed, or stay constant, so that they are only numerically suppressed and shift the $\mathcal O(1)$ coefficients but not the scalings. Along the D3 flows above, the instanton actions presumably behave in the same way, and we expect these flows to survive quantum corrections as well. It would be interesting to verify this explicitly.
\section{Quantum corrections}\label{sec:quantumCorrections}

So far, the instantons and strings dual to black holes have been described in the classical, two-derivative approximation. Near their cores, the type IIB hypermultiplet moduli run off. For the instantons, the dilaton runs to strong coupling, although the core lies at zero distance along a null trajectory in the complexified moduli space. For the strings, the core always lies at infinite distance. For the bound states dual to finite-size black holes, the four-dimensional dilaton runs to weak coupling while the K\"ahler moduli approach their attractor values, whereas for the strings dual to small black holes, the Kähler moduli also explore infinite distances. Further limits can be probed by rescaling the charges of the bound states at fixed radius. In these regimes the hypermultiplet metric receives perturbative ($\alpha'$ and string-loop) and non-perturbative (world-sheet, D-brane and NS5-brane instanton) corrections, which can become unsuppressed precisely along these trajectories. Such corrections can qualitatively change some classical infinite-distance limits of the hypermultiplet moduli space, e.g.\ by freezing volumes that classically shrink. In this section we analyze whether the infinite-distance limits identified for the instantons and the strings at the classical level survive and how they are modified.

Let us stress how the corrections are treated. The trajectories of the previous sections solve the classical BPS flow equations. One could instead solve the same equations with the corrected hypermultiplet metric, which in general deforms the trajectory itself. We do not do this here. Instead, we use the classical trajectory to identify which instanton corrections can become relevant along it. When NS5-instantons remain suppressed and the relevant D-instantons are mutually local, the corrected metric is determined by a single function, the contact potential, which can be computed exactly; from it we read off the tension and the mass of the towers. When instead NS5-instantons or mutually non-local D-instantons become relevant, as near the core of the instantons, we do not know how to compute their effect and cannot conclude anything as to whether these infinite distance limits survive at the quantum level. For simplicity, we mostly consider how the solutions of section \ref{sec:torus} on $T^6/(\mathbb Z_2\times\mathbb Z_2)$ are affected, although several of the arguments, in particular those for the strings dual to finite-size black holes, apply to a generic Calabi--Yau. We use the results of \cite{Marchesano:2019ifh,Baume:2019sry,Alvarez-Garcia:2021pxo}, for other works on this see \cite{Ooguri:1996me,deWit:1999fp, deWit:2001brd, deWit:2006gn, Robles-Llana:2006vct, RoblesLlana:2006is, Saueressig:2007dr, Alexandrov:2008gh, Alexandrov:2009zh, Alexandrov:2009qq, Pioline:2009ia, Alexandrov:2011va, Alexandrov:2013yva, Alexandrov:2014sya}. We now introduce all the ingredients which are necessary for the subsequent discussion: the actions of the instantons correcting the hypermultiplet metric, the contact potential encoding them and the corrected tensions and masses of the tower of states.

\paragraph{Instanton actions.} In terms of the volumes \eqref{eq:intclass-volumes} of a curve $C$, a divisor $D$ and of $X$, the real parts of the instanton actions scale as\footnote{All the instantons are assumed to be Euclidean and with the whole world volume wrapped; from now on we stop stressing it and instead of EDp we just use Dp.} 
\begin{equation}
\begin{aligned}
    &{\rm Re}\,S_{{\rm WS}|C}\sim v_C\,,
    && \quad {\rm Re}\,S_{{\rm D}(-1)}\sim\frac{1}{g_s}\,,
    && \quad{\rm Re}\,S_{{\rm D1}|C}\sim\frac{v_C}{g_s}\,,\\
    &{\rm Re}\,S_{{\rm D3}|D}\sim\frac{v_D}{g_s}\,,
    && \quad {\rm Re}\,S_{\rm D5}\sim\frac{\mathcal V}{g_s}\,,
    && \quad {\rm Re}\,S_{\rm NS5}\sim\frac{\mathcal V}{g_s^2}\,.
    \label{eq:non-perturbative-contributions-general}
    \end{aligned}
\end{equation}
To track them along a given trajectory, we introduce a scaling parameter $\lambda$. Assuming an isotropic scaling of the K\"ahler moduli we write  $g_s\sim\lambda^\mu$ and $t^i\sim\lambda^\nu$. The actions then scale as 
\begin{equation}
\begin{aligned}
    &{\rm Re}\,S_{{\rm WS}|C}\sim \lambda^{\nu}\,,
    && \quad {\rm Re}\,S_{{\rm D}(-1)}\sim \lambda^{-\mu}\,,
    && \quad{\rm Re}\,S_{{\rm D1}|C}\sim \lambda^{\nu-\mu}\,,\\[2mm]
    &{\rm Re}\,S_{{\rm D3}|D}\sim \lambda^{2\nu-\mu}\,,
    && \quad {\rm Re}\,S_{\rm D5}\sim \lambda^{3\nu-\mu}\,,
    && \quad {\rm Re}\,S_{\rm NS5}\sim \lambda^{3\nu-2\mu}\,.
    \label{eq:non-perturbative-contributions}
    \end{aligned}
\end{equation}
 Only sectors whose action does not grow along the limit can modify its asymptotics.

\paragraph{Contact potential.} When NS5-instantons are negligible and the unsuppressed D-instantons are mutually local, the quaternionic-K\"ahler metric is determined by a single function, the contact (or tensor) potential $\chi$ \cite{deWit:2006gn,RoblesLlana:2006is,Alexandrov:2008gh}. Classically it reduces to the four-dimensional dilaton
\begin{equation} \label{eq:contactPotentialCLassic}
    \chi_{\rm cl}=\frac{1}{12}\,g_s^{-2}\,\mathcal{K}_{ijk}t^it^jt^k=\frac12 e^{-2\varphi_4}\,.
\end{equation}
The corrections from $\alpha'$, string loops, world-sheet instantons and D1--D$(-1)$-instantons take the $SL(2,\mathbb Z)$-covariant form
\begin{equation}
    \chi_{\rm corr}=\frac{g_s^{-2}}{8(2\pi)^3}\sum_{k\geq0}n_k^{(0)}
    \sum_{(m,n)\neq(0,0)}
    \frac{1+2\pi|m\tau+n|\,k_it^i}{|m\tau+n|^3}\,e^{-S^k_{m,n}}\,,
    \label{eq:chiSL2Z}
\end{equation}
with $S^k_{m,n}=2\pi k_i\left(|m\tau+n|\,t^i-imc^i-inb^i\right)$, where $\tau$ is the axio-dilaton, $k_i$ runs over effective curve classes, $n^{(0)}_k$ are the genus-zero Gopakumar--Vafa invariants, and $n^{(0)}_0=-\chi_E(X)/2$.

A Poisson resummation over $n$ separates the $k=0$, $m=0$ terms, i.e.\ the $\alpha'^3$ and one-loop corrections,
\begin{equation}
    \chi_{\rm pert}=-\frac{\zeta(3)\,\chi_E}{8(2\pi)^3}\,g_s^{-2}-\frac{\chi_E}{192\pi}\,,
    \label{eq:chipert}
\end{equation}
the $k\neq0$, $m=0$ terms, i.e.\ the world-sheet instantons,
\begin{equation}
    \chi_{\rm WS}=\frac{g_s^{-2}}{4(2\pi)^3}\sum_{k>0}n_k^{(0)}\,
    {\rm Re}\left[{\rm Li}_3\!\left(e^{2\pi ik_iz^i}\right)
    +2\pi k_it^i\,{\rm Li}_2\!\left(e^{2\pi ik_iz^i}\right)\right],
    \label{eq:WScorrections}
\end{equation}
and the $m\neq0$ terms, i.e.\ D$(-1)$ and D1--D$(-1)$ instantons of charge $(k_0,k_i)$, weighted by $K_1(2\pi m\,g_s^{-1}|k_0+k_iz^i|)$ \cite{Marchesano:2019ifh}
\begin{equation}
    \chi_{\rm D1,D(-1)}=\frac{g_s^{-1}}{8\pi^2}\sum_{k_I\neq0}n_{k_I}\sum_{m>0}
    \frac{|k_I z^I|}{m}\cos\!\left(2\pi m\,k_I\zeta^I\right)
    K_1\!\left(2\pi m\,g_s^{-1}|k_I z^I|\right),
\end{equation}
where $z^I=(1,z^i)$ and $\zeta^I=(\tau_1,\tau_1b^i-c^i)$.

In case of mutually local instantons, in particular when the relevant corrections come from D1, D$(-1)$ and D3, one can write the contact potential as ~\cite{Alexandrov:2011va, Alexandrov:2013yva, Baume:2019sry}
\begin{equation}
\chi=\frac{g_s^{-2}}{12}\mathcal{K}_{ijk}t^it^jt^k-\frac{\chi_E}{192\pi}
+\frac{g_s^{-2}}{8 \pi^2}\sum_\gamma \Omega(\gamma)\sum^\infty_{m=1}\frac{{\rm Re}(S_\gamma)}{m}K_1(2\pi m{\rm Re}(S_\gamma))
\label{eq:chi-withD3}
\end{equation}
where $\gamma$ is the instanton charge vector, $\Omega(\gamma)$ the generalized Donaldson--Thomas invariants. 

\paragraph{Towers of states.} Once perturbative and non-perturbative corrections are included, the classical relation $M_p^2\sim e^{-2\varphi_4}M_s^2$ is modified. The four-dimensional dilaton is replaced by the contact potential $\chi$ and \cite{Baume:2019sry}
\begin{equation}\label{eq:defChi}
    \left(\frac{M_s}{M_p}\right)^2 \sim \frac{1}{\chi} \,.
\end{equation}
The relation among the tension in Planck units and the tension in string units of any four-dimensional string is then
\begin{equation}
    \frac{\mathcal{T}}{M_p^2}=\frac{\mathcal{T}}{M_s^2} \frac{M_s^2}{M_p^2}\sim\frac{\mathcal{T}}{M_s^2} \frac{1}{\chi}\,.
\end{equation}
The tension $\mathcal{T}$ may itself receive quantum corrections. Classically, for D$p$-brane wrapping a $(p-1)$-cycle $\Sigma$, the tension in string units is given by the action of a Euclidean D$(p-2)$-brane on the same cycle
\begin{equation}
    \mathcal{T}_{{\rm D}p|\Sigma}=M_s^2\,{\rm Re}\,S_{{\rm D}(p-2)|\Sigma}\,.
    \label{eq:T-ReS}
\end{equation}
As discussed in \cite{Baume:2019sry}, this relation is not spoiled by quantum corrections. Thus, we find
\begin{equation}
    \frac{\mathcal{T}_{\rm F1}}{M_p^2}\sim\frac{1}{\chi}\,,\qquad
    \frac{\mathcal{T}_{{\rm D}p|\Sigma}}{M_p^2}\sim\frac{{\rm Re}\,S_{{\rm D}(p-2)|\Sigma}}{\chi}\,,\qquad
    \frac{\mathcal{T}_{{\rm NS5}|D}}{M_p^2}\sim\frac{{\rm Re}\,S_{{\rm D3}|D}}{g_s\,\chi}\,,
    \label{eq:QC-tensions}
\end{equation}
where the first one is just \eqref{eq:defChi}, the second combines \eqref{eq:T-ReS} and \eqref{eq:defChi}, and the third is the S-dual of the relation for the D5-brane. The instanton actions are both corrected by perturbative and non-perturbative effects. However they can be constrained and determined by dualities.\footnote{See eqs.~(3.35)--(3.38) of \cite{Baume:2019sry} and the discussion below and also App. A} For instance, S-duality can be used to argue that the ten-dimensional dilaton, viewed as a coordinate on $\mathcal{M}_{HM}$, does not get corrected. Thus, given that ${\rm Re}\,S_{{\rm D}(p-2)|\Sigma}\propto g_s^{-1}v_\Sigma$, the corrections can then only enter through the volume of $\Sigma$. In particular, $S_{{\rm D}(-1)|\Sigma}\propto g_s^{-1}$ does not receive corrections. Similarly, KK and winding modes on a two-torus of volume $t$ have 
\begin{equation}
    \frac{M^2_{\rm KK}}{M_p^2}\sim\frac{1}{t\,\chi}\,,\qquad \frac{M^2_{\rm w}}{M_p^2}\sim \frac{t}{\chi}\,.
\end{equation} 

\subsection{Instantons dual to black holes}
We start with the D$(-1)$--D3--D3--D3 instantons of section \ref{sec:instantonstorus}. Near the core the volumes are attracted to constant values while $g_s\sim r^{-2}$ run to strong coupling, i.e. $\lambda=r\to0$ with $\mu=-2$ and $\nu=0$. The world-sheet instanton actions stay finite, while all D-brane instanton actions vanish as $g_s^{-1} \sim r^2$ and the NS5 instanton action as $g_s^{-2} \sim r^4$. NS5-instantons are unsuppressed, and so are the mutually non-local D$(-1)$- and D5-instantons. No duality frame is under control near the core, and the known corrections do not allow us to follow the instantons into this region.

Nevertheless, in Appendix~\ref{sec:appendixB}, we take an exploratory perspective and attempt to follow how such corrections might deflect flows in the complexified moduli space, by integrating in, one sector at a time, the instantons whose actions become small along the flow, using a hierarchy among the instanton actions at large K\"ahler moduli together with S-duality and mirror symmetry. This analysis does not apply directly to our saddles, but we find it an interesting direction to look into, and leave further exploration for future work.

\subsection{Strings dual to black holes}
We now turn to the strings, whose cores lie at weak coupling for most of the solutions, so that we can expect to have more control over the corrections. We first consider the bound states dual to finite-size black holes, then the limits obtained by rescaling their charges, and finally the strings dual to small black holes.

\subsubsection{The F1 limit dual to black hole attractors}\label{sec:QCtoF1}
The first string we consider is the D1--D5--D5--D5 bound state of section \ref{sec:EFTminimal}, dual to a finite-size black hole. Along its flow, the K\"ahler moduli approach their attractor values, $z^i\to z^i_h$, while $g_s\sim L^{-1}$, with $L=\log(r_0/r)\to\infty$. With $\lambda=L$, i.e.\ $\mu=-1$ and $\nu=0$, all D-brane and NS5-brane actions grow as $L$ or $L^2$, and only the perturbative and world-sheet instanton corrections survive. The latter are not parametrically suppressed, but are numerically small, $\sim e^{-2\pi k_it^i_h}$, since the attractor analysis assumes large volume. Since $g_s^2\chi_{\rm WS}$ only depends on $z^i$,
\begin{equation}
    \chi=g_s^{-2}\left[\frac{1}{12}\mathcal{K}_{ijk}t^i_ht^j_ht^k_h
    -\frac{\zeta(3)\chi_E}{8(2\pi)^3}+g_s^2\chi_{\rm WS}(z_h)\right]
    -\frac{\chi_E}{192\pi}
    +\mathcal O\!\left(e^{-2\pi|k_0+k_iz^i_h|/g_s}\right).
\end{equation}
The $\alpha'^3$ and world-sheet instanton corrections only shift the coefficient of $g_s^{-2}$, and the one-loop term is subleading. Hence $\mathcal{T}_{\rm F1}/M_p^2\sim1/\chi\sim L^{-2}$, as classically, and the F1 limit is unaffected. This is precisely the limit discussed in \cite{Marchesano:2019ifh}, in which all D-instanton corrections are suppressed, the trajectory can be described classically, and the distance conjecture is satisfied by the tower of fundamental string states. This is also the trajectory they obtain in \cite{Baume:2019sry} after deflecting a trajectory exploring weak coupling and shrinking volumes.

Notice that this argument only uses that the K\"ahler moduli are finite and at large-volume point while $g_s\to0$. In a generic Calabi--Yau compactification (section \ref{sec:CYStringsLimits}) all strings dual to finite-size black holes are bound states with this property and the universal F1 limit they probe is not obstructed by quantum corrections.

\subsubsection{D1 limit after rescaling charges}\label{sec:QCtoD1}
The F1 limit is reached following the scalar field profile of a configuration dual to a finite-size black hole till its core. In section \ref{sec:BHlimitRescalingCharges} we considered a different procedure. We stopped at a fixed radius $r_{UV}$, where the string probes the UV cutoff, and blew up some of the charges of the bound state, showing that other limits can be probed classically: rescaling $q_0$ at fixed $p^i$ gives an emergent D1 limit, while rescaling $q_0=p^1$ at fixed $p^2, p^3$ gives a decompactification of $T^2_1$ to six dimensions. We now check whether the former survives quantum corrections; the latter has the same scalings, $g_s\sim1$ and $t^1\sim\lambda$, as the D1--D5 string of section \ref{sec:QCtoSmallEFT}, and is treated there. For $q_0\to\infty$ at fixed $p^i$ and $r_{UV}$, one has
\begin{equation}
    g_s=\hat g\,q_0^{1/2}\,,\qquad t^i=\hat t^i\,q_0^{1/2}\,,
    \label{eq:D1-scaling}
\end{equation}
i.e.\ $\lambda=q_0$ with $\mu=\nu=\frac12$. The D3-, D5-, NS5- and world-sheet instanton actions grow, the D$(-1)$ actions vanish and the D1 actions are constant. Since the D$(-1)$ actions vanish, it is easier to work in the S-dual frame. At $C_0=0$ we get \eqref{eq:SdualMap}
\begin{equation}
    g_s'=\frac{1}{g_s}\,,\qquad t'^i=\frac{t^i}{g_s}\,,\qquad
    \chi'=\frac{\chi}{|\tau|}=g_s\,\chi\,,
    \label{eq:S-duality-map}
\end{equation}
so that $\mathcal{T}_{\rm D1}/M_p^2\sim1/\chi'$. The dual frame is at weak coupling, $g'_s\sim q_0^{-1/2}$, and fixed volume, $t'^i=\mathcal O(1)$. The world-sheet instantons are unsuppressed, while the D$(-1)$- and D1-instantons, with actions $\sim g_s$ and $\sim t^i$, are exponentially suppressed. As in the F1 limit,
\begin{equation}
    \chi=g_s\,\tilde\chi(t',c)-\frac{\chi_E}{192\pi\,g_s}
    +\mathcal O\!\left(e^{-2\pi g_s},\,e^{-2\pi k_it^i}\right),
    \label{eq:chi-D1-limit}
\end{equation}
where $\tilde\chi(t',c)=\frac{1}{12}\mathcal{K}_{ijk}t'^it'^jt'^k-\frac{\zeta(3)\chi_E}{8(2\pi)^3}+g'^2_s\chi_{\rm WS}(z')$, with $z'^i=c^i+it'^i$, contains the classical, $\alpha'^3$ and world-sheet terms of the dual frame, and the second term is its one-loop correction.\footnote{Equivalently, in the original frame, the $k=0$ sector of \eqref{eq:chiSL2Z} is $-\frac{\chi_E}{16(2\pi)^3}g_s^{-1/2}E_{3/2}(\tau)$ \cite{RoblesLlana:2006is}, and $E_{3/2}(i/g_s)=E_{3/2}(ig_s)=2\zeta(3)g_s^{3/2}+4\zeta(2)g_s^{-1/2}+\dots$ reproduces the $\alpha'^3$ term in $g_s\tilde\chi$ and the one-loop term of \eqref{eq:chi-D1-limit}.} The leading term has the same homogeneity as $\chi_{\rm cl}$: the $\alpha'^3$ and world-sheet corrections can only shift the value of $t'^i$ by a $q_0$-independent amount, while the one-loop term is suppressed by $q_0^{-1}$. Thus $\chi\sim q_0^{1/2}$, and by \eqref{eq:QC-tensions}
\begin{equation}
    \frac{\mathcal{T}_{\rm D1}}{M_p^2}\sim\frac{M^2_{\rm KK}}{M_p^2}\sim q_0^{-1}\,,\qquad
    \frac{\mathcal{T}_{\rm F1}}{M_p^2}\sim\frac{\mathcal{T}_{{\rm D3}|C}}{M_p^2}\sim\frac{\mathcal{T}_{{\rm NS5}|D}}{M_p^2}\sim q_0^{-1/2}\,,\qquad
    \frac{\mathcal{T}_{{\rm D5}|D}}{M_p^2}\sim q_0^0\,,
\end{equation}
with $M^2_{\rm KK}/M_p^2\sim1/(\chi\,\mathcal V^{1/3})$. The limit remains an equi-dimensional emergent D1 string limit.\footnote{Notice that this is not the D1 limit of \cite{Marchesano:2019ifh,Baume:2019sry}, along which the volumes in D1 units, $t'^i=t^i/g_s$, shrink classically and are frozen by quantum corrections: here $t'^i=\hat t^i/\hat g$ is constant already classically.}

\subsubsection{Limits of strings dual to small black holes}\label{sec:QCtoSmallEFT}
We finally consider the strings of table~\ref{tab:EFTsmall}, dual to small black holes, i.e.\ the elementary D1 and D5 strings and their bound states with fewer than four charges. Unlike for the bound states discussed so far, not all the K\"ahler moduli approach a finite value at the core, and the volumes no longer scale homogeneously. The scalings of the instanton actions are collected in table~\ref{tab:small-actions}.

\begin{table}[t]
    \centering
    \setlength{\tabcolsep}{2.5pt}
    \begin{tabular}{l|c|c|c|c|c|c|c|c|c}
       String & $\mu$ & $\nu_i$ & $ \nu_{\mathcal{V}}$ & ${\rm WS}|_{C_i}$ & D$(-1)$ & ${\rm D1}|_{C_i}$ & ${\rm D3}|_{D_i}$ & D5 & NS5 \\ \hline
      D1 & $\frac12$ & $(\frac12,\frac12,\frac12)$ & $\frac32$ & $(\frac12,\frac12,\frac12)$ & $\textcolor{dark-blue2}{-\frac12}$ & $(\textcolor{dark-blue2}{0},\textcolor{dark-blue2}{0},\textcolor{dark-blue2}{0})$ & $(\frac12,\frac12,\frac12)$ & $1$ & $\frac12$ \\
      D5 & $-\frac12$ & $(\frac12,-\frac12,-\frac12)$ &  $-\frac12$ & $(\frac12,\textcolor{dark-blue2}{-\frac12},\textcolor{dark-blue2}{-\frac12})$ & $\frac12$ & $(1,\textcolor{dark-blue2}{0},\textcolor{dark-blue2}{0})$ & $(\textcolor{dark-blue2}{-\frac12},\frac12,\frac12)$ & $\textcolor{dark-blue2}{0}$ & $\frac12$ \\
      D1--D5 & $0$ & $(1,0,0)$  & $1$ & $(1,\textcolor{dark-blue2}{0},\textcolor{dark-blue2}{0})$ & $\textcolor{dark-blue2}{0}$ & $(1,\textcolor{dark-blue2}{0},\textcolor{dark-blue2}{0})$ & $(\textcolor{dark-blue2}{0},1,1)$ & $1$ & $1$ \\
      D5--D5 & $-1$ & $(0,0,-1)$  & $-1$ & $(\textcolor{dark-blue2}{0},\textcolor{dark-blue2}{0},\textcolor{dark-blue2}{-1})$ & $1$ & $(1,1,\textcolor{dark-blue2}{0})$ & $(\textcolor{dark-blue2}{0},\textcolor{dark-blue2}{0},1)$ & $\textcolor{dark-blue2}{0}$ & $1$ \\
      D1--D5--D5 & $-\frac12$ & $(\frac12,\frac12,-\frac12)$ & $\frac12$ & $(\frac12,\frac12,\textcolor{dark-blue2}{-\frac12})$ & $\frac12$ & $(1,1,\textcolor{dark-blue2}{0})$ & $(\frac12,\frac12,\frac32)$ & $1$ & $\frac32$ \\
      D5--D5--D5 & $-\frac32$ & $(-\frac12,-\frac12,-\frac12)$ &  $-\frac32$ & $(\textcolor{dark-blue2}{-\frac12},\textcolor{dark-blue2}{-\frac12},\textcolor{dark-blue2}{-\frac12})$ & $\frac32$ & $(1,1,1)$ & $(\frac12,\frac12,\frac12)$ & $\textcolor{dark-blue2}{0}$ & $\frac32$
    \end{tabular}
    \caption{Exponents of $L=\log(r_0/r)\to\infty$ along the trajectories of table~\ref{tab:EFTsmall}, with $g_s\sim L^\mu$, $t^i\sim L^{\nu_i}$ and $\mathcal{V}\sim L^{\nu_1+\nu_2+\nu_3}$. Columns 5--10 give the exponents of the real parts of the instanton actions \eqref{eq:non-perturbative-contributions-general}; blue entries are unsuppressed sectors. The $\mathcal{O}(1)$ constant coefficients are not relevant here, thus every $t^i\sim L^0$ in this table should be interpreted as a large but finite number.}
    \label{tab:small-actions}
\end{table}

Three general facts simplify the analysis. First, ${\rm Re}\,S_{\rm NS5}\sim e^{-2\varphi_4}$ grows in all rows. Second, the D-instanton corrections can be computed along all these flows. A D-instanton carries a charge $\gamma^I=(q_0,q_i,p^i,p^0)$, with $q_0$ the D$(-1)$ charge, $q_i$ the D1 charge on $T^2_i$, $p^i$ the D3 charge on $D_i=T^2_j\times T^2_k$ and $p^0$ the D5 charge on $X$. Two D-instantons are mutually non-local when their symplectic pairing $\gamma^I\Omega_{IJ} \gamma^J{}' =q_I p'^I-p^I q'_I$ is non-zero. Since $T^2_i\cdot D_j=\delta_{ij}$, the only non-local pairs are D$(-1)$--D5 and D1$|_{T^2_i}$--D3$|_{D_i}$. For each of the strings, at most one member of each pair is unsuppressed, so the unsuppressed D-instantons are mutually local and their contribution to $\chi$ is known (cf. equations \eqref{eq:contactPotentialCLassic}-\eqref{eq:WScorrections}). Third, T-duality along both circles of $T^2_i$,
\begin{equation}
    t^i\to\frac{1}{t^i}\,,\qquad t^j\to t^j\,,\qquad g_s\to\frac{g_s}{t^i}\,,\qquad j\neq i\,,
    \label{eq:double-T-duality}
\end{equation}
is an exact symmetry of type IIB on this geometric orbifold, mapping it to another geometric orbifold \cite{Vafa:1994rv,Pradisi:2014fqa}. It leaves $g_{4,s}=g_s\mathcal V^{-1/2}$, and hence $\chi_{\rm cl}$, invariant, maps the four-dimensional F1 to itself, and only relabels the D-instantons, mapping D1$|_{T^2_i}\leftrightarrow$ D$(-1)$, D3$|_{D_i}\leftrightarrow$ D5 and D3$|_{D_j}\leftrightarrow$ D1$|_{T^2_k}$ with their actions and weights. Hence $\chi$ is invariant, and we can always go to a frame in which no two-torus shrinks.

\paragraph{One charge.} The D1 row correspond to the scaling \eqref{eq:D1-scaling} with $q_0\to L$, so section \ref{sec:QCtoD1} applies: $\chi\sim L^{1/2}$ and the limit remains an emergent D1 limit. For the D5 on $T^2_2\times T^2_3$, double T-duality along $T^2_2$ and $T^2_3$ gives $g_s^{(T)}=g_s/(t^2t^3)\sim L^{1/2}$ and $t^{(T)}=(t^1,1/t^2,1/t^3)\sim L^{1/2}$, i.e.\ the D1 trajectory, with the D5 mapped to the D1. The limit remains an emergent string limit, with the wrapped D5-brane as the emergent string.

\paragraph{Two charges.} In the D1--D5 row only $t^1\sim L$ grows. The unsuppressed sectors are world-sheet and D1-instantons on $T^2_{2,3}$, D$(-1)$-instantons and D3-instantons on $D_1=T^2_2\times T^2_3$, and double T-duality along $T^2_2$ or $T^2_3$ maps this set to itself. None of them depends on $z^1$: their central charges, $Z=q_0+q_2z^2+q_3z^3+p\, z^2z^3$ for a D3 on $D_1$ bound to D1- and D$(-1)$-instantons, only involve $z^{2,3}$. We use  eq.~\eqref{eq:chi-withD3}, in particular the classical term will give the term linear in $L$ in~\eqref{eq:chi-D1D5}, since $\mathcal V=L\,v_{D_1}$, while all other instanton contributions, such as world-sheet and D1-instantons on classes with $k_1\neq0$ and D3-instantons on $D_2$ and $D_3$, are exponentially suppressed in $L$. Therefore
\begin{equation}
    \chi=\frac12 g_s^{-2}\,v_{D_1}\,t^1+\chi_{(0)}(\tau,z^2,z^3)+\mathcal O\big(e^{-2\pi t^1}\big)\,,
    \label{eq:chi-D1D5}
\end{equation}
with $\chi_{(0)}$ independent of $L$: the corrections only shift the volume of $T^2_1$ by an $\mathcal O(1)$ amount. Thus $\chi\sim L$, and \eqref{eq:QC-tensions} gives $\mathcal{T}_{\rm F1},\mathcal{T}_{\rm D1},\mathcal{T}_{{\rm D5}|D_1}\sim L^{-1}M_p^2$, while the lightest tower is the KK tower of $T^2_1$, $M^2_{{\rm KK},1}/M_p^2\sim1/(t^1\chi)\sim L^{-2}$. The limit remains a decompactification to six dimensions. In the D5--D5 row, $T^2_3$ shrinks, and double T-duality along it maps the trajectory to the D1--D5 one with $T^2_1\leftrightarrow T^2_3$.

\paragraph{Three charges.} In the D1--D5--D5 row, $T^2_3$ shrinks. After double T-duality along it, $g_s^{(T)}=g_s/t^3\sim L^0$ and $t^{(T)}=(t^1,t^2,1/t^3)\sim L^{1/2}$, and all world-sheet, D1-, D3-, D5- and NS5-instantons are suppressed. Only the $k=0$ sector of \eqref{eq:chiSL2Z} of the dual frame survives, i.e.\ the $\alpha'^3$ and one-loop terms and the D$(-1)$-instantons of the dual frame, which are the D1-instantons on $T^2_3$. It only depends on $\tau^{(T)}$, with $\tau^{(T)}_2=t^3/g_s=\mathcal O(1)$, so that
\begin{equation}
    \chi=\frac12\,g_s^{-2}\mathcal V+\mathcal O(L^0)\,,
    \label{eq:chi-n3}
\end{equation}
with a classical term growing as $L^{3/2}$. The corrections enter at relative order $L^{-3/2}$ and can only change the coefficients of the classical scalings. The lightest towers are therefore still the KK modes on $T^2_{1,2}$ and the winding modes on $T^2_3$, with $M^2/M_p^2\sim L^{-2}$, while $\mathcal{T}_{\rm F1}/M_p^2\sim L^{-3/2}$: six dimensions decompactify, and the limit remains a decompactification to ten dimensions. In the D5--D5--D5 row all tori shrink, and double T-duality along all three maps it to $g_s^{(T)}\sim L^0$ and $t^{(T)i}\sim L^{1/2}$, with the D5-instanton mapped to the D$(-1)$; the same argument applies, and the limit is again a decompactification to ten dimensions.

\subsubsection{Summary} Along the string flows dual to finite-size black holes on a generic Calabi--Yau, the unsuppressed corrections only shift the $\mathcal O(1)$ coefficients of the contact potential, i.e.\ the attractor values of the moduli, and do not change the scalings: the F1 and D1 limits remain emergent string limits. The same is true for string flows dual to small black holes on $T^6/(\mathbb Z_2\times\mathbb Z_2)$. The one charge limits remain emergent string limits, the two-charge limits decompactifications to six dimensions, and the three-charge limits decompactifications to ten dimensions, in agreement with table~\ref{tab:EFTsmall}. The small black hole results do not extend to the general Calabi--Yau case. Since the four-dimensional dilaton always runs to weak coupling at the core (cf. appendix \ref{app:singleDbranes}) 
and ${\rm Re}\,S_{\rm NS5}\sim e^{-2\varphi_4}$ the NS5-instantons do not become relevant along the string flows. However, the D-instantons are more delicate. On $T^6/(\mathbb Z_2\times\mathbb Z_2)$, the unsuppressed ones are always mutually local, and double T-duality brings every flow to a frame with no shrinking cycle. In a general Calabi--Yau, D1-instantons on curves and D3-instantons on divisors with non-vanishing intersection numbers can become unsuppressed simultaneously, for instance along flows on which some curves shrink. Their contribution to the contact potential is then not of the simple form used above, and a duality frame in which they are mutually local need not exist. Whether such flows change the nature of the limits is left for future work.

\section{Conclusions \& Outlook}\label{sec:conclusions}

In this work we have asked whether the attractor mechanism, which allows black holes to probe infinite-distance limits of the vector-multiplet moduli space, has a counterpart in the hypermultiplet sector. We T-dualized BPS black holes of four-dimensional $\mathcal N=2$ type IIA compactifications along a Euclidean time circle and along a spatial circle, realizing the temporal and spatial c-map, and obtaining D-instantons and strings of the type IIB hypermultiplet sector. To do so, we used  the dictionary provided by the duality transformations and we properly smeared/unsmeared the D-brane sources. We verified that the resulting configurations solve the equations of motion of type IIB, establishing a solution generating technique. We have constructed the solutions explicitly for a generic Calabi--Yau threefold in the two-derivative, large-volume approximation, and illustrated them on $T^6/(\mathbb Z_2\times\mathbb Z_2)$, where double T-duality gives us additional control. We kept track of the structure inherited by instantons and strings from black holes. The instanton action and the string tension are set by the ADM mass of the dual black hole, while its entropy reappears in the instanton measure through the BPS index. The warp factor of the black hole becomes the profile of the four-dimensional type IIB dilaton. The dual flows therefore always drive the dilaton, and cannot realize purely K\"ahler infinite-distance limits of the hypermultiplet moduli space.

The instantons have a flat Euclidean metric, imaginary axions and a core at strong coupling, which lies at zero distance along a null trajectory in the complexified moduli space. They can not be used to probe infinite distance limits in the same sense as black holes. If we insist in following their classical profile in the complexified moduli space to their core we observe a breakdown of the theory due to a tower of instantons becoming actionless. One might be able to make sense of such instanton corrections in complexified moduli space, and we explore this in Appendix \ref{sec:appendixB}. 

In the Lorentzian world, instantons are complex saddles of the path integral. We argue that their action is bounded from below by the square root of the entropy of the dual black hole. Hence, only instantons dual to small black holes can become unsuppressed along infinite-distance limits at fixed four-dimensional dilaton: these are the elementary instantons that correct the hypermultiplet metric and lower the string scale like a species scale \cite{Marchesano:2019ifh,Baume:2019sry}. Instantons dual to minimal black holes remain suppressed, but their measure carries the same degeneracy as their dual black hole, $\log|\Omega|\sim S_{BH}\sim N_{\rm sp}$. This mirrors the black-hole side, where small black holes are the species themselves, while minimal black holes are macroscopic objects whose entropy counts them.

The strings are axionic strings, in general bound states of EFT strings \cite{Lanza:2021udy}, with no asymptotically flat region and a core that always lies at infinite distance and at weak four-dimensional coupling. All strings dual to finite-size black holes, minimal ones included, probe the same equi-dimensional emergent F1 limit, irrespective of the limit probed by the dual black hole, with the F1 tension at a given radius set by the black-hole entropy. We have argued that this limit survives quantum corrections in a generic Calabi--Yau. Strings dual to small black holes probe other limits: on $T^6/(\mathbb Z_2\times\mathbb Z_2)$, emergent D1 and D5 string limits and decompactifications to six and ten dimensions, which also survive quantum corrections, since double T-duality maps every flow to a frame with no shrinking cycle. The scaling weights of these flows agree with the integral scaling relation of \cite{Lanza:2021udy} and with the additivity of \cite{Etheredge:2026jun} for bound states. Stopping the flow at a fixed radius and rescaling the charges of the bound states gives access to further limits, such as an emergent D1 limit, which also survives quantum corrections. On a generic Calabi--Yau, instead, most strings dual to small black holes classically leave the large-volume regime at fixed or weak ten-dimensional coupling, with the exception of the D1 string and of certain D3 strings, which we expect to survive quantum corrections.

Some interesting open directions are:
\begin{itemize}
    \item \textbf{Quantum-corrected flows.} We have used the classical trajectories to identify which corrections become relevant, rather than solving the BPS flow equations with the corrected hypermultiplet metric, which would in general deform the trajectories themselves. Near the core of the instantons, NS5-instantons and mutually non-local D-instantons become unsuppressed, and no duality frame is under control, unless there is a hierarchy among the instanton actions and the corrections due to the leading instantons are enough to deflect the trajectory. On a generic Calabi--Yau, the strings dual to small black holes along which some cycles shrink are in a similar situation: there is no analogue of the double T-duality of the toroidal orbifold, and mutually non-local D1- and D3-instantons can become relevant simultaneously. Following these solutions, and verifying explicitly that the D3 flows of section \ref{sec:CYStringsLimits} survive, requires the full stringy corrections to the hypermultiplet moduli space.
  \item \textbf{Observables at singular cores.} Unlike the black holes, both the instantons and the strings are singular at their core, and the horizon, with its macroscopic entropy, is lost. Even defining observables that are sensitive to the physics at the core is difficult. The string tension vanishes at the core and blows up at finite distance from it, where the solution breaks down, so that the strings are confined. It would be interesting to see whether they can be continued all the way out to an asymptotic region, like the strings of \cite{Green:1993zr} (perhaps along the lines of \cite{Marchesano:2022avb}). The core tension evaluated at a UV cutoff radius $r_{\rm UV}$ \cite{Lanza:2021udy} could know about the entropy, but it cannot be computed with EFT data alone, and there is no universal prescription for $r_{\rm UV}$. For the instantons, the action is fixed by the asymptotic data, and whether an analogue of the core tension exists is unclear; the entropy only reappears in the measure, which is not a property of the classical saddle. In particular, this makes it difficult to rescale the charges as one does for black holes, whose horizon provides a finite, non-singular locus at which the attractor values are measured: for the strings and instantons there is no such locus, and in section \ref{sec:BHlimitRescalingCharges} we had to stop the flow at an arbitrary radius. It would be interesting to find observables that overcome this, and to understand whether some notion of degeneracy survives for the strings.

    \item \textbf{Microscopic degeneracies.} For the instantons, the microscopic degeneracy of the dual black hole reappears in the measure: an instanton of charge $\gamma$ is weighted by the BPS index $\Omega(\gamma)$ of the dual black hole \cite{Pioline:2009ia} (section \ref{sec:instUVprobes}), which for large charges grows as $e^{S_{BH}(\gamma)}$. The strings should also carry some notion of degeneracy, depending on how many branes of each type they are made of, but we do not know how to define it macroscopically, nor which observable is sensitive to the microscopic counting. The only place where we found the black hole entropy is the running of the four-dimensional dilaton near the core: $e^{-\varphi_4}\sim\sqrt{S_{BH}}\,|\log(r/r_0)|$ \eqref{eq:dilatonSbh}, so that $S_{BH}$ sets the rate at which the string runs to weak coupling and the tension of the associated fundamental string \eqref{eq:TF1entropy}. Since $S_{BH}$ is the two-derivative entropy, this can only capture the degeneracy at large charges, where $\log\Omega(\gamma)\simeq S_{BH}(\gamma)$. It would be interesting to understand the degeneracy of the strings microscopically, e.g.\ from the world-volume theory of the bound state as in \cite{Maldacena:1997de}, and to identify the physical quantity that is sensitive to it.

    \item \textbf{Swampland bounds.} The cores of our instantons lie at zero distance along null trajectories, outside the scope of both the Distance Conjecture and the imaginary distance bound \cite{Maldacena:2026jqd,DiUbaldo:2026rly}, although the classical description breaks down there. It would be interesting to understand which principle controls this breakdown. Similarly, our results suggest that the instantons dual to small black holes are emergent in the sense of \cite{Artime:2026kfq}, while those dual to finite-size black holes are not; it would be interesting to verify this with eight supercharges. Finally, in a generic Calabi--Yau, ED3-instantons on effective but non-nef divisors are not EFT instantons \cite{Lanza:2021udy,Martucci:2024trp}, and it would be interesting to clarify whether the EFT condition is the T-dual counterpart of the attractor flow of the dual black hole remaining within the K\"ahler cone.
    \item \textbf{Less supersymmetry and potentials.} Going down to four-dimensional $\mathcal N=1$ with an orientifold projection would presumably lead to EFT strings similar to those of \cite{Lanza:2021udy,Martucci:2024trp}, with an O3/O7 or O5/O9 projection keeping different halves of the square of dualities of section \ref{sec:stringbounds}. More generally, probing moduli space with attractor flows would be more realistic in the presence of a scalar potential, which would compete with the charges in driving the moduli \cite{Angius:2023xtu,Delgado:2025crl}.

    \item \textbf{Trajectories in complexified moduli space.}
    In Appendix \ref{sec:appendixB}, we have taken a first step towards understanding how quantum corrections modify the limits probed by euclidean path integral saddles. We do not expect that all the trajectories in a complexified moduli space should be meaningful due to the issues of M-theory with different signatures \cite{Dijkgraaf:2016lym}. But the trajectories realized by the saddle of a gravitational euclidean path integral should encode meaningful information about quantum gravity \cite{Lin:2026ztf, Licciardello:2026roi}. For the future, we would like to explore the landscape of  complexified trajectories appearing on path integral saddles, keep track of their compatibility with swampland conjectures and their differences with trajectories not realized by a saddle. For instance, our analysis of section \ref{sec:quantumCorrections} gives evidence that along the quantum-corrected flow of a complexified instanton the lightest states are those of a critical tensionless string, showing compatibility with the Emergent String Conjecture.
\end{itemize}
We hope to come back to these and other interesting questions in the future.

\bigskip

\acknowledgments
We thank Rafael \'Alvarez-Garcia, Ralph Blumenhagen, Jos\'e Calder\'on-Infante, Alberto Castellano, Muldrow Etheredge, Alvaro Herr\'aez, Fernando Marchesano, Luca Martucci, Luca Melotti, Miguel Montero, Vinicius Nevoa, Boris Pioline, Ignacio Ruiz, Cumrun Vafa, Irene Valenzuela, Timo Weigand and Max Wiesner for helpful discussions.  M.D., G.T. and M.Z. acknowledge the hospitality of the Simons Center for Geometry and  Physics during the 2026 Summer Workshop, where part of the research presented here was completed. D.L., G.T. and M.Z. are grateful to Harvard University and the Harvard Swampland Initiative for the hospitality during the final stages of this work. The work of M.D. is supported in part by a grant from the Simons Foundation (602883,CV) and the DellaPietra Foundation. The work of D.L. is supported by the German-Israel-Project (DIP) on ``Holography and the Swamplan''. The work of M.Z. is supported by the Alexander von Humboldt Foundation. M.Z. is grateful to Miriam Gori for continuous encouragement and support.

\section*{Declaration of AI Use}

The authors acknowledge the use of Codex and Claude for language refinement, LaTeX formatting assistance, figure drawing and writing of Mathematica code. The computation reported have been done entirely by the authors without AI assistance, mostly before summer, except for the numerical integration of the Picard-Fuchs equations in appendix \ref{sec:appendixB}. The authors take full responsibility for the final content.

\appendix

\section{Strings from a single type of D-brane}\label{app:singleDbranes}

In this appendix we construct the classical flows of section \ref{sec:CYStringsLimits} for each type of D-brane source. We denote by $C_s$ and $D_s$ the total classes of the curves and divisors wrapped by the D3- and D5-branes, respectively, and by $C$ and $D$ fixed non-zero effective probe curves and divisors. In each case, we identify the harmonic function $H(r)=h+Q\log(r_0/r)$ of section \ref{sec:CYStringsLimits} as follows. For the D1 and D7 sources, only one harmonic function is non-constant, and we identify it with $H$ (up to sign). For the D3 and D5 sources, several harmonic functions can be non-constant, and we set $h=0$ and $Q=1$, i.e.\ $H=\log(r_0/r)$, so that $H_i=h_i+q_iH$ and $\tilde H^i=\tilde h^i+p^iH$.

For each flow we give the scaling with $H$ of $g_s$, of the volumes, and of the probe tensions \eqref{eq:intclass-tensions} and KK masses \eqref{eq:intclass-KK}. Since the flows are classical, we do not know which infinite-distance limits survive once corrections are included, and we only record naively which of these objects become light. Two caveats apply throughout. First, whenever some cycle shrinks, the flow leaves the large-volume regime in which \eqref{eq:prepotentialCY} is valid, and the classical description breaks down near the core. On $T^6/(\mathbb Z_2\times\mathbb Z_2)$, double T-duality \eqref{eq:double-T-duality} maps such flows to a frame with no shrinking cycle, but no such duality is available on a generic Calabi--Yau. Second, when $g_s$ grows, S-duality, $g_s'=1/g_s$ and $t'^i=t^i/g_s$, may provide a better description. However, along all the flows below with shrinking cycles, $g_s$ is constant or vanishes, so that the S-dual frame either has the same shrinking cycles or is strongly coupled.

Along the D1 and D5 flows, $t^i/g_s=\sqrt2\,\tilde H^i$ and $\mathcal V/g_s=-\sqrt2\,H_0$, while along the D3 and D7 flows, $1/g_s=\sqrt2\,\tilde H^0$ and the Einstein-frame divisor volumes $\frac1{2g_s}\mathcal K_{ijk}t^jt^k=\sqrt2\,H_i$, are linear in $H$. Restricted to these saxions and to the RR axions they pair with, as in section \ref{sec:stringbounds}, these flows therefore coincide with 4d $\mathcal{N}=1$ EFT string flows of \cite{Lanza:2021udy}: the D1 and D5 flows with the D1- and D5-string flows of type I compactifications \cite[sec.~6.2]{Lanza:2021udy}, and the D3 and D7 flows with the D3- and D7-string flows of type IIB/F-theory compactifications \cite[sec.~6.3]{Lanza:2021udy}. The classification of the D5 flows in \cite[sec.~6.1--6.2]{Lanza:2021udy}, by the intersection numbers of the charge $J_+$, coincides with ours for $\nu=3,2,1$, and the scalings of the volumes agree with ours where they overlap. In \cite[sec.~6.3]{Lanza:2021udy} the D3 flows are on the base $B$ of an elliptically fibred Calabi--Yau fourfold. Since $c_1(B)\neq0$, $B$ is not itself Calabi--Yau ($B=\mathbb P^3$ in their simplest example), but the flow equations are the same. Below, we compute the tensions of all the probes, beyond that of the string generating the flow, and point out where \cite{Lanza:2021udy} sheds light on our flows.

\paragraph{D1 source:}
We select $\tilde H^0=H_i=0$, with $\tilde H^i=\tilde h^i$ constant, and identify $H=-H_0$. Then $X^i=\frac{i}{2}\tilde h^i$ is constant and purely imaginary, and $X^0$ is real and scales as $X^0\sim H^{-1/2}$. The saxions are $t^i={\rm Im}\,(X^i/X^0)=H^{1/2}t^i_+$, with $t^i_+\propto\tilde h^i$, so that $J=H^{1/2}J_+$ with $J_+=t^i_+\omega_i$ fixed by the constants $\tilde h^i$. Since $J$ is exactly proportional to $J_+$, the constants must be chosen such that $J_+$ lies in the interior of the K\"ahler cone. We obtain the scaling $v_C\sim H^{1/2}$, $v_D\sim H$ and $\mathcal V\sim H^{3/2}$. Substituting into \eqref{eq:intclass-tensions} gives
\begin{equation}
\begin{aligned}
 \tau_{\rm F1}&\sim H^{-1/2},&
 \tau_{\rm D1}&\sim H^{-1},&
 \tau_{{\rm D3}[C]}&\sim H^{-1/2},\\
 \tau_{{\rm D5}[D]}&\sim1,&
 \tau_{{\rm NS5}[D]}&\sim H^{-1/2},&
 \tau_{{\rm D7}[X]}&\sim H^{1/2}\,, \quad  \frac{M_{\rm KK}^{\rm light}}{M_p} \sim H^{-1/2} \,.
\end{aligned}
\label{eq:intclass-D1-probes}
\end{equation}
Naively, the lightest objects are the KK modes along the whole Calabi--Yau, since all curves grow as $\mathcal R_C\sim H^{1/4}$, and the D1-string, with $(M_{\rm KK}^{\rm light}/M_p)^2\sim\tau_{\rm D1}\sim H^{-1}$. However, $g_s\sim H^{1/2}$ grows, and the classical description breaks down near the core. Since all volumes grow, we can hope that S-duality provides a better description: in the S-dual frame $g_s'\sim H^{-1/2}\to0$, while $t'^i=t^i/g_s\propto\tilde h^i$ are constant and can be chosen large. This is the D1 limit discussed in section \ref{sec:QCtoD1}. As for the type I D1-string \cite[sec.~6.2]{Lanza:2021udy}, the flow is thus described in the S-dual frame, where it is a weak-coupling limit for the F1.

\paragraph{D3 source:}
We select $\tilde H^i=H_0=0$, with $\tilde H^0$ constant, and identify $H$ via $H_i=h_i+Hq_i$. Then $X^0=\frac{i}{2}\tilde H^0$ is constant and purely imaginary, $X^i$ is real, and the stabilization equations give
\begin{equation}
 \frac12\mathcal K_{ijk}t^jt^k=\frac{H_i}{\tilde H^0}\,.
 \label{eq:intclass-D3-D7-solution}
\end{equation}
The string coupling $g_s\sim|\tilde H^0|^{-1}$ is constant, and can be chosen small. Equivalently, in cohomological notation, after absorbing the fixed normalization into a constant $c>0$, one may write
\begin{equation}
 \frac12 J^2=\Sigma_0+cH\,\gamma_s ,
 \label{eq:intclass-D3-cohomological} 
\end{equation}
where $\gamma_s, \Sigma_0 \in H^{2,2}(X,\mathbb R)$. $\gamma_s$ is the Poincar\'e dual of
the total class $C_s$ of the source curves, instead $\Sigma_0$ encodes the constants $h_i$. Since $v_D=\frac12J^2\cdot D=\Sigma_0\cdot D+cH\,D\cdot C_s$, the flow remains in the K\"ahler cone at large $H$ only if $D\cdot C_s\geq0$ for every effective divisor $D$, i.e.\ if $C_s$ is movable. We assume this in the following. In \cite{Lanza:2021udy}, the same condition characterizes the D3-strings that are EFT strings. It is necessary but not sufficient for the flow to stay in the K\"ahler cone: the map $J\mapsto\frac12J\wedge J$ is injective but not surjective, and the flow sourced by a movable curve can reach a wall of the K\"ahler cone, where a curve, but no divisor, collapses at finite distance \cite[sec.~6.3]{Lanza:2021udy}. We assume that this does not happen, or that the flow is continued into the corresponding flopped phase, with different intersection numbers.

\begin{itemize}
    \item \emph{Growing direction, $\nu=3$.} We can solve \eqref{eq:intclass-D3-D7-solution} for $t_+^i$ such that $\frac12\mathcal K_{ijk}t_+^jt_+^k\propto q_i$. We obtain $t^i \sim H^{1/2} t_+^i $ and we can introduce $J_+=t_+^i\omega_i$. If $J_+$ lies in the interior of the K\"ahler cone, then
$J_+^3>0$ and the leading scalings are controlled by $J_+$. One finds
\[
J\sim H^{1/2}J_+,
\qquad
\mathcal V\sim H^{3/2},
\qquad
v_C\sim H^{1/2}\,, \qquad v_D\sim H \,,
\]
for every fixed nonzero effective curve $C$ and effective divisor $D$. Hence
\begin{equation}
\begin{aligned}
 \tau_{\rm F1}&\sim\tau_{\rm D1}\sim H^{-3/2},
 &
 \tau_{{\rm D3}[C]}&\sim H^{-1},
 \\[1mm]
 \tau_{{\rm D5}[D]}&\sim\tau_{{\rm NS5}[D]}\sim H^{-1/2},
 &
 \tau_{{\rm D7}[X]}&\sim1,
 \qquad
  \frac{M_{\rm KK}^{\rm light}}{M_p} \sim H^{-1}.
\end{aligned}
\label{eq:intclass-D3nu3}
\end{equation}
Naively, the KK modes along the whole Calabi--Yau, with all curves growing as $\mathcal R_C\sim H^{1/4}$, are the lightest objects, $(M_{\rm KK}^{\rm light}/M_p)^2\sim H^{-2}$, compared with $\tau_{\rm F1}\sim H^{-3/2}$. No cycle shrinks and $g_s$ is constant, so this flow stays at large volume and fixed, possibly small, coupling. These scalings agree with the D3-string flow on $\mathbb P^3$ of \cite[sec.~6.3]{Lanza:2021udy}.

\item\emph{Growing direction, $\nu=2$.}
Consider instead $J_+^3=0$, $J_+^2\neq0$. The condition \eqref{eq:intclass-D3-cohomological} implies $2c \gamma_s=J_+^2$ and the class $C_s$ of the wrapped curves has zero intersection with $J_+$. The leading terms in the large-H expansion of $J^3 $ are proportional to $H^{1/2} J_+ \cdot \Sigma_0 $ and provide the scaling $\mathcal V \sim H^{1/2}$.\footnote{Since $\frac{d}{dH}\frac12J^2=c\,\gamma_s$, one has $\frac{d}{dH}J^3=3c\,J\cdot\gamma_s$. With $J=H^{1/2}J_++H^{-1/2}J_b+\dots$, and $J_+^3=0$, this gives $\frac{d}{dH}J^3\simeq\frac32H^{-1/2}J_+^2J_b$, so that $J^3\simeq3H^{1/2}J_+^2J_b$. Comparing with the direct expansion gives $J_+\cdot\Sigma_0=J_+^2J_b$, which is non-zero by assumption.} Any probe curve $C$ satisfying $J_+\cdot C>0$ grows 
\begin{equation}
    v_C\sim H^{1/2},
\qquad
\tau_{{\rm D3}[C]}\sim1.
\end{equation}
By contrast, the source curves have zero intersection with $J_+$, $J_+\cdot C_s = 0$, and shrink
\begin{equation}
 v_{C_s}\sim H^{-1/2},
 \qquad
 \tau_{{\rm D3}[C_s]}\sim H^{-1}.
 \label{eq:intclass-D3nu2-source}
\end{equation}
The behavior of a general curve with $J_+\cdot C=0$ depends on the subleading terms in the large-$H$ expansion of $J$ and is not fixed by the leading class $J_+$ alone. We can consider an ansatz $t^i \sim H^{1/2} (t_+^i + H^{-1} t^i_b + \dots )$ and solve for the $t^i_b$ expanding \eqref{eq:intclass-D3-D7-solution} in powers of $H$. If we assume then that $J_b = t_b^i\omega_i$ is in the interior of the K\"ahler cone, $J_b \cdot C > 0$ and  we recover the scaling \eqref{eq:intclass-D3nu2-source}. For $J_b$ only nef, we can have curves whose volume is controlled by a lower power of $H$ and it is shrinking faster than \eqref{eq:intclass-D3nu2-source}. For divisors, the decomposition \eqref{eq:intclass-D3-cohomological} allow us to conclude
\begin{equation}
 (v_D,\tau_{{\rm D5}[D]},\tau_{{\rm NS5}[D]})\sim
 \begin{cases}
 (H,H^{1/2},H^{1/2}),
    & D\cdot C_s>0,
 \\[1mm]
 (1,H^{-1/2},H^{-1/2}),
    & D\cdot C_s=0\,,
 \end{cases}
 \label{eq:intclass-D3nu2-divisors}
\end{equation}
where in the second line we used that $J^2 \cdot D > 0 $ and the positivity  of the divisor volume to conclude that $D\cdot\Sigma_0>0$ . The remaining tensions obey
\begin{equation}
 \tau_{\rm F1}\sim\tau_{\rm D1}\sim H^{-1/2},
 \qquad
 \tau_{{\rm D7}[X]}\sim1 \,, \qquad  \frac{M_{\rm KK}^{\rm light}}{M_p}
 \sim H^{-1/2} \,,
 \label{eq:intclass-D3nu2-other-tensions}
\end{equation}
where we used that, since $J_+\neq0$ is nef, there is a curve with $J_+\cdot C>0$, so that the largest radius in string units is $\mathcal R_{\max}\sim H^{1/4}$. Naively, the lightest objects are D3-strings on curves with $J_+\cdot C=J_b\cdot C=0$, if such curves exist. Otherwise, the KK modes on curves with $J_+\cdot C>0$, the source D3-string and the D3-strings on curves with $J_+\cdot C=0$, $J_b\cdot C>0$ all scale as $m^2/M_p^2\sim\tau\sim H^{-1}$. In both cases the source curves shrink, so the large-volume description breaks down near the core, and, since $g_s$ is constant, S-duality does not help.

\item \emph{Growing direction, $\nu = 1$.}
We now assume $J_+^2\equiv0$. Then $\gamma_s \propto J_+\cdot J_b$ and the solutions of \eqref{eq:intclass-D3-D7-solution} requires the ansatz $t^i \sim H t_+^i + t_b^i + \dots$. In this branch
$\mathcal V\sim H$. For probe curves and
divisors one obtains
\begin{equation}
\begin{aligned}
 (v_C,\tau_{{\rm D3}[C]})&\sim
 \begin{cases}
 (H,1),&J_+\cdot C>0,\\
 (1,H^{-1}),&J_+\cdot C=0, \,\, J_b \cdot C > 0 \,,
 \end{cases}
 \\[2mm]
 (v_D,\tau_{{\rm D5}[D]},\tau_{{\rm NS5}[D]})&\sim
 \begin{cases}
 (H,1,1),&D\cdot C_s>0,\\
 (1,H^{-1},H^{-1}),&D\cdot C_s=0\,.
 \end{cases}
\end{aligned}
\label{eq:intclass-D3nu1-probes}
\end{equation}
A curve with $J_+ \cdot C = J_b \cdot C = 0$ is not excluded. Its volume is then set by the next term in the large-$H$ expansion of $J$, and shrinks at least as $v_C\sim H^{-1}$. Moreover,
\begin{equation}
    \tau_{\rm F1}\sim\tau_{\rm D1}\sim H^{-1},
\qquad
\tau_{{\rm D7}[X]}\sim1 \,, \qquad \frac{M_{\rm KK}^{\rm light}}{M_p}\sim H^{-1} \,,
\end{equation}
where for the KK scale we used a curve with $J_+\cdot C>0$, which exists since $J_+\neq0$ is nef. Naively, the lightest objects are the KK modes on curves with $J_+\cdot C>0$, with $\mathcal R_C\sim H^{1/2}$, and, if they exist, the D3-strings on curves with $J_+ \cdot C = J_b \cdot C = 0$, with $m^2/M_p^2\sim\tau\lesssim H^{-2}$. If instead no such curve exists, i.e.\ if $J_b\cdot C>0$ for all curves with $J_+\cdot C=0$, no cycle shrinks: the curves either grow, $v_C\sim H$, or have fixed volumes, $v_C\sim1$, which can be taken large by choosing the constants $h_i$, and $g_s$ is constant. As for $\nu=3$ with $J_+$ ample, the flow then stays at large volume and fixed, possibly small, coupling.

\end{itemize}

\paragraph{D5 source:}
We select $\tilde H^0=H_i=0=q_0$, with $H_0<0$ constant. Then $X^i$ is imaginary and $X^0$ real. We identify the positive harmonic function $H$ via $\tilde H^i=\tilde h^i+Hp^i$. Let $J_+=p^i\omega_i=[D_s]\in H^{1,1}(X,\mathbb R)$ be the Poincar\'e dual of the total class $D_s$ of the divisors wrapped by the D5-branes, and define $J_b=\tilde h^i\omega_i$. We assume that $J_+$ is non-zero and nef, and that $HJ_++J_b$ lies in the interior of the K\"ahler cone along the flow, which holds e.g.\ if $J_b$ is ample.  The attractor equations imply $t^i = \tilde{H}^i/(2X^0)$ and
\begin{equation}
    (X^0)^2 \propto  \mathcal{K}_{ijk} \tilde{H}^i \tilde{H}^j \tilde{H}^k \sim H^{\nu} \,,
\end{equation}
where $\nu = \nu(J_+)$ and we assumed $J_+^\nu J_b^{3-\nu}>0$. This gives
\begin{equation}
    J = \frac{1}{2 X^0} (H J_+ + J_b)\sim H^{1-\nu/2}J_+ \,, \qquad g_s  \sim  H^{-\nu/2} \,.
\end{equation}
Only the components $t^i$ with $p^i\neq0$ scale as $H^{1-\nu/2}$, while the others scale as $H^{-\nu/2}$. Since $g_s\to0$ and some cycles shrink in all cases below, the caveats of the introduction apply. In the S-dual frame, $t'^i=t^i/g_s\propto\tilde H^i$ no longer shrink, but $g'_s\sim H^{\nu/2}$ is strongly coupled.
For any fixed nonzero effective probe curve $C$ and divisor $D$,
the exact classical volume formulas are
\begin{equation}
\begin{aligned}
 v_C& \propto \frac{1}{X^0} \bigl(H\,J_+ \cdot C+J_b\cdot C\bigr),
 \\
 v_D& \propto \frac{1}{(X^0)^2}
       \bigl(H^2 J_+^2 \cdot D +2H J_+J_b \cdot D + J_b^2 \cdot D \bigr) \,,
    \\
 \mathcal{V}& \propto \frac{1}{(X^0)^3}
       \int H^3 J_+^3 + 3H^2 J_+^2 J_b  +3H J_+J_b^2 + J_b^3 \,.
\end{aligned}
\label{eq:intclass-D5-probe-volumes}
\end{equation}
\begin{itemize}
    \item \emph{Interior source class $\nu=3$.}
If $J_+$ lies in the interior of the K\"ahler cone, then $J_+^3>0$
and every fixed nonzero effective probe scales homogeneously
\begin{equation}
 J\sim H^{-1/2}J_+,
 \qquad
 \mathcal V\sim g_s\sim H^{-3/2},
 \qquad
 v_C\sim H^{-1/2},
 \qquad
 v_D\sim H^{-1}.
 \label{eq:intclass-D5nu3-volumes}
\end{equation}
All curves have radii $\mathcal R_C\sim H^{-1/4}$ in string units, and the KK modes along any of them have the same scaling.
The reduced tensions and KK scale then obey
\begin{equation}
\begin{aligned}
 \tau_{\rm F1}&\sim\tau_{{\rm D7}[X]}\sim H^{-3/2},
 &
 \tau_{\rm D1}&\sim1,
 &
 \tau_{{\rm D3}[C]}&\sim H^{-1/2},
 \\[1mm]
 \tau_{{\rm D5}[D]}&\sim H^{-1},
 &
 \tau_{{\rm NS5}[D]}&\sim H^{1/2},
 &
 \frac{M_{\rm KK}}{M_p}&\sim H^{-1/2}.
\end{aligned}
\label{eq:intclass-D5nu3-probes}
\end{equation}
Naively, the F1 and the wrapped D7 are the lightest objects. However, all cycles shrink and the classical description breaks down near the core.

\item \emph{Nef and big boundary source $\nu=3$.}
If instead $J_+^3>0$ but $J_+$ lies on the boundary of the K\"ahler cone,
the subleading term $J_b$ in \eqref{eq:intclass-D5-probe-volumes}
must be retained. The powers of $\mathcal V$, $g_s$,
$\tau_{\rm F1}$, $\tau_{\rm D1}$ and $\tau_{{\rm D7}}$ are
the same as in \eqref{eq:intclass-D5nu3-volumes} and \eqref{eq:intclass-D5nu3-probes}, but \eqref{eq:intclass-D5-probe-volumes} gives
\begin{subequations}
\begin{align}
 (v_C,\tau_{{\rm D3}[C]})&\sim
 \begin{cases}
 (H^{-1/2},H^{-1/2}),&J_+\cdot C>0,\\
 (H^{-3/2},H^{-3/2}),&J_+\cdot C=0,
 \end{cases} \label{eq:intclass-D5nu3-boundary-probesV1}
 \\[2mm]
 (v_D,\tau_{{\rm D5}[D]},\tau_{{\rm NS5}[D]})&\sim
 \begin{cases}
 (H^{-1},H^{-1},H^{1/2}),&J_+^2 \cdot D >0,\\
 (H^{-2},H^{-2},H^{-1/2}),&J_+^2 \cdot D=0,\ J_+J_b \cdot D>0,\\
 (H^{-3},H^{-3},H^{-3/2}),&J_+^2 \cdot D =J_+J_b \cdot D=0.
 \end{cases} \label{eq:intclass-D5nu3-boundary-probesV2}
\end{align}
\end{subequations}
Since $HJ_++J_b$ is K\"ahler, curves with $J_+\cdot C = 0$ have $J_b \cdot C > 0$. The KK modes on curves with $J_+\cdot C>0$, with $\mathcal R_C\sim H^{-1/4}$, remain at $H^{-1/2}M_p$.
Curves with $J_+\cdot C=0$ give D3 scales tied to F1 and wrapped D7. If effective divisors satisfying $J_+^2 \cdot D = 0$ exist, naively their D5-strings are the lightest objects, the lightest ones being those on divisors with $J_+J_b\cdot D=0$, with $\tau\sim H^{-3}$. As in the interior case, all cycles shrink and the classical description breaks down near the core.

\item \emph{Source divisor $\nu=2$.}
Consider $J_+^3=0$ and $J_+^2\not\equiv0$, with
$J_+^2J_b>0$. Then $\mathcal{V}\sim H^{-1}$. Probe volumes are 
\begin{subequations}
\begin{align}
 (v_C,\tau_{{\rm D3}[C]})&\sim
 \begin{cases}
 (1,1),&J_+\cdot C>0,\\
 (H^{-1},H^{-1}),&J_+\cdot C=0,
 \end{cases}
 \\[2mm]
 (v_D,\tau_{{\rm D5}[D]},\tau_{{\rm NS5}[D]})&\sim
 \begin{cases}
 (1,1,H),&J_+^2\cdot D>0,\\
 (H^{-1},H^{-1},1),
   &J_+^2\cdot D=0,\ J_+J_b\cdot D>0,\\
 (H^{-2},H^{-2},H^{-1}),
   &J_+^2\cdot D=J_+J_b\cdot D=0.
 \end{cases} \label{eq:intclass-D5nu2-probes}
\end{align}
\end{subequations}
The total source class $D_s$ belongs to the second divisor class, since
$[D_s]=J_+$, $J_+^3=0$ and $J_+^2J_b>0$.
The remaining tensions obey
\begin{equation}
 \tau_{\rm F1}\sim\tau_{{\rm D7}[X]}\sim H^{-1},
 \qquad
 \tau_{\rm D1}\sim1\,, \qquad \frac{M_{\rm KK}^{\rm light}}{M_p}
 \sim H^{-1/2} \,,
 \label{eq:intclass-D5nu2-other-tensions}
\end{equation}
where we used a curve with $J_+\cdot C>0$, which exists since $J_+\neq0$ is nef, so that $\mathcal R_{\max}\sim1$. The KK modes on curves with $J_+\cdot C>0$ and the source-D5, F1 and wrapped-D7 scales are parametrically
equal. Depending on the structure of the CY, they match certain D3, and D5 scales. If an effective divisor in the last row of \eqref{eq:intclass-D5nu2-probes} exists, naively the D5-strings wrapped on it are the lightest objects. However, curves with $J_+\cdot C=0$ shrink, $v_C\sim H^{-1}$, and the classical description breaks down near the core.

\item \emph{Source divisor $\nu=1$.}
We now assume $J_+^2\equiv0$, with $J_+J_b^2>0$. $\mathcal{V} \sim H^{-1/2}$. For probe curves and divisors one obtains
\begin{subequations}
\begin{align}
 (v_C,\tau_{{\rm D3}[C]})&\sim
 \begin{cases}
 (H^{1/2},H^{1/2}),&J_+\cdot C>0,\\
 (H^{-1/2},H^{-1/2}),&J_+\cdot C=0,
 \end{cases}
 \\[2mm]
 (v_D,\tau_{{\rm D5}[D]},\tau_{{\rm NS5}[D]})&\sim
 \begin{cases}
 (1,1,H^{1/2}),&J_+J_b\cdot D>0,\\
 (H^{-1},H^{-1},H^{-1/2}),&J_+J_b\cdot D=0.
 \end{cases}
\end{align}
\label{eq:intclass-D5nu1-probes}
\end{subequations}
The total source class $D_s$ lies in the second divisor class because
$[D_s]=J_+$ and $J_+^2J_b=0$.
Moreover,
\begin{equation}
 \tau_{\rm F1}\sim\tau_{{\rm D7}[X]}\sim H^{-1/2},
 \qquad
 \tau_{\rm D1}\sim1 \,, \qquad  \frac{M_{\rm KK}^{\rm light}}{M_p}
 \sim H^{-1/2} \,,
 \label{eq:intclass-D5nu1-other-tensions}
\end{equation}
where we used a curve with $J_+\cdot C>0$, which exists since $J_+\neq0$ is nef, so that $\mathcal R_{\max}\sim H^{1/4}$. Naively, the KK modes on curves with $J_+\cdot C>0$, with $\mathcal R_C\sim H^{1/4}$, and the D5-strings on divisors with $J_+J_b\cdot D=0$, including the source, are the lightest objects, with $m^2/M_p^2\sim\tau\sim H^{-1}$. Curves with $J_+\cdot C=0$ shrink, $v_C\sim H^{-1/2}$, so the classical description again breaks down near the core.
\end{itemize}

\paragraph{D7 source:}
We set $H_0 = \tilde{H}^i = 0$, with $H_i=h_i$ constant, and identify $H=\tilde H^0$. If we can solve \eqref{eq:intclass-D3-D7-solution} with $H_i=h_i$, we obtain $J = H^{-1/2} J_b$ with $J_b$ in the interior of the K\"ahler cone. Then $\mathcal V\sim H^{-3/2}$ and $g_s\sim H^{-1}$ and we get
\begin{equation}
\begin{aligned}
 \tau_{\rm F1}&\sim H^{-1/2},&
 \tau_{\rm D1}&\sim H^{1/2},&
 \tau_{{\rm D3}[C]}&\sim1,\\
 \tau_{{\rm D5}[D]}&\sim H^{-1/2},&
 \tau_{{\rm NS5}[D]}&\sim H^{1/2},&
 \tau_{{\rm D7}[X]}&\sim H^{-1}.
\end{aligned}
\label{eq:intclass-D7-probes}
\end{equation}
The KK masses along any curve, all with $\mathcal R_C\sim H^{-1/4}$, remain finite in Planck units, while, naively, the source D7-string is the lightest object among the probes. However, all cycles shrink, and the classical description breaks down near the core. S-duality gives $t'^i=t^i/g_s\sim H^{1/2}$, but $g'_s\sim H$ is strongly coupled. This is the weak-coupling D7-string flow of \cite[sec.~6.3]{Lanza:2021udy}, along which the Einstein-frame volumes stay fixed while all string-frame volumes vanish. 

\section{Deflected flows in complexified moduli space}\label{sec:appendixB}

In this appendix we take an exploratory approach to how quantum corrections might deflect flows in the complexified hypermultiplet moduli space. The idea is to integrate in the instantons whose actions become small along the classical flow, one sector at a time, and to ask whether the resulting corrections can prevent the flow from reaching the regions in which no duality frame is under control. We do not solve the BPS equations with the corrected metric: we only use the classical flow to order the instanton sectors, and estimate the effect of the first ones through dualities.

This analysis does not apply directly to the instanton saddles of section~\ref{sec:instantonstorus}. There, the RR axions are required to be real at infinity, so that the saddles are T-dual to Lorentzian black holes, and they acquire imaginary parts along the flow, see \eqref{eq:instantonaxions}. Since the holomorphic combinations $T_I$ are constant along the flow, see \eqref{eq:complexaction}, while $s_I\to0$ at the core, the real parts of the actions of an instanton and of an anti-instanton of charge $N_I$, which add up to $4\pi N_I s_I(r)$, cannot both remain positive all the way to the core. With real asymptotic axions, the instanton action stays at its asymptotic value, while the anti-instanton action, $2\pi N_I\left(2s_I(r)-s_{I,\infty}\right)$, vanishes already at $H_I=2h_I$, i.e.\ at finite radius and, for $I=0$, at $g_s=2g_s^\infty$, before the hierarchy discussed below can develop. Relaxing the boundary conditions, by allowing complex asymptotic values of the axions, can allow one to move this radius. In the rest of this appendix we therefore ignore the imaginary parts of the RR axions, except in the last paragraph.

For K\"ahler moduli finite but large, we can introduce a hierarchy among the instanton actions along the flow. The idea is that, starting at a point away from the core, where $g_s$ is already large but still finite, some instantons may have actions parametrically smaller than others. As we move further along the trajectory towards the core, we can then identify successive regions in which more and more instantons become relevant. Instead of following the trajectory all the way to the core, we stop at the first regions, where only the first set of instantons contribute significantly. We study the deflections caused by these instantons, and then continue along the corrected trajectory. In this sense, we assume that we can safely integrate in the effects of the first tower of instantons and neglect the others. This is in the spirit of the emergence proposals criterion of \cite{Blumenhagen:2023xmk, Blumenhagen:2023yws, Blumenhagen:2023tev}, where we integrate out full tower of states whose mass scale is below a threshold, there identified with the species scale.\footnote{Note that it necessarily makes sense to only consider the first tower of instantons in the first region \eqref{eq:regions1} when $T\to \infty$, via a possible rescaling of charges.} Here two possibilities may arise. First, the corrections in this initial region may be small, in which case we would need to proceed to subsequent regions. As discussed in section~\ref{sec:quantumCorrections}, however, we would eventually encounter the same difficulty in evaluating the corrections. Alternatively, the deflection may be sufficiently large to prevent the corrected trajectory from entering the subsequent regions altogether. The possibility of such deflections was discussed also in \cite{Baume:2019sry}, where it was pointed out that regions accessible along the classical trajectory may no longer be, once corrections are taken into account.

Let us keep track of the corrections in the only case in which the mechanism could be applied, i.e. for flows with the same profiles of $g_s$ and $t^i$ as the instantons T-dual to finite-size black holes. We assume a generic Calabi--Yau and that the K\"ahler moduli are all of order $t^i \sim T \gg 1$. If we ignore the RR axion the actions \eqref{eq:non-perturbative-contributions-general} imply that classically we have a first region corresponding to $g_s < 1$ in which all corrections are suppressed. We then have five more regions each one of them associated with a new instanton with action becoming parametrically small $S_{\rm inst} \lesssim 1 $\footnote{More precisely, we assume the tower of instanton corrections to have a finite radius of convergence. An instanton is becoming unsuppressed in a region in which the contribution $e^{-S_{\text{inst}}}$ of the lightest instanton of the tower is bigger than the radius of convergence. We identify such threshold with 1 because for large enough $T$ the hierarchy is preserved.\label{fn:footnoteSuguale1}}

\begin{equation}\label{eq:regions1}
	\mathrm{IIB}:\hspace{1cm}
	\begin{aligned}
		1 \lesssim g_s \lesssim T: & \quad 
		\text{D}(-1) \\
		T \lesssim g_s \lesssim T^{3/2}: & \quad  
		\text{D}(-1),\, \text{D1} \\
		T^{3/2} \lesssim g_s \lesssim T^{2}: & \quad  
		\text{D}(-1),\, \text{D1},\, \text{NS5} \\
		T^{2} \lesssim g_s \lesssim T^{3}: & \quad  
		\text{D}(-1),\, \text{D1},\, \text{NS5},\, \text{D3} \\
		T^{3} \lesssim g_s: & \quad  
		\text{D}(-1),\, \text{D1},\, \text{NS5},\, \text{D3},\, \text{D5}
	\end{aligned}
\end{equation}
For an Euclidean instanton, the axions are running along the imaginary direction, decreasing monotonically and eventually reaching a constant value.\footnote{This can be easily determined replacing into $\xi^I$ of \eqref{eq:dictionary2} the harmonic functions dependence \ref{sec:explicitBHs} and noticing that $\xi^I \sim \mathcal{O}(H^{-1})$. A constant value can be approached as in \eqref{eq:instantonaxions} recalling that $\xi^I$ is defined up to an integration constant.} In presence of a non-vanishing imaginary RR axion, the instanton actions are allowed to vanish at a finite value of $g_s$. We assume that the imaginary parts of the axions are small enough not to modify the hierarchy of \eqref{eq:regions1} (as we mentioned at the beginning of this appendix, this requires boundary conditions different from those of our the instanton we built). To understand which region can provide a significant deflection, we consider the S-dual frame. Under S-duality  
\begin{equation} \label{eq:SdualMap}
    g_s' = C_0^2 g_s +g_s^{-1}\,, \qquad t^i{}' = \sqrt{C_0^2 + g_s^{-2}} t^i \,.
\end{equation}
We start analyzing the case with $C_0 = 0$ (see later for a non-vanishing $C_0$). In the S-dual frame the classical trajectory explores weak ten-dimensional coupling $g_s' \sim g_s^{-1}$, shrinking volumes $t^i{}' \sim T/g_s$ and strong four-dimensional coupling $g_{4,s}'\sim g_s^{1/2} T^{-3/2} $. Using $g_s \rightarrow \infty$ to parametrize the S-dual classical trajectory, the regions \eqref{eq:regions1} become
\begin{equation}\label{eq:regions2}
	\mathrm{IIB}':\hspace{1cm}
	\begin{aligned}
		1 \lesssim g_s \lesssim T: & \quad
		\text{nothing} \\
		T \lesssim g_s \lesssim T^{3/2}: & \quad
		\text{WS} \\
		T^{3/2} \lesssim g_s \lesssim T^{2}: & \quad
		\text{WS},\, \text{D5} \\
		T^{2} \lesssim g_s \lesssim T^{3}: & \quad
		\text{WS},\, \text{D5},\, \text{D3} \\
		T^{3} \lesssim g_s: & \quad
		\text{WS},\, \text{D5},\, \text{D3},\, \text{NS5}
	\end{aligned}
\end{equation}
If the imaginary axions were not affecting the hierarchy of \eqref{eq:regions1}, they will not affect the one of \eqref{eq:regions2}. The absence of unsupressed instantons in the region $1 \lesssim g_s \lesssim T$ in the S-dual description strongly suggests that we need to enter at least the second region to have a strong deflection, i.e. take into account world-sheet instantons effects. In order to understand what happens we change once more duality frame  performing a mirror symmetry. Our trajectory now is contained within type IIA complex structure moduli space. The four-dimensional dilaton $g_{4,s}'$ is left invariant and the K\"ahler moduli $t^i{}'$  are related to the algebraic complex-structure coordinate $\zeta^i$ via the logarithmic map $z^i{}' = \frac{1}{2\pi i} \log \zeta^i $. The latter is valid in the classic approximation, close to the large volume point $\zeta^i = 0$. The endpoint of the trajectories corresponding to vanishing K\"ahler moduli is $\zeta^i = 1$. 

Away from the large volume point the mirror map gets corrected. As long as the only relevant instantons effects on the IIB side are due to D1,D$(-1)$ and WS instantons, we have the corrected map $z^i{}' = X^i(\zeta)/X^0(\zeta)$, where $X^I(\zeta)$ are periods of the 3-cycles of the mirror Calabi--Yau $Y$
\begin{equation}
	X^I(\zeta) = \int_{A^I} \Omega_Y \,,
\end{equation}
with $\Omega_Y$ the holomorphic 3-form of the mirror CY and $A^I \in H_3 (Y,\mathbb{Z})$ form one half of a symplectic basis of integral 3-cycles. One can then use the corrected mirror map to determine what happens to the  $t^i{}'$ along the trajectory $g_s \rightarrow \infty$, i.e. approaching $\zeta^i = 1$. What one can learn from examples is that there is no unique answer. In general, the $X^I$ can be determined as solution of Picard-Fuchs (PF) equations and will provide a correction of the form
\begin{equation}
	z^i{}' = \frac{X^i}{X^0} = \frac{1}{2 \pi i } \left[\log \zeta^i +  f^i(\zeta)\right] \,,\qquad 
	f^i(\zeta)
	=
	\sum_{\substack{n_i \geq 0}}
	c^i_{n_1,\ldots,n_r}
	\prod_{j=1}^{r}(\zeta^j)^{n_j}\,,
\end{equation}
where $c^i_{0,\ldots,0} = 0 $ and $r = h^{2,1}(Y)$. However, which is the behaviour of $t^i{}'(1)$ depends on the details of the system we are studying. Let us illustrate this with some examples \cite{Walcher:2006rs, Candelas:1993dm, Hosono:1993qy}.  The simplest model we can consider is the one-parameter quintic, whose mirror dual has $h^{2,1} = 1$ \cite{Walcher:2006rs}. The series one gets solving the PF equations has a radius of convergence smaller than $1$ due to a conifold point. In order to reach $\zeta^1 = 1$ we have to perform an analytic extension and choose a trajectory within the fundamental region continued from large volume. Along every trajectory the volume encoded in the K\"ahler moduli is always positive. Indeed, the quintic admits a point of minimal (positive) volume corresponding to the Gepner point. The existence of such minimal volume point is not always guaranteed. For instance, the two-parameter octic $X_8(1,1,2,2,2)$ \cite{Candelas:1993dm, Hosono:1993qy} does not have a positive minimal volume point and one has to verify explicitly the value of the K\"ahler moduli. To avoid singularities (conifold points and gauge enhancement) we select the path with $\zeta^1 = \zeta^2 = \zeta$ which from the large volume point $\zeta = 0$ reaches the unitary circle along $e^{-i\pi/4}$ and then follows it till $\zeta =1$. Integrating numerically the PF equations one can verify that along the whole trajectory the volumes are positive, i.e. $t^1 > 0$, $t^2 > 0$. Finally, it is instructive to consider the two-parameter, degree-14 CY hypersurface $X_{14}(1,2,2,2,7)$ \cite{Candelas:1993dm, Hosono:1993qy}. Following the same trajectory of before and performing a numerical evaluation one can verify that we reach a negative volume in $\zeta = 1$  and we are outside of the regime of validity of our approximation. We conclude that depending on the details of the Calabi--Yau, a classical trajectory with shrinking K\"ahler moduli can stop running at a finite quantum volume but there is no universal behaviour for these flows as we have for the strings.

Let us continue our analysis for the trajectories of instantons which eventually reach finite volume. In \cite{Baume:2019sry, Castellano:2023jjt} having a finite volume as endpoint is simple to interpret. Their classical trajectories originally explore weak $g_s$ and $g_{4,s}$ regimes; the presence of a finite volume is modifying the scaling of the light states at infinite distance, but it is not changing the nature of the limit.  In our case, the analysis on the K\"ahler moduli does not fix the scaling of the dilaton. The original classical trajectory has weak $g_s$ and strong $g_{4,s}$ in IIB'; depending on the parametrization we are following when computing the quantum correction, i.e. requesting a weak $g_s$ or a strong $g_{4,s}$, we explore opposite regimes at the core approaching the finite quantum volume. Which is the correct infinite distance limit realized by an instanton can be determined only analyzing the asymptotic scalings in the corrected EOMs and goes beyond the scope of this work.

We now comment briefly on the two possibilities. In case the IIB’ trajectory surviving is the one with strong $g_{4,s}$ coupling, $g_s$ is deflected to strong coupling and we have essentially exchanged IIB and IIB’. The corrections computed with mirror symmetry do not prevent us from entering a region where instantons corrections such as NS5 or D5 instanton effects become relevant. No duality frame is under control near the core, and the known machinery does not allow us to follow the instanton into this region. In case the trajectory realized in IIB’ is the one with weak $g_s$ coupling, the finite volume deflects $g_{4,s}$ to weak coupling in the core. The IIB’ limit is then an emergent F1 string limit along which all instantons are suppressed except for WS instantons which are $\mathcal{O}(1)$.\footnote{Notice that given that the instanton actions are growing large and the RR axions are finite, they will not vanish approaching the core.} These effects do not introduce further deflection\footnote{More precisely, they introduce at most a finite rescaling of the quantities, not modifying whether something vanishes, diverges or it is finite at the core.}(see section~\ref{sec:QCtoF1}). The regions we were originally encountering along the classical trajectory have been removed. In the initial IIB frame we have instead a trajectory exploring large volume and strong $g_s$, with $t^i \sim g_s$ and weak $g_{4,s}$. This is an emergent D1 string limit with D$(-1)$ and D1 instantons unsuppressed near the core. These corrections modify the scalings of the tensions (see section~\ref{sec:QCtoD1}), but do not change the nature of the limit.

To conclude, let’s briefly discuss the case with non-vanishing $C_0$.  Picking $C_0 = i |C_0|$  we get that the instantons carrying D$(-1)$ charge are proportional to  $g_s^{-1} \pm |C_0|$ and some of them vanish for $g_s = 1/|C_0|$.  Even before reaching the strong coupling regime we can expect a large deviation at $g_s = 1/|C_0|$, which is realized at finite spacetime distance. To maintain computational control, we assume that the imaginary axions boundary conditions are selected in such a way there is no other instanton action vanishing before $g_s = 1/|C_0|$. The S-dual trajectory explores weak coupling $g_s' =  \lambda^2 g_s$, shrinking volume $t^i{}' =  \lambda t^i$ and divergent axion $C_0' = - C_0 \lambda^{-2}$ with $\lambda = \sqrt{g_s^{-2} + C_0^2} \rightarrow  0$; the other RR axions are reshuffled but they keep being finite. WS and D5 instantons have a vanishing action along this trajectory. Again we can setup a hierarchy and study the effect of the WS correction in the mirror dual. As before, there are some cases in which the K\"ahler moduli should stop at a finite volume, depending on the CY we are considering. However, now we have weak coupling in IIB' for both $g_{4,s}$ and $g_s$, therefore there is no ambiguity and we can safely assume that the corrected trajectory still has it. We have then an F1 limit and no instanton has vanishing action. Usually we only have WS instanton which have non-divergent action; in this case we also have certain D$(-1)$ branes, due to the divergent $C_0$ axion. Indeed the combination $g_s^{-1}{}' - i C_0' = \lambda^{-2}\left(g_s^{-1} - |C_0|\right) $ entering in one of the D$(-1)$ instanton action is finite. In the original IIB frame the trajectory has finite $g_s < 1/|C_0|$, finite $C_0$, divergent volumes $t^i$ and weak $g_{4,s}$ coupling. Along this trajectory all instanton actions diverge except for the D$(-1)$ which were vanishing also originally. However, they will not induce further deflections now because $t^i \sim 1/|\tau|$ and the cubic scaling in $t^i$ of the classic piece of the contact potential \eqref{eq:contactPotentialCLassic} is matched by the cubic scaling in $1/|\tau|$ of the correction term \eqref{eq:chiSL2Z}. One can finally verify that the lightest string is the D1 extracting the dependence on the K\"ahler moduli from \eqref{eq:intclass-tensions}, but recalling that in the presence of a $C_0$ axion we have
\begin{equation}
    \mathcal{T}_{p,q} = \mathcal{T}_{\rm F1} |p - q \tau| \,.
\end{equation}
So the D1 string is tensionless and has tension smaller than the fundamental string in the limit $\tau \rightarrow 0$.

\bibliographystyle{JHEP}
\inputencoding{latin2}
\bibliography{references2}
\inputencoding{utf8}
\end{document}